\documentclass[twocolumn,twocolappendix]{aastex701}

\usepackage{amsmath}
\usepackage{xspace}
\usepackage{pifont}
\usepackage{array}
\usepackage{grffile}

\graphicspath{{./Images/}}
\DeclareGraphicsExtensions{.bmp,.png,.jpg,.pdf}

\definecolor{lightred}{RGB}{255,153,153}
\definecolor{darkgreen}{RGB}{60,120,80}
\newcommand{\cmark}{\ding{51}}
\newcommand{\xmark}{\ding{55}}

\newcommand{\PaperMP}{Park et al. (under review)\xspace}
\newcommand{\PaperMPcitep}{(Park et al. under review)\xspace}
\newcommand{\rhalf}{r_{\text{half}}}

\newcommand{\Mhalo}{M_{\text{halo}}}
\newcommand{\kappaCR}{\kappa_{\text{CR}}}
\newcommand{\dx}{\Delta x}
\newcommand{\dxres}{\Delta x_{\text{min}}}
\newcommand{\Lya}{{\rm Ly\alpha}}
\newcommand{\Mstar}{\mathrm{M}_{*}}
\newcommand{\magUV}{\mathrm{M}_{\text{UV}}}
\newcommand{\rhoUV}{\rho_{\text{SFR, UV}}}
\newcommand{\SFR}{\mathrm{SFR}}
\newcommand{\SFRHalpha}{\mathrm{SFR}_{\text{H}\alpha}}
\newcommand{\SFRUV}{\mathrm{SFR}_{\text{UV}}}
\newcommand{\SFRthree}{\mathrm{SFR}_{\text{3 Myr}}}
\newcommand{\SFRten}{\mathrm{SFR}_{\text{10 Myr}}}
\newcommand{\SFRfifty}{\mathrm{SFR}_{\text{50 Myr}}}
\newcommand{\SFRhundred}{\mathrm{SFR}_{\text{100 Myr}}}
\newcommand{\Bburstup}{3}
\newcommand{\Bburstdown}{50}
\newcommand{\Bburst}{B_{\text{\Bburstup/\Bburstdown}}}

\newcommand{\alphaM}{\alpha_{\Mstar}}
\newcommand{\alphaUV}{\alpha_\text{UV}}
\newcommand{\JWST}{\textit{JWST}\xspace}
\newcommand{\azahar}{\textsc{Azahar}\xspace}
\newcommand{\Azahar}{\textsc{Azahar}\xspace}

\iffalse
\newcommand{\HDtext}{\textcolor{\usecolour}{\textit{Standard HD}}}
\newcommand{\HDBoosttext}{\textcolor{\usecolour}{\textit{Calibrated HD}}}
\newcommand{\RTnsCRiMHDtext}{\textcolor{\usecolour}{\textit{Non-Thermal}}}
\else
\newcommand{\HDtext}{\textit{Standard HD}}
\newcommand{\HDBoosttext}{\textit{Calibrated HD}}
\newcommand{\RTnsCRiMHDtext}{\textit{Non-Thermal}}
\fi
\newcommand{\HDrun}{\HDtext\xspace}
\newcommand{\HD}{\HDtext\xspace}
\newcommand{\HDBoostrun}{\HDBoosttext\xspace}
\newcommand{\HDBoost}{\HDBoosttext\xspace}
\newcommand{\RTnsCRiMHDrun}{\RTnsCRiMHDtext\xspace}
\newcommand{\RTnsCRiMHD}{\RTnsCRiMHDtext\xspace}
\newcommand{\fullphys}{\textit{full-physics}\xspace}
\newcommand{\Via}{\textsc{Via}\xspace}

\newcommand{\pandora}{Pandora\xspace}

\newcommand{\ramses}{\textsc{ramses}}
\newcommand{\Zsun}{\,\mathrm{Z_\odot}}
\newcommand{\Msun}{\,\mathrm{M_\odot}}
\newcommand{\erg}{\,\mathrm{erg}}
\newcommand{\cm}{\,\mathrm{cm}}
\newcommand{\km}{\,\mathrm{km}}
\newcommand{\pc}{\,\mathrm{pc}}
\newcommand{\ckpc}{\,\mathrm{ckpc}}
\newcommand{\kpc}{\,\mathrm{kpc}}
\newcommand{\cMpc}{\,\mathrm{cMpc}}
\newcommand{\Mpc}{\,\mathrm{Mpc}}
\newcommand{\Kelvin}{\,\mathrm{K}}
\newcommand{\K}{\Kelvin}
\newcommand{\s}{\,\mathrm{s}}

\newcommand{\yr}{\,\mathrm{yr}}
\newcommand{\Myr}{\,\mathrm{Myr}}
\newcommand{\Gyr}{\,\mathrm{Gyr}}

\newcommand{\Gauss}{\mathrm{G}}

\newcommand{\magab}{\,\mathrm{mag}}
\newcommand{\kmps}{\,\km\,\s^{-1}}

\shorttitle{Bursty First Galaxies in the Azahar Simulations}
\shortauthors{Martin-Alvarez et al.}

\begin{document}

\title{The Azahar Project: Non-Thermal Physics Drives Star Formation Burstiness and the Evolution of the UV Luminosity Density at Cosmic Dawn}

\author[sname=Martin-Alvarez,gname=Sergio,orcid=0000-0002-4059-9850]{Sergio Martin-Alvarez}
\affiliation{Kavli Institute for Particle Astrophysics \& Cosmology (KIPAC), Stanford University, Stanford, CA 94305, USA}
\email[show]{martin-alvarez@stanford.edu}
\correspondingauthor{Sergio Martin-Alvarez}

\author[sname=Wechsler,gname=Risa,orcid=0000-0003-2229-011X]{Risa H. Wechsler}
\affiliation{Kavli Institute for Particle Astrophysics \& Cosmology (KIPAC), Stanford University, Stanford, CA 94305, USA}
\affiliation{Department of Physics, Stanford University, Stanford, CA 94305, USA}
\affiliation{SLAC National Accelerator Laboratory, Menlo Park, CA 94025, USA}
\email{rwechsler@stanford.edu} 

\author[sname=Sijacki,gname=Debora,orcid=0000-0002-3459-0438]{Debora Sijacki}
\affiliation{Kavli Institute for Cosmology, Cambridge (KICC), University of Cambridge, Madingley Road, Cambridge CB3 0HA, UK}
\affiliation{Institute of Astronomy, University of Cambridge, Madingley Road, Cambridge CB3 0HA, UK}
\email{deboras@ast.cam.ac.uk}

\begin{abstract}
JWST observations, which have pushed the discovery and characterization of galaxies to cosmic dawn, have revealed significant deficiencies in state-of-the-art galaxy formation simulations, motivating the need for novel, physically grounded models. We present the first results from the \azahar suite of ten high-resolution ($\sim$20~pc), large-volume zoom-in cosmological simulations, which follow the formation of thousands of galaxies and progressively incorporate radiative transfer (RT), cosmic rays (CRs), and magnetohydrodynamics in addition to the `standard' baryonic physics. Our \fullphys model, which simultaneously includes RT and CRs on the fly, reproduces the observed UV luminosity function from high redshifts ($z \sim 14$) to cosmic noon ($z \sim 3$), as well as the evolution of the stellar mass function, the galaxy main sequence, and observed gas metallicities. It does so through a combination of burstier star formation, with enhanced variability on both long ($\gtrsim 50$~Myr) and short ($\lesssim 10$~Myr) timescales, and a transition from high to low outflow mass-loading factors over cosmic time as CR pressure builds up in the ISM. We find that simple (boosted) thermal SN feedback models fail to capture the cosmic evolution of the very first galaxies because they lack realistic feedback channels that operate on different timescales and have distinct thermodynamical properties, thereby either overproducing stellar mass or driving overly explosive outflows. Our results indicate that non-thermal galaxy formation physics is crucial for providing a robust theoretical framework with which to interpret the high-redshift galaxy populations now being uncovered by JWST.

\end{abstract}

\keywords{\uat{Cosmic rays}{329} --- \uat{Galaxy formation}{595} --- \uat{High-redshift galaxies}{734} --- \uat{Hydrodynamical simulations}{767} --- \uat{Magnetohydrodynamics}{1964} --- \uat{Reionization}{1383}}

\section{Introduction}
\label{s:Introduction}
The \textit{James Webb Space Telescope} (\JWST) has revealed a surprisingly bright and massive population of galaxies at $z \gtrsim 8$ \citep{Finkelstein2023, Harikane2023, Leung2023, Harikane2024, Hegde2024}, challenging theoretical predictions for early galaxy formation \citep[e.g.,][]{Boylan-Kolchin2023, Labbe2023, Steinhardt2023, Whitler2025}. 
Several explanations can justify this abundance of UV-bright galaxies, from stochastic and bursty star formation \citep{Mason2023, Shen2023}, to feedback-free starbursts \citep{Dekel2023} and reduced dust attenuation \citep{Ferrara2023}.
These observations require us to build a new theoretical framework to explain: \textit{(i)} highly bursty star formation histories \citep{Endsley2023, Sun2023a, 
Asada2023, Carvajal-Bohorquez2025}, \textit{(ii)} ionized metal-poor interstellar media (ISM; \citealt{Curti2023, Sanders2024, Endsley2024}), \textit{(iii)} growing evidence for abundant and over-massive supermassive black holes (SMBHs; \citealt{Maiolino2024, Greene2024, Ubler2025}), and \textit{(iv)} accelerated galaxy morphological and evolutionary growth revealed through early disk formation and galaxy quenching only a few hundred million years after the Big Bang \citep{Nelson2023, Looser2024, Robertson2024, Weibel2025, Danhaive2025}.

Interpreting this new observational landscape requires innovative galaxy formation models that can simultaneously reproduce galaxy population statistics (e.g., UV luminosity, stellar mass functions) and the detailed physical conditions of the ISM and circumgalactic media (CGM) across the first few billion years of cosmic history. In fact, the developing, often shallow gravitational potential wells of high-redshift galaxies render such systems especially sensitive to the physics of galaxy formation \citep{Geha2012}, therefore requiring a comprehensive understanding of the baryon cycle to fully characterize galaxy formation in the early Universe. The regulation of the growth of these systems, driven by stellar feedback, such as ionizing radiation, stellar winds, and supernova (SN) explosions \citep[e.g.,][]{Efstathiou1992, Rosdahl2015b, Hayward2017}, not only suppresses star formation through preventive processes, but also drives multi-phase galactic outflows into the CGM and intergalactic medium \citep[IGM;][]{Tumlinson2017, Concas2022, Thompson2024}. From their early onset, these baryonic outflows shape the thermodynamical state and recycling timescales of the CGM/IGM \citep[e.g.,][]{Ford2014}, set their chemical enrichment \citep[e.g.,][]{Tumlinson2017, Chisholm2018}, drive their ionization state \citep[e.g.,][]{Werk2014, Werk2016, Roca-Fabrega2019, Cadiou2025}, and modulate the redshift evolution of matter clustering \citep{Chisari2019, vanDaalen2020, Martin-Alvarez2025}.

Observations are now beginning to detect galactic outflows at high redshifts, revealing the existence of extended and ionized winds already in place within the first billion years of cosmic history \citep[e.g.,][]{Llerena2023, Carniani2024, Saldana-Lopez2025}. These detections hint at how the interplay between star formation, feedback, and galactic outflows may differ compared to lower-redshift galaxies \citep[e.g.,][]{Chisholm2017, Marasco2023}. High-redshift outflows exhibit mass-loading factors largely exceeding those of present-day star-forming galaxies, suggesting outflows played an enhanced role in regulating galaxy growth in the early Universe. Under this self-regulation scheme \citep{Somerville2015}, low- and intermediate-mass galaxies may have been predominantly governed by strongly ejective feedback, efficiently expelling gas from shallow potential wells, before transitioning toward more preventive, self-regulating modes of feedback at later times \citep{Heckman2015, McQuinn2019, Kado-Fong2024}.

Despite extensive progress in our modeling of galaxy formation, most hydrodynamical simulations still calibrate their feedback prescriptions to reproduce key observed galaxy properties and self-regulate star formation \citep[e.g.,][]{Pillepich2018b, Dave2019, Vogelsberger2020, Pakmor2023b, Schaye2023}. In fact, the limited dynamical range in spatial scales that can be reached by current state-of-the-art galaxy formation simulations means that `sub-grid' prescriptions must be used to capture unresolved physics, often requiring artificially enhanced feedback strengths or hydrodynamically decoupled outflows \citep[e.g.,][]{Springel2003a, Oppenheimer2010, Agertz2013, Crain2015, Weinberger2017, Kannan2021, Kugel2023, Chaikin2026}. In fact, many of these limitations are frequently encountered even in high-resolution cosmological zoom-in simulations capable of resolving the thermodynamical structure of the ISM \citep[e.g.,][]{Hopkins2014, Rosdahl2018, Smith2019, Marinacci2019, Dubois2021, Pallottini2022, Kannan2025, MaxRey2025}.
One of the frequent consequences of this calibration is single-phase galactic winds that struggle to simultaneously reproduce observations of outflow properties, galaxy stellar masses, star formation rates, and metal enrichment patterns \citep[e.g.,][]{Muratov2015, Somerville2015, Concas2022, Smith2023}.

These shortcomings highlight the necessity for more physically complete models. These models should incorporate the additional physics and processes we expect to be at play in observed galaxies, such as stellar winds \citep{Hopkins2012a, Agertz2013, Fichtner2024, Deng2024}, Pop III modeling \citep{Abel2002, Katz2025, Brauer2025}, and dust evolution \citep{Choban2022, Dubois2024, Trayford2026, Curro2026}.
Here, we focus on three non-thermal physical processes that are well known to influence fundamental phenomena such as star formation, stellar feedback, and galactic outflows: the physics of magnetism, radiation, and cosmic rays. These three energy components are also notable due to their expected equipartition with thermal and turbulent energies, or even their dominance of the total energy budget, across various galaxy environments.

Magnetic fields influence the structure and dynamics of the ISM by modifying gas-phase distributions and supporting molecular clouds against gravitational collapse \citep{Federrath2012, Shukurov2018, Martin-Alvarez2020, Robinson2024}. They also shape the topology and mixing layers of galactic outflows \citep{Evirgen2019}.
They further affect the thermal stability of the CGM and the coherence of large-scale outflows \citep{Ji2018, vandeVoort2021, Zhang2024}, even for the low-mass regime of galaxies \citep{Taziaux2025}.  

Stellar radiation, particularly in the hydrogen-ionizing regime, provides an early channel of feedback by photoheating and photoevaporating gas around newly formed stars. This affects giant molecular cloud lifetimes and significantly modulates star formation prior to the onset of the first supernova explosions \citep{Geen2015b, Rosdahl2015b, Sartorio2021, Kim2023Tigress, Andersson2024, Katz2026}. Ionizing radiation also shapes the ionization structure of the ISM and CGM and may influence large-scale inflows and starvation in low-mass systems \citep{Katz2020}.

While already proposed in prior decades as important components of galaxy formation \citep{Pfrommer2007, Hanasz2013}, cosmic rays have recently gained popularity due to their ability to drive galactic outflows and suppress star formation (e.g., \citealt{Girichidis2018, Hopkins2021a, Curro2024, Sike2025}; see \citealt{Ruszkowski2023} for a recent review). As a result, a body of work has emerged addressing their transport physics in the ISM and CGM \citep{Armillotta2021, Hopkins2022a, Girichidis2022} to improve the accuracy of their treatment in simulations.
The importance of cosmic rays at high redshifts also extends to processes such as reionization, as they significantly modulate the escape of ionizing photons \citep{Farcy2022, Yuan2024, Farcy2025}.

Our prior work using the \pandora\ simulations \citep{Martin-Alvarez2023, Martin-Alvarez2026} demonstrated that the \emph{combined} impact of these non-thermal processes in a cosmological simulation of a dwarf galaxy leads to notable differences with respect to models relying solely on thermal feedback. Importantly, the interplay between these processes is highly non-linear, leading to effects beyond the simple addition of their individual contributions. This combination naturally improves agreement with various observational relations, ranging from galaxy growth, colors, and sizes, to outflows and metal enrichment.
This foundation motivates expanding from a single galaxy to a larger cosmological sample, providing both a more statistically robust understanding of non-thermal processes and a comprehensive picture of their impact across different galaxy masses and their cosmological environments.

Here, we present the first results from the \azahar simulation suite, a new set of high-resolution cosmological simulations focused on a comoving volume of approximately $(10\ \mathrm{cMpc})^3$, with a maximum physical resolution of $\sim 20$\,pc, sufficient to resolve the internal structure of galaxies in detail.
The main suite is comprised of ten models that progressively incorporate non-thermal physical components, systematically exploring the impact of radiation transport, cosmic rays, and magnetohydrodynamics on galaxy formation from $z \sim 14$ to $z \sim 3$.  
In this introductory study, prior to the discussion of the full simulation suite, we focus only on three representative models that bracket the importance of these non-thermal processes: a non-calibrated hydrodynamical model (\HDrun), a calibrated supernova-boosted model (\HDBoostrun), and a `\textit{full-physics}' model including radiative transfer, cosmic rays, and magnetic fields (\RTnsCRiMHDrun). We investigate how these processes affect some of the most challenging new questions in galaxy formation revealed by \JWST, such as the rapid emergence of the first galaxies and their evolution all the way to cosmic noon ($z \sim 3$).

Our manuscript is structured as follows. In Section~\ref{s:Methods}, we introduce our numerical setup and the physical models explored in the \azahar simulations. In Section~\ref{s:Results}, we present our main results, spanning galaxy population statistics, star formation burstiness, outflow properties, and metal enrichment distributions. Finally, we summarize our main findings in Section~\ref{s:Conclusions}.

\section{Numerical Methods}
\label{s:Methods}
In this work we present the \azahar simulation suite, focusing on three illustrative models out of the ten main models of the suite. This section provides a description of the \azahar simulation setup, the adopted galaxy formation model, and the details of galaxy catalogs. A detailed description for each of the ten main models is reserved for our upcoming publication. 
Table~\ref{tab:azahar_summary} provides a summary of the numerical setup of the \Azahar simulations, and of the galaxy formation physics included for the three models studied here.

\azahar is generated using the \textsc{ramses} code \citep{Teyssier2002}, which couples an Eulerian treatment of the fluid component and a Lagrangian treatment of the stellar and dark matter components. All of these mass components are coupled through a gravitational solver, taking advantage of the grid decomposition of the simulation domain. Through the use of Adaptive Mesh Refinement (AMR), we recursively increase the spatial resolution of our simulations in regions of interest. To mark cells for refinement, we employ two criteria: i) a quasi-Lagrangian refinement strategy requiring a cell to refine whenever its total contained mass $(\Omega_m / \Omega_b)\,m_{\text{baryons}} + m_{\text{DM}}$ surpasses $8 \, m_{\text{DM}}$, and ii) a Jeans criterion that requires the local Jeans length, $\lambda_J$, to be refined by at least four resolution elements ($\Delta x \leq \lambda_J / 4$). 

The magnetohydrodynamical (MHD) solver of \ramses~employs a constrained transport (CT) implementation \citep{Teyssier2006, Fromang2006}. It makes use of the electromotive forces to compute the evolution of the magnetic field through the induction equation, which ensures by construction that no spurious magnetic divergence is generated down to numerical precision. This guarantees that the solenoidal constraint, $\vec{\nabla} \cdot \vec{B} = 0$, is preserved if no additional magnetic divergence is included in the simulations. This avoids the generation of spurious modifications of the thermodynamical quantities of the simulation \citep{Toth2000}.
We also make use of the radiative transfer (RT) implementation by \citet{Rosdahl2013, Rosdahl2015a}. While resolved RT through the interstellar medium requires even higher resolutions ($\lesssim 5$~pc; \citealt{Kimm2014}), our models are capable of resolving the approximate structures of the ISM, approaching the regime of radiative transfer convergence \citep{Yuan2024}. 
Finally, we model the impact of hadronic cosmic rays in the GeV regime as a fluid energy-density component, evolved through an implicit solver \citep{Dubois2016}. This implementation allows for anisotropic diffusion with a constant diffusion coefficient, and is extended to account for CR streaming and streaming heating by \citet{Dubois2019}. The solver accounts for streaming losses, a relativistic adiabatic scaling, and both Coulomb and hadronic losses \citep{Guo2008}. The models studied here do not include CR streaming (Section~\ref{ss:Azahar_models}); the streaming models will be presented alongside the full \azahar suite in future work. The only sources of CRs in our simulations are SNe, as described in the galaxy formation model section below.

\subsection{The \Azahar galaxy formation models}
\label{ss:Azahar_models}

\begin{table}[t!]
\centering
\caption{Summary of the numerical and physical setup of the \Azahar simulation suite (top table) and the three illustrative models (bottom table) analyzed in this work.}
\label{tab:azahar_summary}

\textit{Full suite configuration}
\begin{tabular}{l | c}
\hline
\hline
Quantity & Value \\
\hline
Cosmology & \citealt{PlanckCollaboration2016cosmo} \\
$H_0$ & $67.9\ \mathrm{km\,s^{-1}\,Mpc^{-1}}$ \\
$\Omega_{\rm m}$ & 0.3065 \\
$\Omega_{\rm b}$ & 0.0483 \\
$\Omega_\Lambda$ & 0.6935 \\
\hline
Initial redshift & $127$ \\
Box size & $25\,\cMpc$ \\
Low-resolution base grid & $256^3$ \\
High-resolution base grid & $1024^3$ \\
\\
High-res.~volume (App.~\ref{ap:hmf_volume}) & $\sim (12\,\cMpc)^3$ equivalent \\
Maximum spatial resolution & $23.8\,\pc$ \\
Dark matter particle mass & $4.5\times10^5\,\Msun$ \\
Stellar particle mass & $4\times10^4\,\Msun$ \\
Refinement algorithm & adaptive mesh refinement \\
Refinement criteria & quasi-Lagrangian + Jeans \\
\hline
MHD solver & Constrained Transport\\
Radiative transfer & moment-based RT\\
Star formation model & magneto-thermo-turbulent \\
Supernova feedback & mechanical feedback \\
SN energy $E_{\rm SN}$ & $\alpha_{\rm SN}\, 10^{51}\,\mathrm{erg}$ \\
IMF & Kroupa (2001) \\
\hline
\end{tabular}

\vspace{0.2cm}

\textit{Physics included in the three studied models}
\begin{tabular}{l | ccc}
\hline
\hline
Model & \textit{Standard} & \textit{Calib.} & \RTnsCRiMHDrun \\
\hline
SN boost $\alpha_{\rm SN}$ & 1.0 & 4.0 & 1.0 \\
MHD & \xmark & \xmark & \cmark \\
Initial field $B_0$ (G) & -- & -- & $3\times10^{-20}$ \\
SN mag. energy $f_{\rm mag}$ & -- & -- & 0.01 \\
RT stellar sources & \xmark & \xmark & \textsc{bpass} v2.0 \\
Reduced speed of light & -- & -- & $\tilde{c} = 0.01\,c$ \\
Cosmic rays & \xmark & \xmark & \cmark \\
Diffusion $\kappa_{\rm CR}$ ($\mathrm{cm^2\,s^{-1}}$) & -- & -- & $3\times10^{28}$ \\
SN CR energy $f_{\rm CR}$ & -- & -- & 0.1 \\
\hline
\end{tabular}
\end{table}

The \Azahar galaxy formation models build on the physical implementation and configuration explored in the \textsc{Pandora} simulations of a dwarf galaxy \citep{Martin-Alvarez2023, Martin-Alvarez2026}, but now for the first time simulating a sufficiently large cosmological volume to capture a statistical sample of galaxies, with a range of masses, assembly histories, and diverse environments. 

\subsubsection{Initial Conditions}
The initial conditions for the simulations are generated with \textsc{music} \citep{Hahn2011}, initialized at redshift $z = 127$, assuming the \citet{PlanckCollaboration2016cosmo} cosmology.
This corresponds to a Hubble constant of $H_0 = 67.9\ \mathrm{km\,s^{-1}\,Mpc^{-1}}$ ($h=0.679$), a total matter density parameter of $\Omega_{\mathrm{m}} = 0.3065$, a baryon density of $\Omega_{\mathrm{b}} = 0.0483$, and a dark energy density of $\Omega_{\Lambda} = 0.6935$.
The $25\,\cMpc$ cosmological box contains a high-resolution sub-volume of effective cubic side $\sim10\,\cMpc$. This zoom size corresponds to the Lagrangian high-resolution region. Following its redshift evolution (Appendix~\ref{ap:hmf_volume}) yields an effective $\sim(12\,\cMpc)^3$ volume, employed for our number-density calculations. The low-resolution region is resolved with 256 elements per side; we impose a maximum cell size of $\dx = 24\,\ckpc$ in the high-resolution region. This region is further refined to a target full-cell width spatial resolution of $\dxres = 23.8\,\pc$, combining both the quasi-Lagrangian and Jeans criteria stated above. Particle masses in the high-resolution region are 
$m_{\text{DM}} \approx 4.5 \times 10^5 \,\Msun$ for the dark matter component and $m_{\text{*}} \approx 4 \times 10^4 \,\Msun$
for the stellar component.

\subsubsection{Cooling and heating processes}
\label{ss:gas_physics}

We include standard galaxy formation cooling and heating processes in all our models. For radiative cooling, we account for primordial and metal cooling of gas at temperatures $T > 10^4$~K through interpolated \textsc{Cloudy} tables \citep{Ferland1998}. Below that temperature, we account for gas cooling including metal-line emission following \citet{Rosen1995}. For the baryonic gas, we assume an ideal monatomic gas with an adiabatic index of $\gamma = 5/3$, and a metallicity floor of $10^{-4} \Zsun$, where we assume a solar metallicity of $\Zsun = 0.012$. This corresponds to the critical metallicity required for gas fragmentation to enable the formation of Population II stellar clusters \citep{Schneider2012}. In addition, all our models include a spatially uniform and time-dependent UV background \citep{Haardt1996}, enabled at $z = 9$.

\subsubsection{Local, magneto-thermo-turbulent star formation}
\label{ss:star_formation}

We model star formation in \azahar using a local, physically motivated magneto-thermo-turbulent (MTT) prescription. This is based on the approach suggested by \citet{Federrath2012}, introduced in \textsc{ramses} by \citet{Kimm2017}, and extended to account for the effects of magnetic fields in \citet{Martin-Alvarez2020}.
This MTT model is characterized by two main features: i) it only allows for star formation in self-gravitating regions of the ISM where gravitational pull is higher than the local pressure support, and ii) it allows for a locally varying star formation efficiency (SFE), $\epsilon_\mathrm{ff}$, determined by the local MTT properties of the gas cells.
When included in the simulation, we account for the presence of magnetic fields through an effective sound speed, $c_{s,\mathrm{eff}} = c_s \sqrt{1 + \beta^{-1}}$, where the plasma $\beta$ is the ratio of the thermal to magnetic pressure, and an isotropic magnetic pressure is implicitly assumed.

To ensure that star formation is only triggered below the resolution limit of the simulations, we only allow cells at the highest level of refinement to form stars \citep{Rasera2006}. Further details on this MTT model can be found in Appendix~B of \citet{Martin-Alvarez2020}.

\subsubsection{Supernova feedback}
\label{ss:supernova}

Supernova (SN) feedback is implemented using the mechanical SN feedback prescription from \citet{Kimm2014, Kimm2015}. Each stellar particle samples the initial mass function (IMF) stochastically over its first 50~Myr. This specifies the number of SN events taking place during each simulation timestep. SN events take place individually, and each particle is allowed to generate multiple SNe according to its mass. When a SN occurs, it injects mass, momentum, and energy into its hosting cell and its neighbors. Each SN has an energy of $E_{\text{SN}} = 10^{51}$~erg, generated by a progenitor of $M_{\text{SN}} = 10\,\Msun$. The specific SN energy injected is given by $\varepsilon_{\text{SN}} = \alpha_{\text{SN}} E_{\text{SN}}/M_{\text{SN}}$; $\alpha_{\text{SN}}$ is the feedback strength calibration parameter, with a fiducial value of $\alpha_{\text{SN}} = 1.0$ in our two models without parameter calibration to galaxy observables, \HDrun and \RTnsCRiMHDrun. The parameter is set to $\alpha_{\text{SN}} = 4.0$ for the model with enhanced SN feedback, \HDBoostrun, calibrated to reproduce the galaxy stellar mass function at $z = 3$.
We assume a Kroupa IMF \citep{Kroupa2001}, with massive stars returning a mass fraction $\eta_{\text{SN}} = 0.213$ to the ISM, and with a subfraction $\eta_{\text{metals}} = 0.075$ of this mass being returned to the ISM in the form of metals. For our \RTnsCRiMHDrun~model, including magnetism and cosmic rays, SN events also inject energy into these non-thermal components. The total energy injected by each SN event remains unchanged and is divided among the thermal, kinetic, and non-thermal channels. 
Approximately 10\% is directed to cosmic rays and 1\% to magnetic fields (Section~\ref{ss:Azahar_models}).
As a result, the \RTnsCRiMHDrun model has a smaller thermal + kinetic energy per SN than \HDrun, but the same total energy per SN.

\subsubsection{The \RTnsCRiMHDrun simulation setup}
\label{ss:non_thermal_model}

In this section, we detail the configuration of the  \RTnsCRiMHDrun~model of the \Azahar simulation suite. This \RTnsCRiMHDrun~model was also explored in the Pandora project \citep{Martin-Alvarez2023} under the label RTnsCRiMHD+SfFb.

The magnetic field configuration is equal to that of the \textsc{MBinj} models in \citet{Martin-Alvarez2024}. In this configuration, the magnetic field has two sources. The first source is an ab initio, weak magnetic field seeded uniformly along the $z$-axis of the simulation box. The magnetic field strength is set to $B_0 = 3\times 10^{-20}$~G, resembling the field strength sourced by a Biermann battery effect \citep{Attia2021}. In practice, due to the numerical limitations in astrophysical simulations with resolutions $\gtrsim 0.1\,\pc$ ($\Gamma_{\text{amp}}\sim 80\,\Gyr^{-1}$), this field strength will not amplify above the kinematic regime in the ISM of galaxies \citep{Martin-Alvarez2022}. 
The second method of seeding injects magnetic energy through SN explosions, injecting small-scale circular loops (guaranteeing $\vec{\nabla} \cdot \vec{B} = 0$) around each explosion \citep{Martin-Alvarez2021}. Each event provides a magnetic energy of $E_{\text{inj,mag}} = 0.01\,E_{\text{SN}} \sim 10^{49}\,\erg$, reproducing typical SN-remnant magnetizations of $\sim10^{-5}\,\Gauss$ at $\sim 10\,\pc$ scales \citep{Parizot2006}. These magnetic fields expand to larger scales as the supernova remnants (SNRs) evolve. The resulting magnetizations are comparable to those inferred from far-infrared and radio observations \citep{Martin-Alvarez2024, Dacunha2025}. Further information on this injection implementation is provided in Appendix~A of \citet{Martin-Alvarez2021}.

In addition to the adopted UV background, here we employ the RT treatment to capture the effects and propagation of Lyman-continuum radiation, separated into three radiation bins, and sourced by stellar particles. Explicitly, the energy bins are $13.6$--$24.59$~eV (H\,\textsc{i} ionization), $24.59$--$54.42$~eV (He\,\textsc{i} ionization), and $> 54.42$~eV (He\,\textsc{ii} ionization). The stellar emission is selected according to the properties (mass, metallicity, and age) of each stellar particle, and following the \textsc{bpass} v2.0 model \citep{Eldridge2008, Stanway2016}. The radiative solver allows up to 500 RT sub-cycles per hydrodynamical timestep. It further assumes a reduced speed of light $\tilde{c} = 0.01\,c$, representing the propagation speed of the ionization fronts, with the objective of alleviating the demanding time-stepping of a full speed-of-light solution. The RT configuration in this model broadly follows that of the SPHINX simulation \citep{Rosdahl2018}.

Finally, SN events are the sole source of cosmic rays in the \RTnsCRiMHDrun model, with each event providing $E_{\text{CR}} = f_{\text{CR}} E_{\text{SN}} = 10^{50} \erg$, where $f_{\text{CR}} = 0.1$, approximately in line with observations \citep{Morlino2012, Helder2013}.
We assume a constant diffusion coefficient $\kappaCR = 3 \times 10^{28} \cm^2 \text{s}^{-1}$, consistent with observations of $\gamma$-rays generated through cosmic ray hadronic losses \citep{Ackermann2012,Salem2016,Pfrommer2017a}, and with estimates for the isotropic coefficient in the Milky Way \citep{Trotta2011,Cummings2016}. No CR streaming effects are included in the \RTnsCRiMHDrun~model studied here.
We note that these choices of $\kappaCR$ and $f_{\text{CR}}$ are well established in previous studies \citep[e.g.,][]{Pfrommer2017b, Wiener2017, Butsky2018, Dashyan2020}. Although these quantities are likely to also vary with both CR spectral energy and local ISM properties, this dependence is beyond the scope of this project.
More generally, none of the non-thermal parameter values described were adjusted to match galaxy property observables; their values are instead motivated by independent observational, theoretical, or numerical constraints, as described above.

\subsection{Galaxy identification, tracking, and measurements}
\label{ss:trackers}

\subsubsection{Galaxy tracking algorithm}
\label{sss:tracking}

Here we provide a brief description of our galaxy finding and tracking method, and provide a full description of the method in Appendix~\ref{ap:galaxy_tracker}.

We identify and follow the evolution of galaxies in our simulations using a particle-tracking post-processing method. Galaxy trackers are seeded in dark matter halos and subhalos identified using the \textsc{AdaptaHOP} \citep{Aubert2004} implementation of \citet{Tweed2009}.
Galaxy trackers are then followed through a representative subset of stellar particles associated with the galaxy, applying distance exclusion criteria to separate systems. Galaxy positions are estimated by computing the center of mass of these stellar particles, and refined using an iterative shrinking-spheres calculation. We allow for galaxy mergers, where low-mass trackers are merged into the most massive galaxy of a host halo when they are persistently recorded at separations smaller than a few resolution elements.

\subsubsection{Property measurements and apertures}
\label{sss:measurements}
Galaxy property measurements are then performed in four spherical regions:
\begin{enumerate}
\item{Galaxy:} with an aperture extending up to twice the half-mass radius of the system.
\item{Inner CGM:} a spherical region extending from the edge of the \textit{galaxy}, up to 0.2 of the virial radius.
\item{Inner CGM boundary:} a narrow boundary of 200~pc (i.e., 10~$\dxres$) around the \textit{inner CGM region}, used to measure gas inflows and outflows across this radius.
\item{Halo:} a region spanning from the \textit{inner CGM} up to the virial radius of the hosting halo.
\end{enumerate}

For each galaxy, we measure $\sim 1000$ quantities. These span from morphological properties derived from the mass profiles, to energy budgets, phase-resolved masses, enrichments and kinematics, inflow and outflow rates, and absolute magnitudes. Before any property is computed, all relevant cell and particle quantities are recast into the frame of the galaxy as inferred from the tracked stellar component. Where relevant, we define the disk direction as the direction of the corresponding angular momentum vector.
We estimate UV luminosities integrating the flux over the 1400--1600~\AA~range, and using all the stellar particles within the \textit{galaxy}. We model the emission of each particle through a single stellar population (SSP), according to its stellar mass, age, and metallicity. Our SSP spectral emission follows \textsc{bpass}~v2.3 \citep{Stanway2018, Byrne2022}. Following a similar approach to \citet{Kannan2025}, we estimate dust attenuation through a simple empirical Meurer-based correction \citep{Meurer1999, Bouwens2016}. At the redshifts and UV magnitudes probed by our galaxy sample, UV attenuation is small, consistent with recent JWST measurements \citep[e.g.,][]{Bouwens2021, Donnan2024, Saxena2026}. 

Previous studies have already used our tracking algorithm \citep[e.g.,][]{Martin-Alvarez2023, Sanati2024, Dacunha2025}, and preliminary versions of the \Azahar galaxy tracker catalogs \citep[e.g.,][]{Yuan2025, Dome2025, Belfiori2026}.
In upcoming work presenting the full simulation suite, we will release the post-processing galaxy-tracking algorithm, followed by the full set of measurements for each galaxy in the \Azahar simulation suite (see Data Availability).

\section{Results}
\label{s:Results}

In this section, we compare our three models using key observational diagnostics: the UVLF evolution, observational tracers of star formation burstiness, outflow scaling relations, and enrichment constraints. Through these, we show that non-thermal physical processes are central to self-regulated galaxy growth and provide a unified explanation of high-redshift galaxy observations.

The pressure support provided by magnetic fields and cosmic rays, combined with early feedback from stellar radiation, reshapes the star formation distribution across short and long timescales. The resulting increase in burstiness, most prevalent at $z \gtrsim 6$, leads to ejective self-regulation at early times, and transitions to preventive regulation at later times.
Cosmic ray-driven winds are sustained and temperate, in good agreement with observational constraints: by entraining a higher proportion of less enriched gas, these winds naturally regulate the mass-metallicity relation.    

\subsection{Qualitative comparison of the high-resolution sub-volume}
\label{ss:qualitative_comparison}


\newcommand{\gwidth}{0.1765\linewidth}
\newcommand{\lwidth}{0.588\linewidth}
\newcommand{\bwidth}{\linewidth}
\newcommand{\bgalwidth}{0.25\linewidth}

\begin{figure*}[t!]
    \centering
    \includegraphics[width=\bwidth]{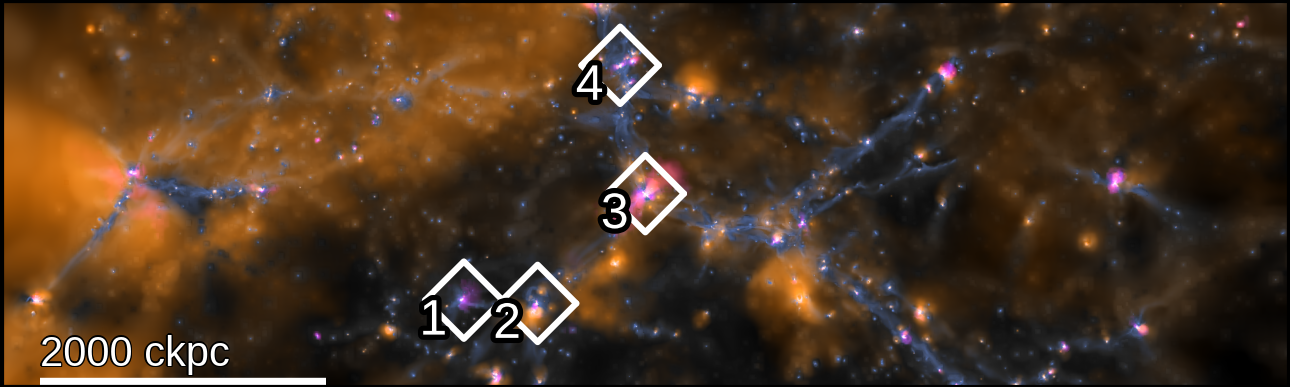}\\
    \includegraphics[width=\bgalwidth]{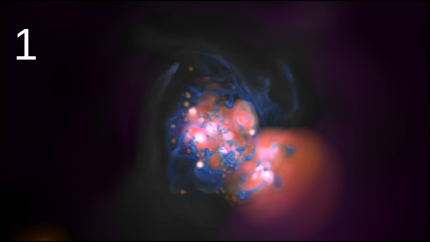}%
    \includegraphics[width=\bgalwidth]{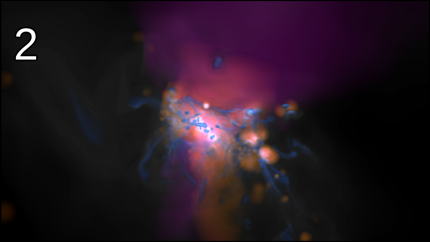}%
    \includegraphics[width=\bgalwidth]{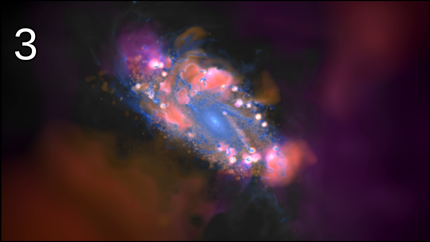}%
    \includegraphics[width=\bgalwidth]{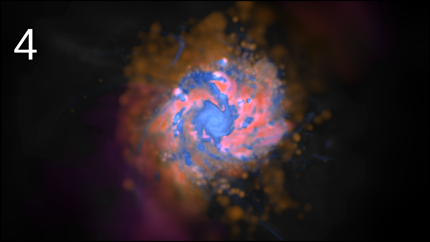}\\
    \includegraphics[width=\bgalwidth]{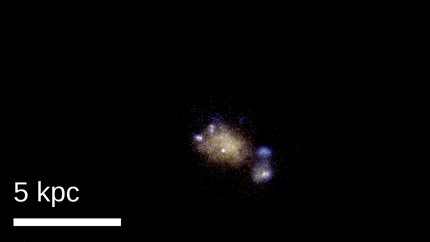}%
    \includegraphics[width=\bgalwidth]{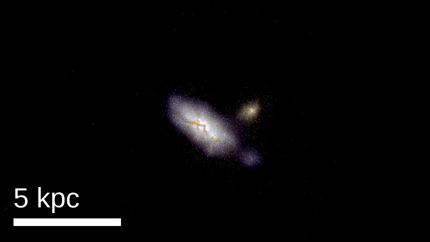}%
    \includegraphics[width=\bgalwidth]{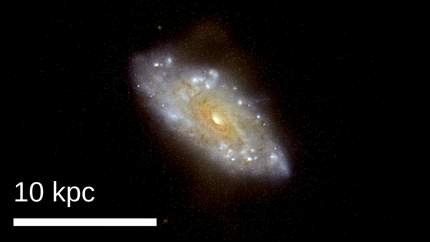}%
    \includegraphics[width=\bgalwidth]{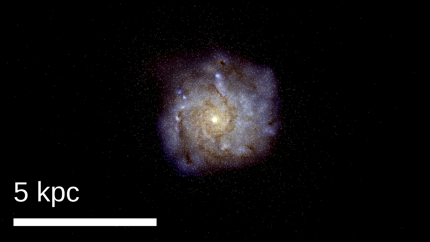}\\
    \includegraphics[width=\bwidth]{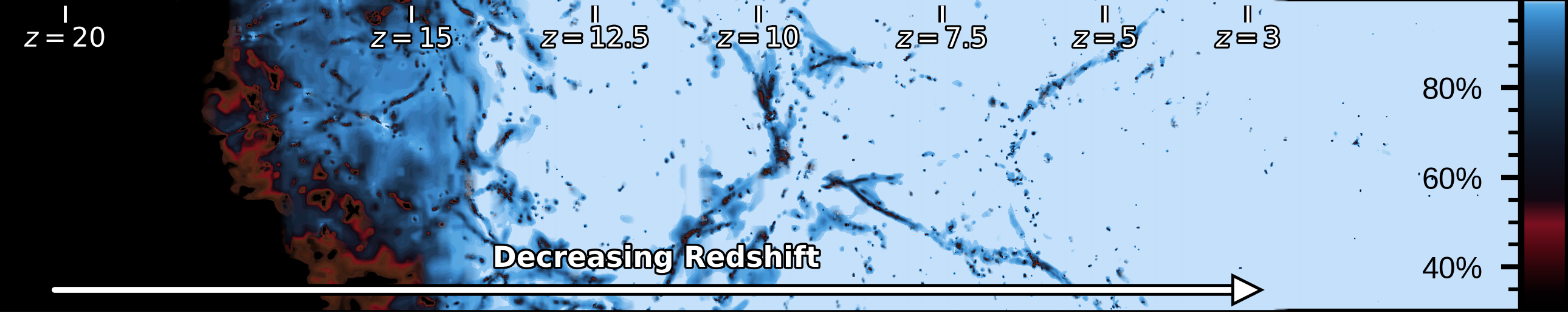}\\
    \caption{{\bf (Top panel)} Projected view showing part of the high-resolution sub-volume ($8\,\cMpc \times 3\,\cMpc \times 8\,\cMpc$) for our \RTnsCRiMHDrun model. Colors show the neutral hydrogen (blue), ionized hydrogen (grey), stellar surface density (white), and ionized radiation (orange for low and purple for high energy) densities. Galaxies appear as bright knots across the cosmic web, with the inter-galactic space permeated by Lyman-continuum radiation. 
    {\bf (Second row)} Zoomed-in view of the galaxies marked with diamonds in the top panel (colors as above), highlighting a highly resolved ISM with complex morphologies and kinematics.
    {\bf (Third row)} Synthetic JWST observations ([F444W, F277W, F150W]; RGB) of the same four galaxies shown in the second row.
    {\bf (Bottom panel)} Redshift-evolution sweep of projected, density-weighted ionized hydrogen percentage for the same region shown in the top panel. From left to right, it evolves from $z \sim 25$  down to $z = 3$, revealing the emergence of the first galaxies and their ionizing `bubbles'.}
    \label{fig:front_presentation}
\end{figure*}

\newcommand{\gwidthcomp}{0.1876\linewidth}
\newcommand{\lwidthcomp}{0.6248\linewidth}
\begin{figure*}[t!]
    \centering
    \includegraphics[width=\lwidthcomp]{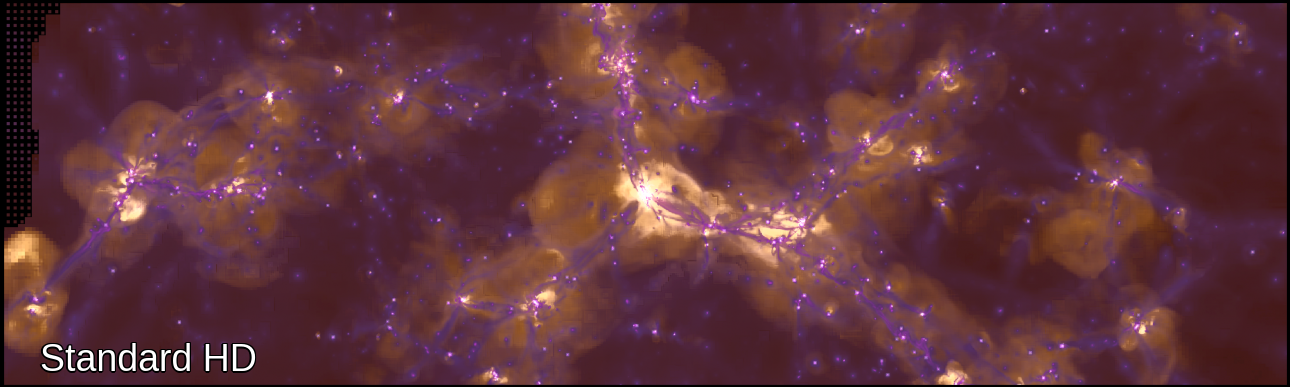}%
    \includegraphics[width=\gwidthcomp]{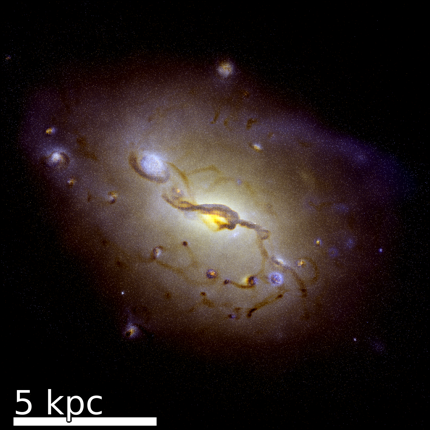}%
    \includegraphics[width=\gwidthcomp]{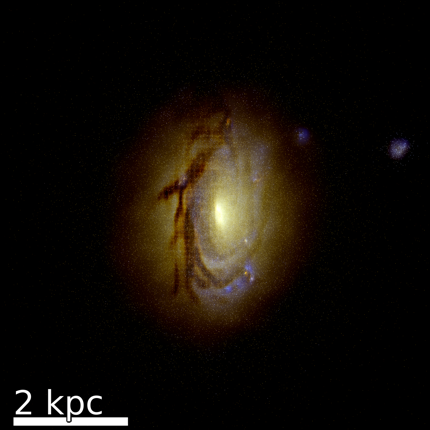}\\
    \includegraphics[width=\lwidthcomp]{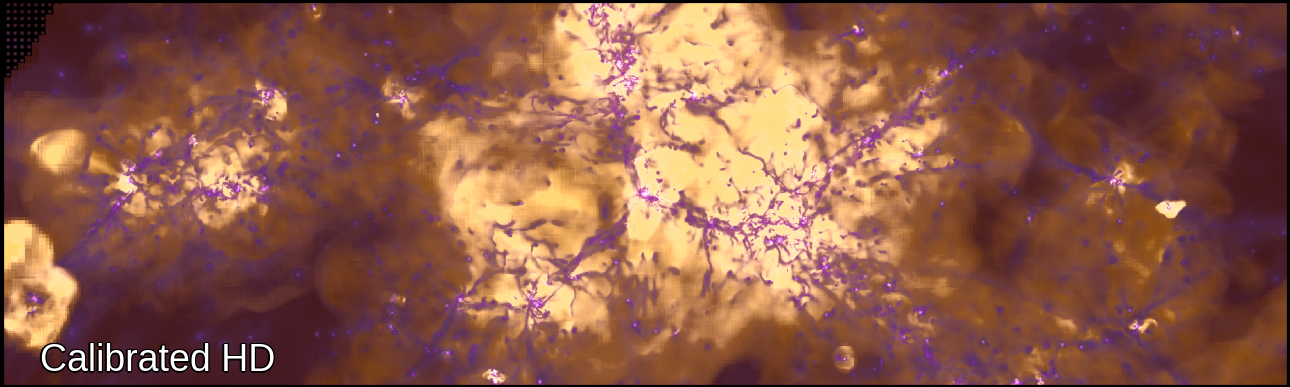}%
    \includegraphics[width=\gwidthcomp]{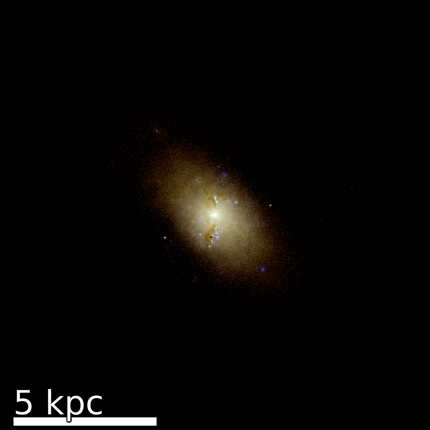}%
    \includegraphics[width=\gwidthcomp]{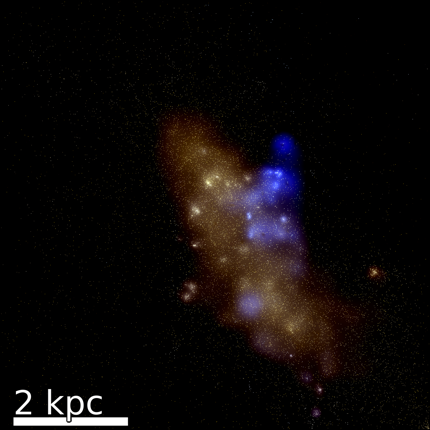}\\
    \includegraphics[width=\lwidthcomp]{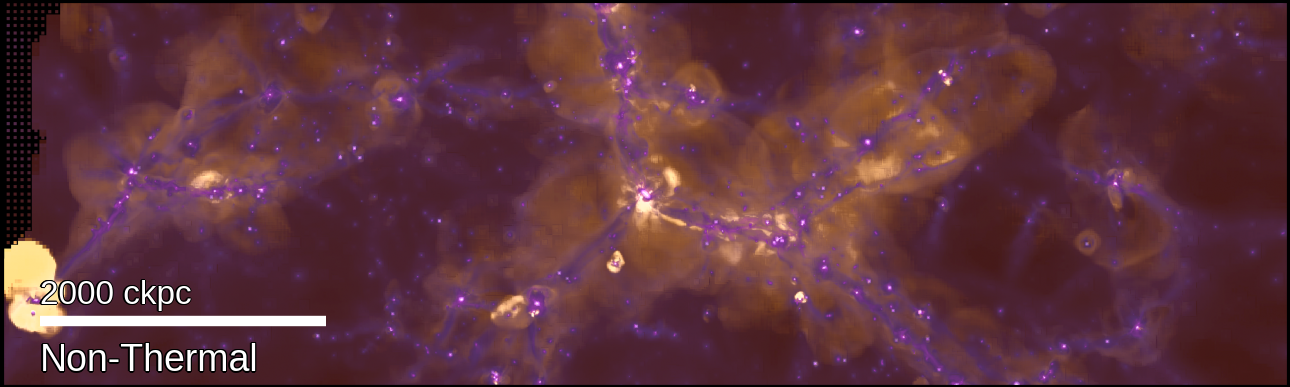}%
    \includegraphics[width=\gwidthcomp]{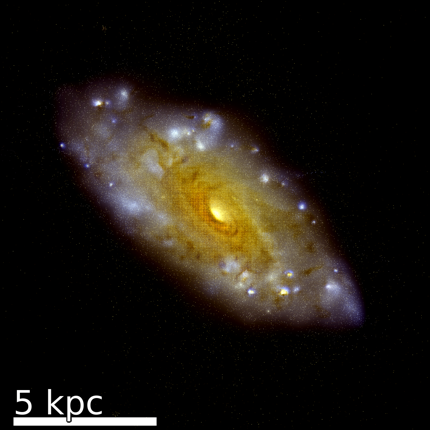}%
    \includegraphics[width=\gwidthcomp]{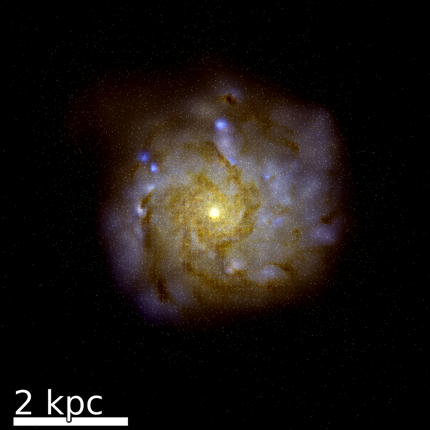}
    \caption{Comparison of our three illustrative simulations. From top to bottom models are standard hydrodynamics (\HDrun), calibrated hydrodynamics (\HDBoostrun), and \RTnsCRiMHDrun.
    {\bf (Left column)} Colors show total gas density (blue/purple), high temperature gas (orange), and stellar surface density (white). Galaxy outflows are very different across models, with \HDBoostrun producing much hotter and extended galactic winds.
    {\bf (Two right columns)} Synthetic JWST observations of two galaxies across the three models, illustrating how galaxy morphologies depend on the physical model assumed.}
    \label{fig:model_comparison}
\end{figure*}

We begin our analysis by qualitatively comparing the high-resolution region across our three simulations (Figs.~\ref{fig:front_presentation} and \ref{fig:model_comparison}).
\iftrue 
The top panel in Fig.~\ref{fig:front_presentation} shows a projected large-scale view ($8\,\cMpc \times 3\,\cMpc \times 8\,\cMpc$) of the \RTnsCRiMHDrun~model. It displays the radiation field, gas (ionized and neutral) and stellar density. The panel highlights the appearance of the large-scale structure in the simulation, with cosmic filaments connecting several hundreds of galaxies. Structures with deeper potentials and filaments with higher densities are dominated by neutral gas. Large-scale ionization bubbles trace the escaping radiation surrounding star-forming systems. A clear example is the intense radiation flux escaping a starburst disk galaxy, which is located toward the leftmost edge of the panel; another is the bi-conical escape of high-energy Lyman-continuum (LyC) radiation from the central disk galaxy, perpendicular to the filaments due to the disk-cosmic web alignment \citep{Dubois2014} and reinforced by filament self-shielding.

The second row panels show zoomed-in views of multiple galaxies, revealing a diversity of structures: from extended gas disks to star-forming clumps driving H\,\textsc{ii} bubbles and leaking LyC radiation into the CGM, as well as complex, merging multi-component systems.
A third row shows synthetic JWST observations of the same four galaxies, illustrating stellar morphologies.

We compare our three models in Fig.~\ref{fig:model_comparison}, which displays composite gas/stellar maps (left column) and synthetic JWST observations of two representative galaxies (two rightmost columns) for our \HDrun, \HDBoostrun, and \RTnsCRiMHDrun simulations. The left column shows a composite map of gas density, stellar density, and hot gas ($T \gtrsim 10^{6}\,\Kelvin$).
\HDBoostrun is the most distinct across the three, with high temperature galactic outflows extending deep into the IGM and disrupting cosmic filamentary inflows onto the galaxies (see e.g., the environment of the galaxy in the image center).
\fi

Differences in the stellar components for two representative galaxies with halo masses at $z = 3$ of $\Mhalo \sim 2 \times 10^{11}\,\Msun$ and $\Mhalo \sim 1 \times 10^{11}\,\Msun$ ($\Mstar = 1.4\times10^{10}$ and $2.5\times10^{9}\,\Msun$ in \RTnsCRiMHDrun) are clearly shown in the panels displaying synthetic JWST observations (two rightmost columns of Fig.~\ref{fig:model_comparison}; [F444W, F277W, F150W] filters)\footnote{Mocks are generated with our modified version of the \textsc{sunset} software. We model dust extinction as a 3D absorption screen, and stellar emission as described in Section~\ref{sss:measurements}.}. 
Unsurprisingly, galaxies are most massive in the \HDrun run, with older stellar components and massive star-forming clumps embedded in the disk. The \HDBoostrun run produces significantly less massive galaxies without rotationally supported stellar disks. Conversely, in the \RTnsCRiMHDrun run galaxies have thinner, rotationally supported disks. These disks contain multiple clumpy, star-forming regions, and have evolved central bulges. While their dust content is lower than in the \HD run due to more realistic stellar masses and metallicities, they still exhibit well-defined dust structures. 
\PaperMP presents a dedicated analysis of galaxy morphologies, and their implications for the detectability of high-redshift galaxies.

Finally, the bottom panel of Fig.~\ref{fig:front_presentation} shows the evolution of the hydrogen fraction, $x_{\rm HII}$, from $z\!\sim\!25$ (left edge) to $z\!=\!3$ (right edge).
It reveals the emergence of ionizing bubbles driven by the first cosmic sources, expanding and eventually coalescing into a fully ionized IGM, with thin self-shielded knots and galaxies along the densest cosmological filaments.

\subsection{Non-thermal physics impact on high-redshift UV luminosities, star formation, and stellar masses}
\label{ss:uv_non_thermal}
\subsubsection{The UV luminosity function across cosmic time}

\begin{figure*}[t!]
    \centering
    \includegraphics[width=0.48\textwidth]{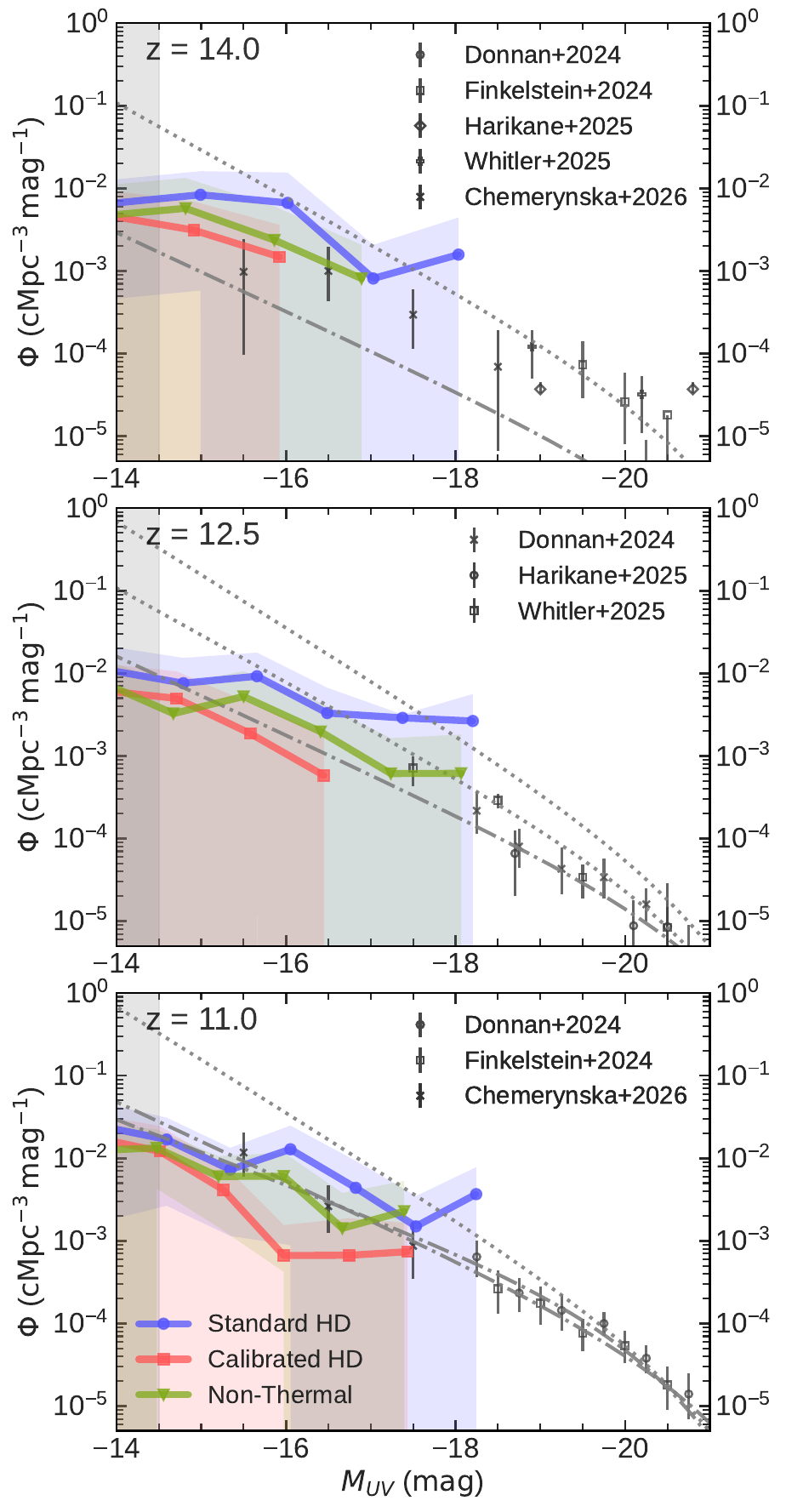}
    \hfill
    \includegraphics[width=0.48\textwidth]{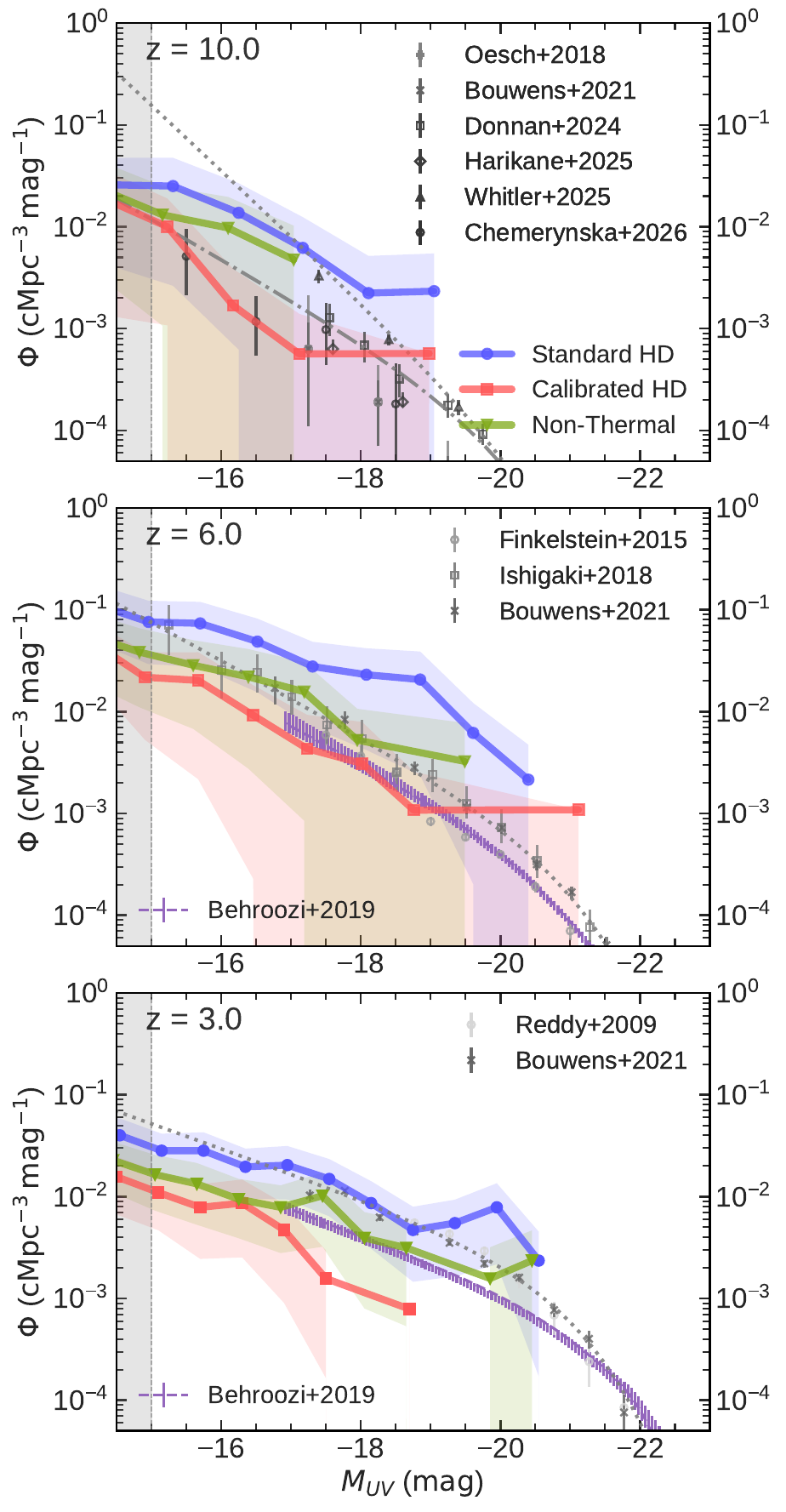}
    \caption{UVLF evolution for \HDrun (blue), \HDBoostrun (red), and \RTnsCRiMHDrun (green) with recent observational constraints. Redshift decreases from top to bottom, spanning $z = 14,\,12.5,\,11$ (left) and $z = 10,\,6,\,3$ (right). Shaded bands combine Poisson noise, cosmic variance, and volume uncertainty. Observations are shown as grey open data points.
    Left panels and top right panel show functional fits from \citet{Whitler2025} as dotted lines, and \citet{Donnan2024} as dot-dashed. Right central and right bottom panels show fits from \citet{Bouwens2021} as dotted lines, and the empirical model of \citet{Behroozi2019} in purple.
    The \RTnsCRiMHD model remains in reasonable agreement with observations at all times, whereas the \HDBoost model underpredicts the UVLF ($z \gtrsim 10$), and the \HD model overpredicts it ($z \sim 6$).
    }
    \label{fig:UVLF}
\end{figure*}

Building on this qualitative overview, we turn to quantifying how non-thermal physics influences the formation of galaxies. First we focus on two key challenges: the UV detections of abundant galaxies at extremely high redshifts ($z \gtrsim 10$), and the persistent discrepancies between traditional simulations and UV luminosity function (UVLF) measurements across cosmic time. 

Fig.~\ref{fig:UVLF} shows the UVLFs of our three simulations alongside observational constraints. This figure compiles measurements from:
\citet{Reddy2009} at $z = 3$;
\citet{Oesch2018} at $z = 10$; 
\citet{Finkelstein2015} at $z = 6$;
\citet{Ishigaki2018} at $z = 6$;
\citet{Bouwens2021} at $z = 10$, 6 and 3;
\citet{Donnan2024} at $z = 14$, 12.5, 11, and 10;
\citet{Finkelstein2024} at $z = 14$ and 11;
\citet{Harikane2025} at $z = 14$, 12.5, and 10;
\citet{Chemerynska2026} at $z = 14$, 11, and 10;
and \citet{Whitler2025} at $z = 14$, 12.5, and 10.
We also include observational functional fits to the UVLFs across our panels as grey lines for reference.
At high redshift ($z > 7$), we combine galaxies across multiple snapshots within $\Delta z = 0.5$ to reduce UVLF stochasticity. At lower redshifts, where statistics improve, we use a single snapshot per redshift.
At faint magnitudes, we show a vertical shaded band indicating galaxies whose star formation is sampled by only a few stellar particles, leading to stochastic UV emission.
Our model error bands combine volume uncertainty, Poisson noise from galaxy counts, and cosmic variance.
Our uncertainties are dominated by cosmic variance
\footnote{We estimate cosmic variance as $\Delta\phi_{\rm cv} = \phi\,b\,(M_h,z)\,\sigma_{\rm DM}(R,z)$, with $b\,(M_h,z)$ the bias of the host halos at the given luminosity and $\sigma_{\rm DM}(R,z)$ the linear matter variance on our volume scale, both from \texttt{Colossus} \citep{Diemer2018}.}, with additional Poisson noise in our most massive/brightest bins.

The left panels of Fig.~\ref{fig:UVLF} show our UVLF at the redshifts of the earliest galaxy detections by JWST ($z \in [11, 14]$). 
Both the \HD and \RTnsCRiMHD models host an abundant population of galaxies with $\magUV < -16$ at these early redshifts, in contrast to the more suppressed \HDBoost. The \RTnsCRiMHDrun model reproduces more consistently the shape and normalization of the observed UVLF. This model has a lower normalization than \HD, which shows a mild but systematic overabundance at $-17 \lesssim \magUV \lesssim -16\,\magab$. Because our volume samples a limited magnitude range, with only partial overlap with current observations, we focus on the relative model behavior rather than on the agreement with observations.

The \HDBoostrun model has a stronger early suppression of star formation, with UVLF measurements below observational constraints by $z \lesssim 12.5$. The degree of boosted SN feedback in this model needed to reproduce late-time stellar masses leads to overly efficient regulation of early star formation, causing tension with the abundant population of galaxies with $\magUV \lesssim -16\,\magab$ inferred by JWST \citep{Finkelstein2023, Harikane2023}.

The differences between the three models are imprinted early, within the first few hundred Myr of galaxy formation, and grow increasingly pronounced toward lower redshift. To investigate variations in the faint-end slope, we fit the UVLF with a single power-law form
\footnote{At these redshifts our volume contains galaxies typically fainter than $\magUV \sim -18$. Therefore, we fit the UVLF faint-end in the $-18 \lesssim \magUV \lesssim -14.5$ range. This is fainter than, and overlaps only partially with, the range probed by observations. As a result, the resulting $z > 10$ slopes provide only rough estimates.}.
At $z \gtrsim 12.5$, we infer similar faint-end slopes for our three models, with $\alphaUV$ in the range from $\sim -1.9$ to $\sim -2.0$.
These are similar or somewhat shallower than the observationally inferred range of $\alphaUV \sim -2.0$ to $-2.3$ \citep[e.g.,][]{Donnan2024, Whitler2025}.

The evolution of the UVLF from $z = 10$ to $z = 3$ is shown in the right panels of Fig.~\ref{fig:UVLF}. 
The \HD model under-regulates star formation, leading to overly bright galaxies. This is most pronounced for $M_{\mathrm UV} < -18\,\magab$, with number densities $\sim$1~dex above the observations\footnote{Note that the bright end of the UVLF is limited by our sampled volume, restricting our sampling of rare, bright galaxies.}.
The \HDBoost model systematically underpredicts the abundance of galaxies especially toward $z = 3$, while the \RTnsCRiMHDrun~model provides a reasonable match to observations across this entire redshift range.

We fit the faint-end slope once again with a single power law, now spanning $-20 \lesssim \magUV \lesssim -15.5$.
The \HDrun model is our shallowest model, flattening from $\alphaUV\,(z = 10) = -1.63 \pm 0.19$ to $\alphaUV\,(z = 3) = -1.46 \pm 0.11$. The \HDBoostrun model has a steeper high-redshift slope, evolving from $\alphaUV\,(z = 10) = -1.68 \pm 0.14$ to $\alphaUV\,(z = 3) = -1.65 \pm 0.35$. The \RTnsCRiMHDrun model is steepest at high redshift, with $\alphaUV\,(z = 10) = -2.16 \pm 0.36$, flattening to $\alphaUV\,(z = 6) = -1.75 \pm 0.25$ and $\alphaUV\,(z = 3) = -1.43 \pm 0.11$.

The inclusion of non-thermal physics moderates the efficiency and timing of star formation. It avoids the early overproduction of stars in massive galaxies while preventing the late-time suppression seen in our more simple physical models that account only for thermal SN feedback. 
We discuss a physical driver of this variation in Section~\ref{ss:outflows_metallicities}, tracking a shift in the energy budget of the neutral ISM from thermal to non-thermal support as redshift decreases.

\subsubsection{The cosmic star formation rate density evolution}

\begin{figure}[t!]
    \centering
    \includegraphics[width=\linewidth]
    {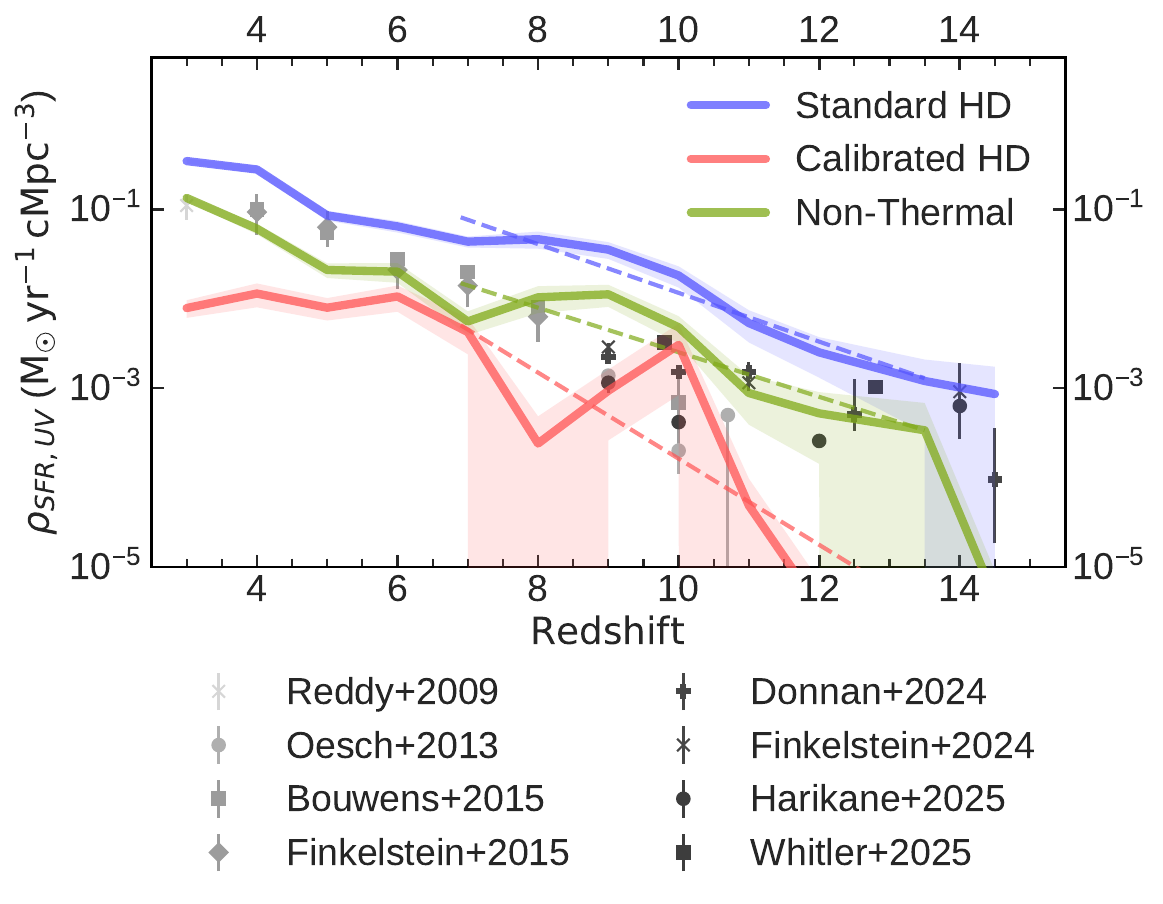}\\
    \caption{Evolution of the cosmic star formation rate density as inferred from the UV luminosity density. We compare \HDrun (blue), \HDBoostrun (red), and \RTnsCRiMHDrun (green) models with various observational estimates (grey data points). Error bands are estimated as for Figure~\ref{fig:UVLF}, also accounting for variance within each redshift interval.
    While the \HDrun model reproduces the early onset of the UVLF at high redshift, it overproduces stars at $z \lesssim 10$. \HDBoost has a lower SFRD at $z \gtrsim 11$, due to thermal SN feedback that is too strong and bursty. In this model, the observed SFRD is recovered somewhat around $z \sim 10$. However at lower redshifts it is too low again due to excessive explosiveness inherent to the model. Instead, \RTnsCRiMHD provides our best match to observations, with self-regulated star formation across cosmic time.}
    \label{fig:UVLD_evolution}
\end{figure}

A more compressed statistic to characterize cosmic star formation is the UV luminosity density (UVLD). Here we examine the cosmic star formation rate density (SFRD), which can be derived from the UVLD. Our results are shown in Fig.~\ref{fig:UVLD_evolution} together with observational estimates, from $z \sim 14$ to $z \sim 3$ \citep{Reddy2009, Oesch2013, Bouwens2015, Finkelstein2015, Donnan2024, Finkelstein2024, Harikane2025, Whitler2025}.
We first compute the UVLD integrating bright galaxies ($\magUV \leq -17\,\magab$), as done by several observational estimates \citep[e.g.,][]{Bouwens2015, Oesch2018, Donnan2024}. Following these studies, we use the UVLD to estimate the SFRD, $\rhoUV$, following a linear conversion
\begin{equation}
    \rhoUV \cdot V = \kappa_{\mathrm{UV}}\, L_{\mathrm{UV}},
\end{equation}
where $V$ is the volume of our simulation, and $\kappa_{\mathrm{UV}} = 1.15 \times 10^{-28}\; (M_\odot~\mathrm{yr}^{-1})/(\mathrm{erg}\ \mathrm{s}^{-1}\ \mathrm{Hz}^{-1})$ is a direct conversion factor \citep{Madau2014}.
For each redshift, we compute the median $\rhoUV$ across snapshots within $\Delta z = 0.5$. In addition to the volume estimate uncertainty, cosmic variance, and Poisson galaxy sampling noise, error bands also show variance within each redshift interval.

As expected, the SFRD evolution is consistent with our UVLF results (Fig.~\ref{fig:UVLF}). The \HDrun model significantly overproduces the SFRD for $z \lesssim 10$. This highlights the need for additional feedback mechanisms capable of suppressing $\rhoUV$ by $\sim0.5$--$1$~dex. The \HDBoost model has the opposite behavior: low SFRD at the highest redshifts, favored only by pre-JWST estimates \citep{Oesch2013}, and at odds with new data. The model matches observations at $z \sim 10$, but over-suppresses star formation by $z \sim 4$.
Combined, the \HD and \HDBoost models highlight an important limitation of simple, calibration-based approaches. These models can reproduce the SFRD for some range of redshifts, but struggle to capture the entire SFRD evolution with cosmic time, especially when taking into account very high redshift JWST data.

In contrast, our \RTnsCRiMHD model provides the best overall match to the cosmic SFRD across our studied redshift range. 
This reflects its bursty, self-regulated star formation histories early on (Section~\ref{ss:burstiness}), which transition from ejective to preventive self-regulation (Section~\ref{ss:outflows_metallicities}). The bursty phases produce compact, high-surface-brightness systems, which are also the most detectable in flux-limited surveys \PaperMPcitep.

Focusing on the high-redshift regime ($z \in [7, 12]$), now probed in detail by JWST, we fit the redshift evolution of the cosmic SFRD with a linear relation in log-space. The best-fit relations for $\log_{10} \left[\rhoUV / (\Msun \, \yr^{-1} \, \cMpc^{-3})\right]$ are:
\[ \begin{array}{l r}
(-0.222 \pm 0.032)\,z + (0.522 \pm 0.330) & \HDrun,\\
(-0.474 \pm 0.138)\,z + (1.097 \pm 1.419) & \HDBoostrun,\\
(-0.183 \pm 0.047)\,z + (-0.518 \pm 0.480) & \RTnsCRiMHDrun.
\end{array} \]
The steeper evolution of the \HDBoostrun model is due to the strong suppression of star formation in the first galaxies. The \RTnsCRiMHDrun model and the \HDrun model have shallower scalings, caused by rapid early star formation followed by a smoother evolution. The behavior and measured slope of the \RTnsCRiMHDrun model ($-0.183 \pm 0.047$) are consistent with recent observational constraints by \citet{Donnan2024}, which find a slope for this relation of $-0.140 \pm 0.068$.

\subsubsection{The galaxy stellar mass function and main sequence}

\begin{figure}[t!]
    \centering
    \includegraphics[width=\linewidth]{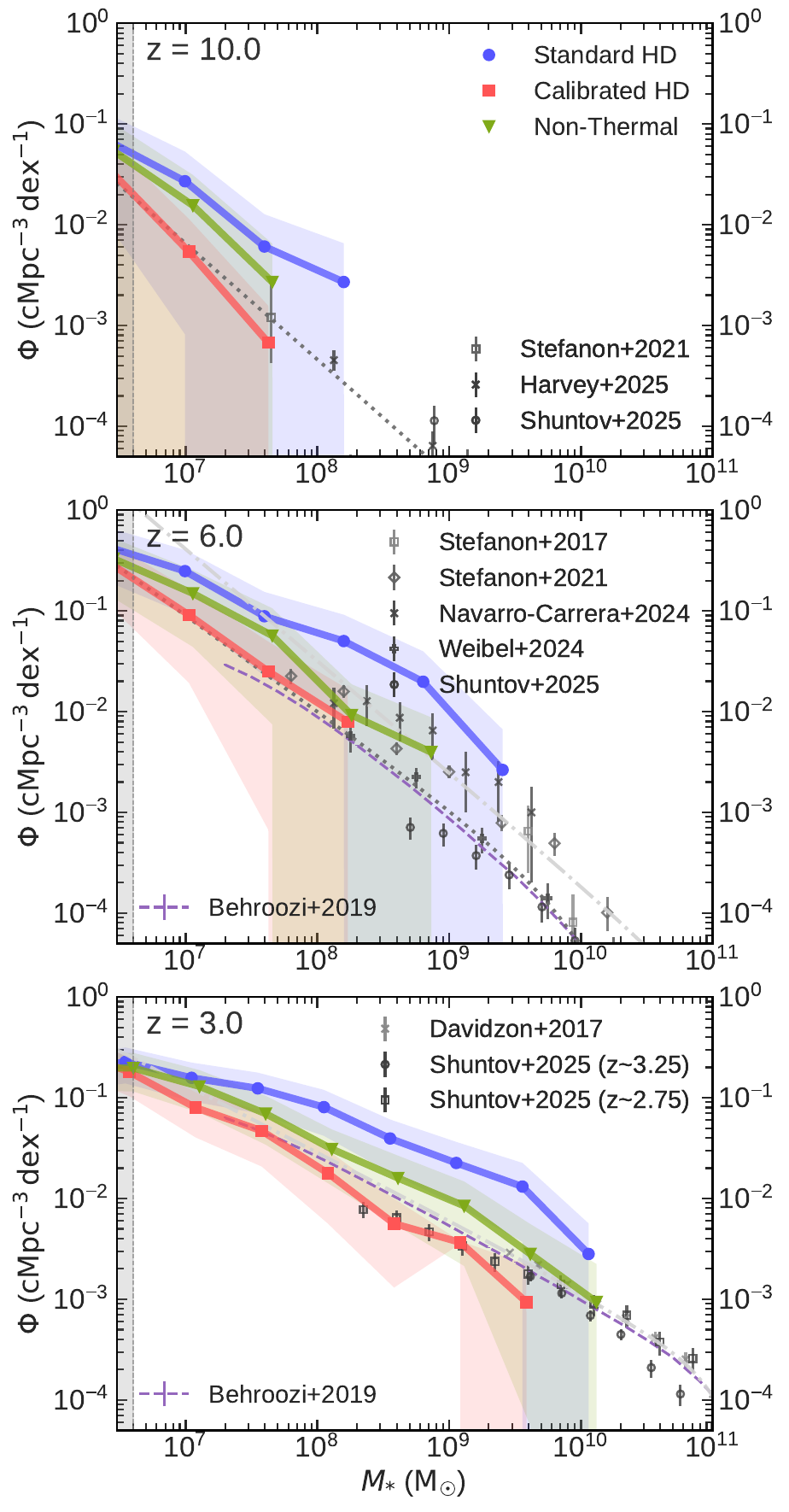}%
    \caption{Galaxy stellar mass function at $z = 10$ (top), $z = 6$ (center), and $z = 3$ (bottom) for our \HDrun (blue), \HDBoostrun (red), and \RTnsCRiMHDrun (green) models. Error bands are computed as for Fig.~\ref{fig:UVLF}. We compare our simulations with theoretical empirical estimates (purple lines; \citealt{Behroozi2019}), observational data (grey data points) and observational Schechter fits (grey lines; dot-dashed for \citealt{Davidzon2017}, and dotted for \citealt{Stefanon2021}).
    At redshift $z = 10$, all our three models have stellar mass distributions comparable or slightly above observations. At later times, \HDBoost provides the best fit to predictions by construction. The next-best agreement is the \RTnsCRiMHDrun\ model, which slightly overpredicts stellar masses. 
    As expected, \HD overpredicts stellar masses due to its star formation under-regulation.
    }
    \label{fig:GSMF}
\end{figure}

\begin{figure*}
    \centering
    \includegraphics[width=1.95\columnwidth]{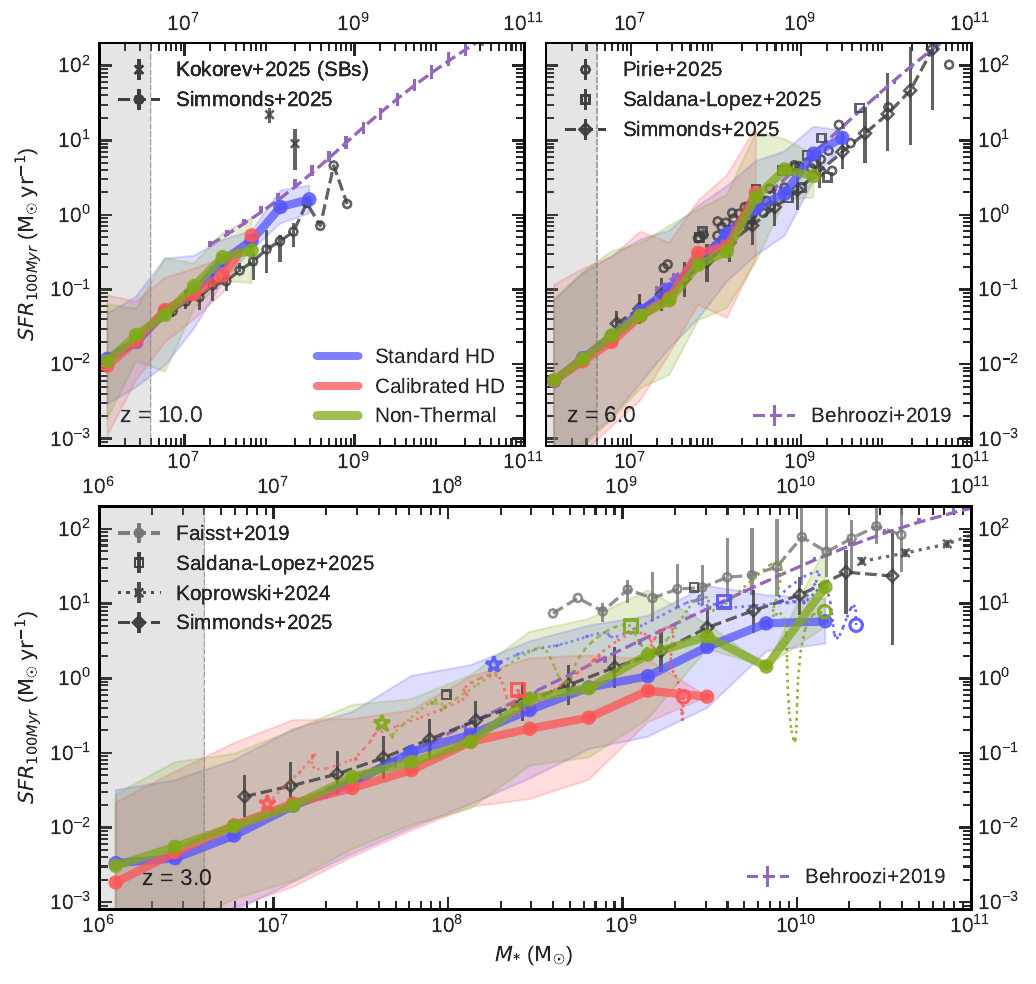}\\
    \caption{Galaxy main sequence of star-forming galaxies at $z = 10$ (top left), $z = 6$ (top right), and $z = 3$ (bottom) comparing our simulations with observational data (grey points), and empirical predictions (purple). Shaded bands display the 2~$\sigma$ variance of our measurements across the simulated galaxies. Observational datasets with more than 100 galaxies are summarized with median and quantiles. 
    Thin dotted lines in the bottom panel trace the redshift evolution (from $z = 10$, open stars; to $z = 6$, open squares; and $z = 3$, open circles) of the most massive galaxy in each simulation.
    All simulations have similar median main sequences, in agreement with the \citet{Simmonds2025} observations.
    }
    \label{fig:main_sequence}
\end{figure*}

To connect these results regarding cosmic star formation with the integrated growth of galaxies, we show the galaxy stellar mass function (GSMF) of our three simulations in Fig.~\ref{fig:GSMF}, with uncertainties computed as in Fig.~\ref{fig:UVLF}.
We compare against empirical model predictions from \citet{Behroozi2019}, a compilation of observational measurements spanning $z \sim 3$--10 \citep{Davidzon2017, Stefanon2017, Stefanon2021, Navarro-Carrera2024, Weibel2024, Harvey2025, Shuntov2025}, and Schechter functional fits \citep{Davidzon2017, Stefanon2021}.

The GSMF for the \HD model behaves as expected from our UVLF findings (Fig.~\ref{fig:UVLF}): it overpredicts stellar masses at all redshifts. 
By contrast, the \HDBoost model is designed to reproduce the GSMF at $z \sim 3$, and it subsequently provides a reasonable match to observations around that redshift. At $z = 6$ it is broadly consistent with the data, although it lies on the lower end of recent observational estimates. At $z = 10$ the \HDBoost model agrees with the masses inferred by \citet{Stefanon2021}, although its mean values lie somewhat below the JWST-inferred stellar mass number densities at the massive end.
The \RTnsCRiMHD model reproduces the observed stellar mass function well within the error band, but the mean values lie somewhat above empirical predictions and observations for all redshifts considered\footnote{By omitting stellar mass loss from stellar winds, our models do not include a channel of stellar mass reduction. For massive galaxies, these winds can return up to $\sim 1/3$ of the stellar mass to the ISM as gas.}.

We quantify the trends of the models by fitting single power-law functions to the GSMF, spanning $\Mstar \sim 10^{7}$--$10^{10}\,\Msun$. We focus on the low-mass end behavior as described by the low-mass-end slope $\alphaM$.
For the \HD model, $\alphaM$ mildly flattens from $\alphaM\,(z = 10) \simeq -1.79 \pm 0.39$ to $\alphaM\,(z = 3) \simeq -1.60 \pm 0.09$, as its low-mass galaxy population becomes overmassive. 
The \HDBoost run initially has a steeper faint-end slope (although relatively unconstrained due to a low number of galaxies), but becomes shallower at later times ($\alphaM\,(z = 3) \simeq -1.62\pm 0.20$).
The \RTnsCRiMHD model has a similarly steep slope at $z \sim 10$, with $\alphaM \simeq -2.16 \pm 0.47$, in agreement with
the inferred high-redshift behavior for the faint end. It progressively flattens down to $\alphaM\,(z = 6) \simeq -2.01 \pm 0.29$, and $\alphaM\,(z = 3) \simeq -1.59 \pm 0.11$. At $z \sim 3$ all of our three models are consistent with the observed low-mass slope ($\alphaM \sim -1.4$ to $-1.6$; \citealt{Davidzon2017, Stefanon2021}). Despite this, we caution that the individual slope fits have large associated uncertainties due to the volume of the simulations, and note that our galaxy populations only overlap with the observations over $\sim1$~dex of stellar masses ($\Mstar \sim 10^{9}$--$10^{10}\,\Msun$).

The main sequence (MS) of star-forming galaxies connects the GSMF and the UVLD by directly exploring whether large systematic variations in star formation efficiency take place. We show this relation in Fig.~\ref{fig:main_sequence}, comparing our three models with observational datasets \citep{Faisst2019,  Koprowski2024, Kokorev2025, Simmonds2025, Pirie2025, Saldana-Lopez2025} and theoretical estimates \citep{Behroozi2019}.
Our SFR values are derived from the total mass of stars formed in a galaxy during the last 100~Myr. Measurements combine all galaxies across snapshots within $\Delta z = 0.5$. 
To illustrate how the evolution of an individual galaxy differs from the intrinsic evolution of the relation, we plot the in-situ SFR--$\Mstar$ redshift evolution of the most massive galaxy in each of our models (thin dotted lines, bottom panel). 

Despite the differences in the UVLF and GSMF discussed above, the three models are in good agreement with observations from \citet{Simmonds2025}. Starburst candidates from \citet{Kokorev2025} are well above the main sequence, as expected for rare and short-lived outliers, not well sampled in our limited simulation volume. 

At $z = 6$, the \RTnsCRiMHD and \HDBoost models exhibit mildly higher MS scatter, suggesting higher star formation burstiness \citep{Cole2025}. 
At this redshift, the median MS relation steepens across our three models, and especially for \RTnsCRiMHDrun and \HDBoostrun, driven by enhanced SFRs at $\Mstar \gtrsim 10^8\,\Msun$. 
The three models are still consistent with observational data by \citet{Simmonds2025}, and with the extrapolation of empirical estimates by \citet{Behroozi2019}. The redshift trajectories of the most massive galaxy in these two models exhibit star formation peaks around $z\sim 6$ (colored open squares), in contrast with the smoother evolution in the \HDrun model. This suggests that differences in the UVLF are related to the duty cycle and burstiness of star formation.

At $z = 3$, our simulated median MS relations lie within the observed scatter \citep{Faisst2019, Simmonds2025, Clarke2025}, with \RTnsCRiMHDrun in closest agreement with data by \citet{Simmonds2025}. 
The most massive galaxy trajectories continue to show clear differences: the \HDrun model maintains a smoother SFR than the other models, with rapid variations around $z \sim 3.5$ driven by multiple mergers. Conversely, \RTnsCRiMHD and \HDBoost grow through intermittent bursts of high specific star formation rate (sSFR).
As our measurements average star formation over the last 100~Myr, further understanding differences between our models motivates a more detailed analysis of star formation burstiness.

\subsection{Quantifying the drivers of star formation burstiness}
\label{ss:burstiness}

\begin{figure}[t!]
    \centering
    \includegraphics[width=\columnwidth]{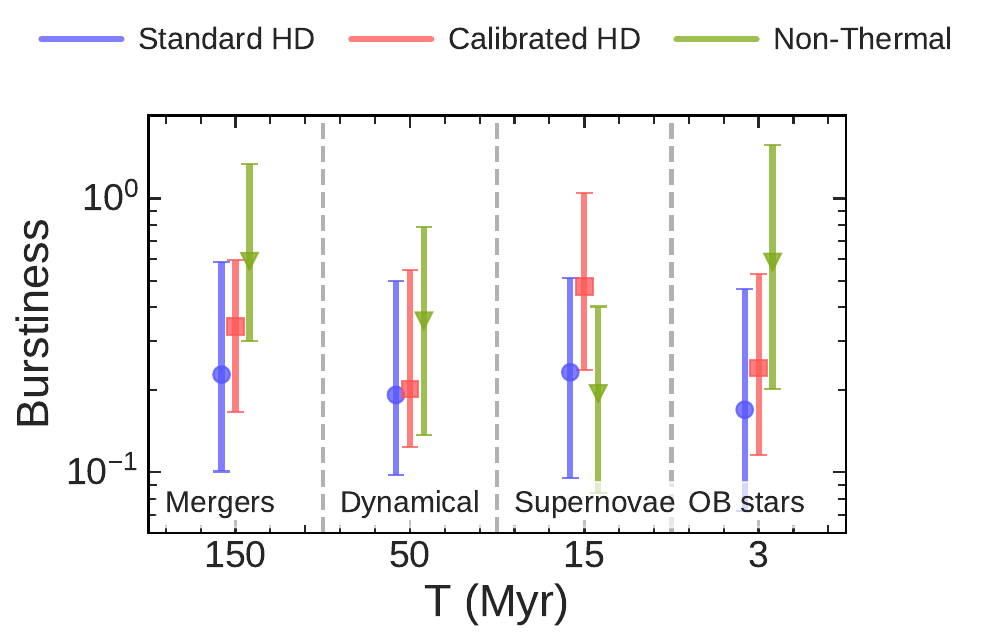}\\
    \caption{Star formation burstiness (defined as the spectral variance integrated across four characteristic timescale bands; see text) for galaxies with stellar mass $\Mstar(z = 3) > 5\times10^7\,\Msun$, separated for each of the three simulations. Error bars show the 1~$\sigma$ variance across the full distribution of these galaxies.
    Timescales are broadly associated with mergers ($T \approx 150\,\Myr$), dynamical cycles ($50\,\Myr$), SN feedback ($15\,\Myr$), and OB-star lifetimes ($3\,\Myr$).
    All models show high variability at long timescales, associated with cosmological accretion and hierarchical growth. 
    The \RTnsCRiMHD model has the highest burstiness, both at long timescales and at the shortest ones ($\lesssim10$~Myr). The \HDBoost model has the highest variability at SN cycling timescales, whereas the \HDrun model has the lowest fluctuations.}
    \label{fig:FFT_insitu}
\end{figure}

In this section, we review how the degree of burstiness of star formation varies across our three simulations. We first present a direct calculation of SFR burstiness, and then connect this measurement with observationally accessible estimates.

Our direct burstiness measurement reflects the temporal clustering of in-situ star formation measured through the power spectrum of the fractional variability in the star formation history (SFH) of galaxies. To ensure well-resolved SFHs, we use for our analysis galaxies with stellar masses at $z = 3$ of $\Mstar> 5 \times 10^7\,\Msun$. Details of how in-situ SFHs are built, our treatment of mergers, and signal and noise processing are provided in Appendix~\ref{ap:burstiness_FFT}.

The result is shown in Fig.~\ref{fig:FFT_insitu}. We quantify burstiness using the spectral star formation variance $\sigma^2$ integrated over four characteristic timescale bands, corresponding to: (i) halo growth and galaxy mergers ($\sim$150~Myr; \textit{Mergers}), (ii) galactic dynamical cycles and gas recycling ($\sim$50~Myr; \textit{Dynamical}), (iii) supernova-driven regulation ($\sim$15~Myr; \textit{Supernovae}), and (iv) short-timescale variability associated with massive stars ($\sim$3~Myr; \textit{OB stars}). Error bars show the variance from the full distribution of galaxies.

At the longest timescales (\textit{Mergers} timescale band), driven by cosmological accretion and hierarchical growth, all models have very high burstiness. The \RTnsCRiMHDrun model exhibits the strongest cyclic variations at these timescales, reflecting a more pronounced star formation response to events such as accretion and galaxy mergers. Non-thermal physics promotes the accumulation of gas and its rapid depletion in bursts of star formation.

At $\sim$50~Myr (\textit{Dynamical} timescale band), the \RTnsCRiMHDrun model again displays the highest burstiness, likely associated with large-scale gas cycling and ISM gas inflows and outflows modulated by non-thermal processes.
At intermediate timescales (\textit{Supernovae} timescale band), the \HDBoost model shows the highest burstiness, reflecting its strong feedback-driven suppression and the subsequent cyclic re-ignition of star formation. This is consistent with previous studies connecting very strong momentum-driven feedback with bursty SFHs \citep[e.g.,][]{Governato2010, Faucher-Giguere2013, Hopkins2014}. 

At the shortest timescales (\textit{OB stars} timescale band), the \HD and \HDBoost models flatten, whereas the \RTnsCRiMHD simulation has a burstiness upturn. This substantial short-timescale variability is again driven by non-thermal and early stellar feedback \citep{Martin-Alvarez2023, Martin-Alvarez2026}. Dedicated studies confirmed this effect for isolated channels such as FUV radiative feedback for metal-poor high-redshift galaxies \citep{Sugimura2024}.

Overall, \HDrun exhibits the lowest burstiness across all timescales. The calibrated model increases variability primarily at SN-regulated scales, while the \RTnsCRiMHDrun model shows enhanced burstiness across most of the  bands, with distinct signatures at the dynamical and shortest timescales. This supports a scenario where non-thermal physics drives additional variability channels, leading to more structured and temporally clustered star formation histories.

\begin{figure*}
    \centering
    \includegraphics[width=1.9\columnwidth]{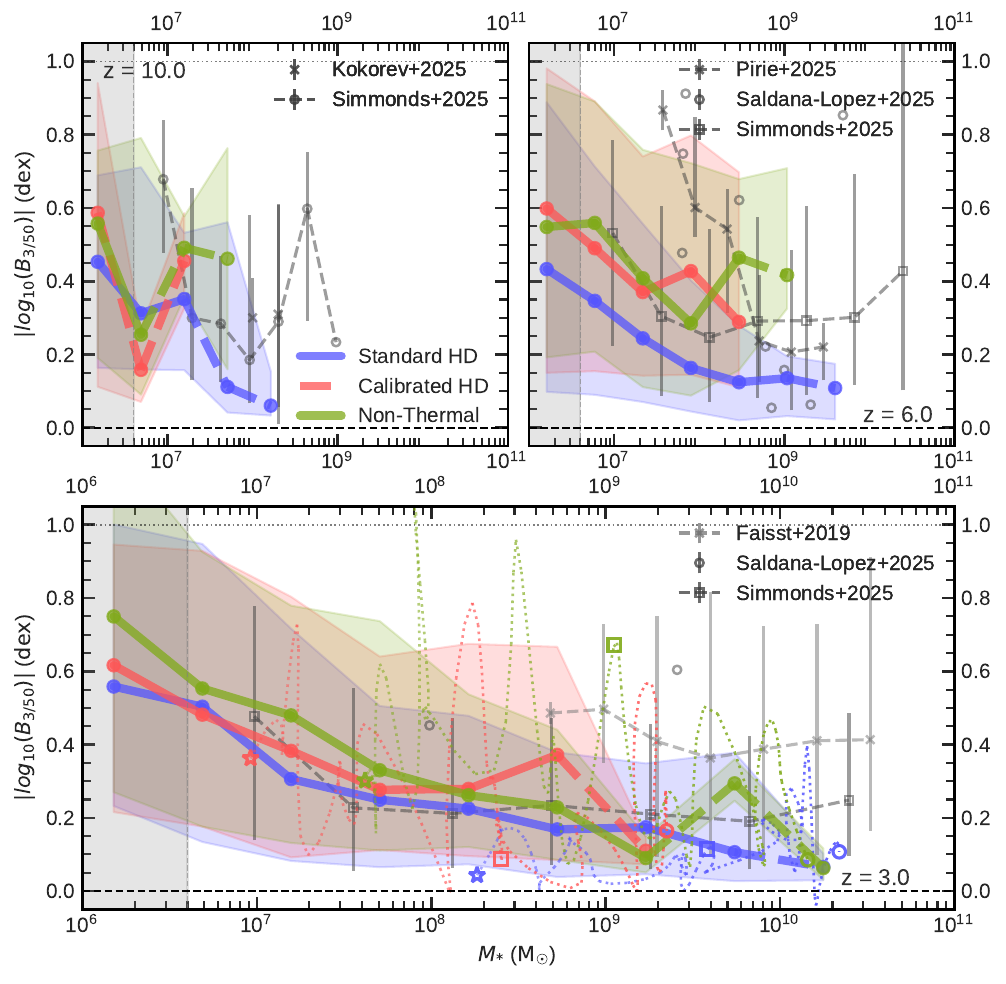}\\
    \vspace{-0.3cm}
    \caption{Burstiness parameter $\Bburst = \SFRthree /\,\SFRfifty$ (H$\alpha$/UV analogue; see Appendix~\ref{ap:burst} for the $10/100$~Myr definition, which yields consistent results). We show $\Bburst$ as a function of stellar mass at $z = 10$ (top left), $z = 6$ (top right), and $z = 3$ (bottom) for our standard hydrodynamical (\HD, blue), calibrated hydrodynamical (\HDBoost, red), and \fullphys (\RTnsCRiMHD, green) models. Lines are shown as dashed for stellar mass bins with fewer than 10 galaxies, and shaded bands display the 1~$\sigma$ galaxy distribution scatter. In the bottom panel, thin dotted lines trace the evolution of the most massive galaxy in each simulation, from $z = 10$ (open stars), to $z = 6$ (open squares), and $z = 3$ (open circles). We include observational burstiness estimates for comparison (grey points).
    %
    %
    We find significantly burstier high-redshift star formation for the \RTnsCRiMHD and \HDBoost models. These differences persist until $z = 3$, when they become less pronounced. The \RTnsCRiMHDrun model is in good agreement with observations by \citet{Simmonds2025}, followed by \HDBoostrun.}
    \label{fig:burstiness_SF}
\end{figure*}

Our (magneto-)turbulence-regulated SFE prescription is similar to that of \citet{Semenov2025}, who argue such prescriptions naturally drive stronger SFR variability, especially in the first galaxies. While their proposed disk-transition mechanism likely contributes to enhanced burstiness, the primary driver of variability in our models is more generic: a stringent local SFE threshold produces gas accumulation and rapid star-formation release \citep[e.g.,][]{Faucher-Giguere2013, Iyer2024}. 
Two results support this: i) burstiness decreases systematically with stellar mass (Fig.~\ref{fig:burstiness_SF}) across both irregular and disk-like systems; and ii) our \HDrun model shows the lowest burstiness at every timescale (Fig.~\ref{fig:FFT_insitu}) despite sharing the SFE prescription with the other two models. In the absence of sufficient feedback, the prescription saturates rather than producing discrete accumulation-release events.
These results are consistent with \citet{Dome2025}, who found that burstiness in the \azahar simulations systematically increased with the inclusion of radiation in addition to magnetism, and was the highest when cosmic rays were combined with these two components. The specific influence of each non-thermal component will be systematically addressed in future work.

To connect these results with observations, we study the ratio of SFR measured on short (3~Myr; traced by H$\alpha$) and long (50~Myr; traced by UV emission) timescales, $\Bburst = \SFRthree/\,\SFRfifty$, which is a simple estimator of short-timescale variability. Appendix~\ref{ap:burst} motivates our specific timescale selection, and shows that adopting a canonical $\SFRten/\,\SFRhundred$ yields very similar results. 
We show in Fig.~\ref{fig:burstiness_SF} the variation of $|\log_{10}(\Bburst)|$ (values of 0 and 1 correspond to low and high burstiness, respectively) as a function of stellar mass. We combine galaxies from multiple snapshots within $\Delta z = 0.5$ of the target redshift, and exclude galaxies with null SFR. We compare our results with observations from \citet{Faisst2019}, \citet{Pirie2025}, \citet{Simmonds2025}, and \citet{Kokorev2025}.

Our models preserve the same ordering at all times: \RTnsCRiMHDrun has the most bursty star formation, closely followed by \HDBoostrun. The \HDrun model is the least bursty. This trend is preserved across stellar masses, with burstiness systematically decreasing toward higher stellar masses at all redshifts \citep{Endsley2025}. Burstiness also decreases with cosmic time, most rapidly after $z \sim 6$, and with a stronger trend for more massive systems. Our smallest galaxies ($\Mstar \lesssim 10^{7}\,\Msun$) have approximately unchanged burstiness across the studied redshifts.
This confirms that the short-to-long star formation ratio is capable of capturing the trends that have been determined through the analysis of the power spectrum of the fractional variability in the SFH of galaxies\footnote{While elevated $\Bburst$ values may indicate recent upturn in star formation, the population scatter reflects more closely the underlying SFH variability, and is a better proxy for intrinsic burstiness (Fig.~\ref{fig:FFT_insitu}).}.
Simulations such as \textsc{thesan-zoom} find comparable increases in short-timescale SFR variability toward high redshift \citep{McClymont2025a}, which in their employed model is the result of rapid gas inflows rather than feedback.
At high redshift ($z = 10$ and $6$), the \RTnsCRiMHDrun and \HDBoostrun models are both consistent with the burstiness observed by \citet{Simmonds2025}. The scatter in these models provides a good explanation for the $z = 10$ starbursts observed by \citet{Kokorev2025}, and the data from \citet{Pirie2025} at $z \sim 6$.
By $z = 3$, all models have reduced burstiness at $\Mstar \gtrsim 10^{8}\,\Msun$ and become more comparable. Despite this, the higher scatter of the \RTnsCRiMHDrun and \HDBoostrun models provides an explanation for observations by \citet{Faisst2019}, which, being shallower than modern JWST surveys, may sample the high-burstiness, more luminous tail of the galaxy population.

In the bottom panel of Figure~\ref{fig:burstiness_SF}, thin dotted lines show the evolution of the most massive galaxy in the three simulations. These tracks demonstrate how the observed $\Bburst$ emerges from the intrinsic variability of the individual systems. While the median of our models lies below the values reported by \citet{Pirie2025} and \citet{Faisst2019}, individual galaxies undergo phases of burstiness in good agreement with these observations, as expected for studies preferentially probing the luminous, high-burstiness tail of the population \citep{Simmonds2025}.

\subsection{Transition from ejective to preventive self-regulation in the presence of non-thermal physics}
\label{ss:outflows_metallicities}

\begin{figure}[t!]
    \centering
    \includegraphics[width=\columnwidth]{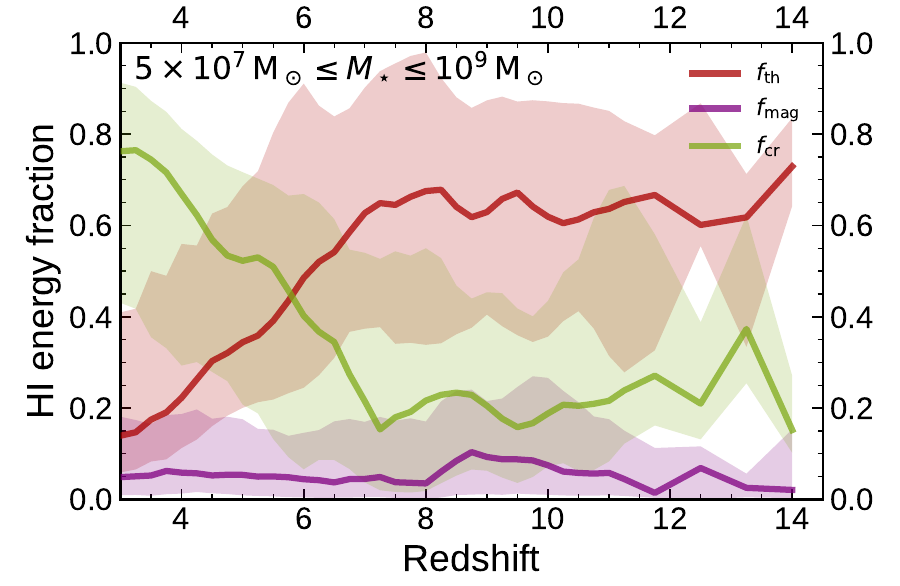}\\
    \caption{Fractional contribution of thermal ($f_{\rm th}$, red), cosmic ray ($f_{\rm cr}$, green), and magnetic ($f_{\rm mag}$, purple) energy to the energy balance of the H\,\textsc{i} gas as a function of redshift, for galaxies in the \RTnsCRiMHDrun model with $5\times10^{7} \leq \Mstar \leq 10^{9}\,\Msun$. Lines show the median, and shaded bands the 1~$\sigma$ scatter across the population.
    The relative contribution of cosmic rays is secondary at early times and increases toward lower redshift, steeply so after $z \sim 7$.}    
    \label{fig:energy_budget}
\end{figure}

To better understand the systematic differences identified above for the \RTnsCRiMHDrun model, we review the energy balance of the neutral gas, the reservoir that feeds star formation. In Fig.~\ref{fig:energy_budget} we show the thermal, cosmic ray, and magnetic contributions to the energy budget of the H\,\textsc{i} gas, for galaxies with $5\times10^{7} \leq \Mstar \leq 10^{9}\,\Msun$ in the \RTnsCRiMHDrun model. At high redshift, the neutral ISM is thermally dominated, with subdominant cosmic ray and magnetic contributions. As redshift decreases, the relative importance of the cosmic-ray contribution grows steadily, and most rapidly after $z \sim 7$. Appendix~\ref{ap:energy_budget} shows the same measurement for galaxies above and below this mass range. The transition is present for all galaxy masses, and becomes more pronounced for more massive galaxies. Interestingly, this transition is weakest in our smallest galaxies, which remain bursty across cosmic time. 
The ionized gas (H\,\textsc{ii}) displays a similar transition, but with a systematically higher thermal fraction (not shown).

As a result, the star formation regulation relies primarily on the energy injected by supernovae at early times, akin to the hydrodynamical models. As cosmic rays accumulate in the ISM, their smooth and sustained pressure contributes to suppressing in-situ star formation. The contribution from cosmic ray and magnetic pressure is most important during star-formation episodes, in rough equipartition with the turbulent energy, and concentrates in regions of elevated star formation \citep{Belfiori2026}.
This transition motivates studying the outflows and metal enrichment of the simulated galaxies, to understand how the interplay of star formation burstiness and non-thermal support affects wind launching and the self-regulation mechanism at play in our galaxies.

\begin{figure*}
    \centering
    \includegraphics[width=1.9\columnwidth]{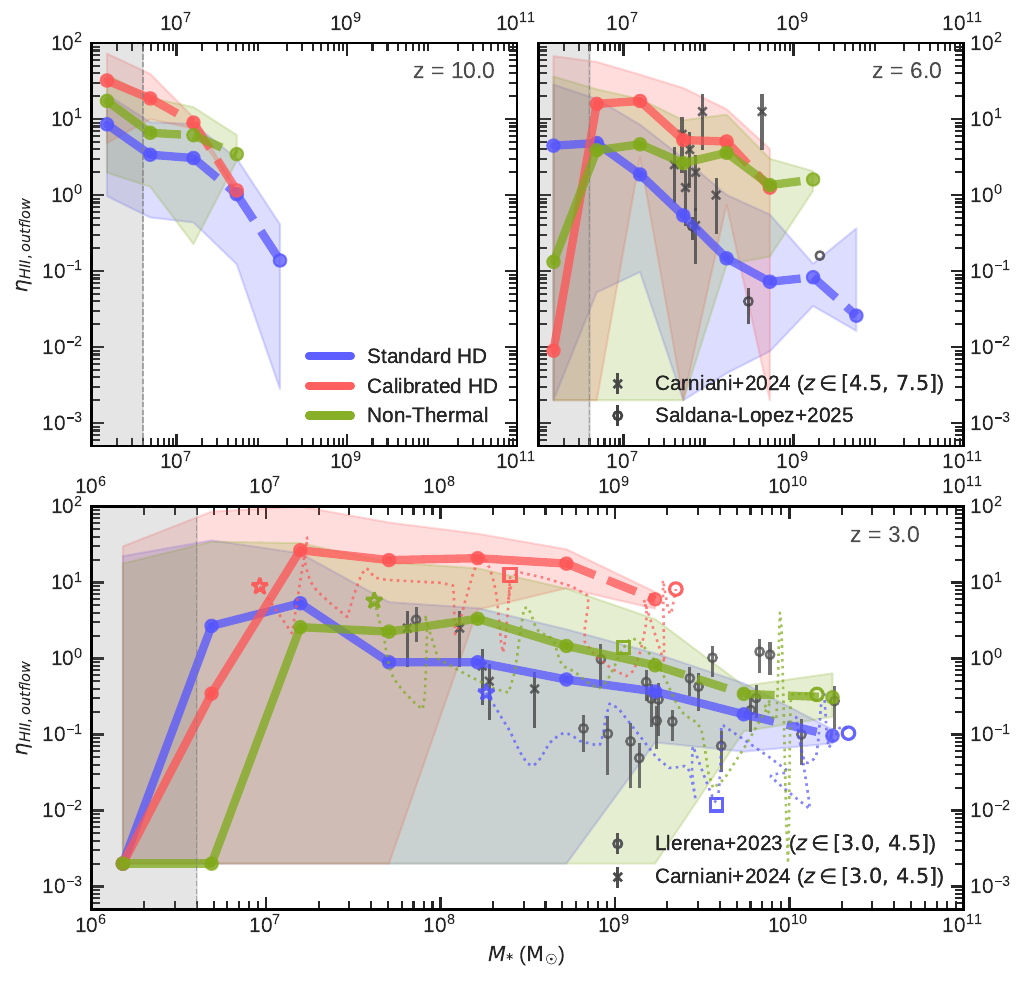}\\
    \caption{Ionized gas outflow mass-loading factor as a function of stellar mass at $z = 10$ (top left), $z = 6$ (top right), and $z = 3$ (bottom) for the \HDrun (blue), \HDBoostrun (red), and \RTnsCRiMHDrun (green) models, compared with observational estimates (grey). Lines are shown as dashed for stellar mass bins with fewer than 10 galaxies, and shaded bands show the 1~$\sigma$ galaxy population scatter. Thin dotted lines in the bottom panel show the evolution of the most massive galaxy from $z = 10$ (open stars), to $z = 6$ (open squares), and $z = 3$ (open circles).
    Increasing SN strength (\HDBoostrun) increases mass-loading factors at all times, matching $z \sim 6$ observations. However, the model disagrees with observations at later times, which are better reproduced by \HDrun. The \RTnsCRiMHD model has a more complex behavior, and transitions between these regimes: from more ejective outflows at high redshift to more preventive outflows at later times, once a sufficient cosmic ray pressure builds up.}
    \label{fig:mass_loading}
\end{figure*}

For a more direct comparison with observations, we show in Fig.~\ref{fig:mass_loading} the ionized gas mass-loading factor as a function of stellar mass at $z = 10$, $6$, and $3$. Results combine all galaxies from snapshots within $\Delta z = 0.5$. We compute mass-loading factors as
\begin{equation}
    \eta_{\rm HII,outflow} = \frac{\dot{M}_{\rm HII,outflow}}{\SFRhundred},
    \label{eq:mass_loading_factor}
\end{equation}
where $\dot{M}_{\rm HII,outflow}$ is the ionized-gas outflow rate measured across the inner-CGM boundary and $\SFRhundred$ is the star formation rate over the last 100~Myr.

Across cosmic time, we identify three main trends: i) mass-loading factors increase toward lower stellar masses, especially at high redshifts; ii) in the \RTnsCRiMHD model $\eta_{\rm HII,outflow}$ decreases with cosmic time, indicating a transition from more ejective outflows to more moderate, preventive outflows at later times; and iii) the models preserve their relative ordering, with \HDBoost yielding the highest mass-loading factors, and \HD the lowest. 

At $z = 10$, all models predict mass-loading factors typically higher than unity, being of the order of $10$ for low mass galaxies. No robust observational outflow constraints are yet available at this redshift for direct comparison.
By $z = 6$, observations by \citet{Carniani2024} become available, spanning a comparable mass range ($\Mstar \sim 10^{7.5}$--$10^{9}\,\Msun$), complemented by the outflow constraints of \citet{Saldana-Lopez2025}. Both the \RTnsCRiMHD and \HDBoostrun models are consistent with the observed $\eta_{\rm HII,outflow} \gtrsim 1$, with the \RTnsCRiMHDrun model exhibiting a larger scatter. From this redshift onward, an increasing fraction of low-mass galaxies ($\Mstar \lesssim 10^{7}\,\Msun$) experience extended periods of quiescence. This causes a large scatter in $\eta_{\rm HII,outflow}$, and shifts the distribution toward our floor ($2\times10^{-3}$). Such quiescence is less prevalent in the \HDrun model, sustaining the $\eta_{\rm HII,outflow}$ trends down to lower masses.

By redshift $z = 3$, observational data \citep{Llerena2023} reveal a likely transition from ejective ($\eta_{\rm HII,outflow} > 1$) to preventive ($\eta_{\rm HII,outflow} \lesssim 1$) self-regulation. By this redshift, and in agreement with observations, the \RTnsCRiMHDrun model has evolved toward much more moderate mass-loading factors across the entire mass range explored. The observations clearly disfavor the highly ejective \HDBoostrun. However, it is worth emphasizing that observational systematics may affect the retrieved mass-loading factors, particularly when estimated from the broad components of H$\alpha$ line profiles, which may bias $\eta_{\rm HII,outflow}$ down \citep{Martin-Alvarez2026}, especially for low-mass galaxies.

Finally, the dotted line tracks show the evolution of the most massive galaxy in the three simulations, highlighting how individual systems transition through stages of both highly ejective and lower outflow activity. Across their evolution, the mass-loading factor of each galaxy decreases on average. Large deviations are caused by episodes of intense star formation and subsequent quiescent phases, as seen for the \RTnsCRiMHDrun galaxy at $\Mstar\sim10^{10}\,\Msun$.

\begin{figure*}
    \centering
    \includegraphics[width=2\columnwidth]{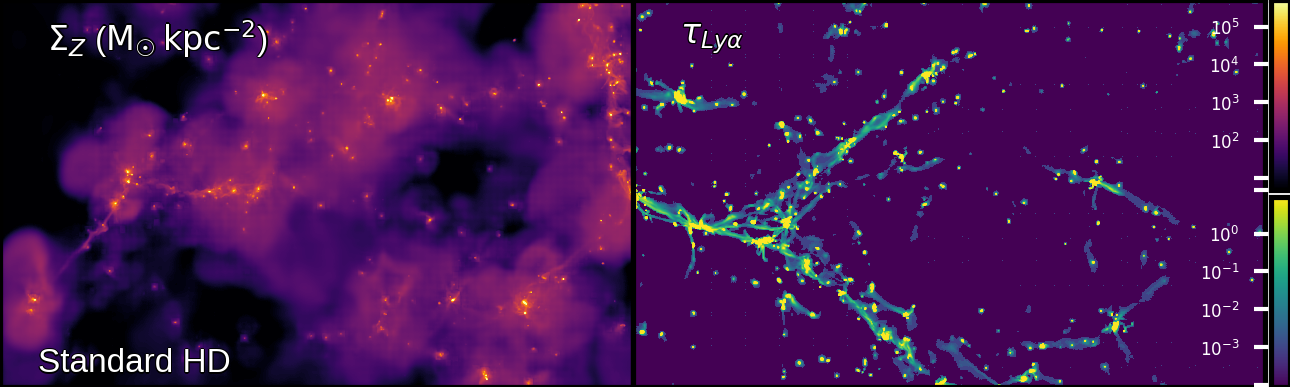}\\    
    \includegraphics[width=2\columnwidth]{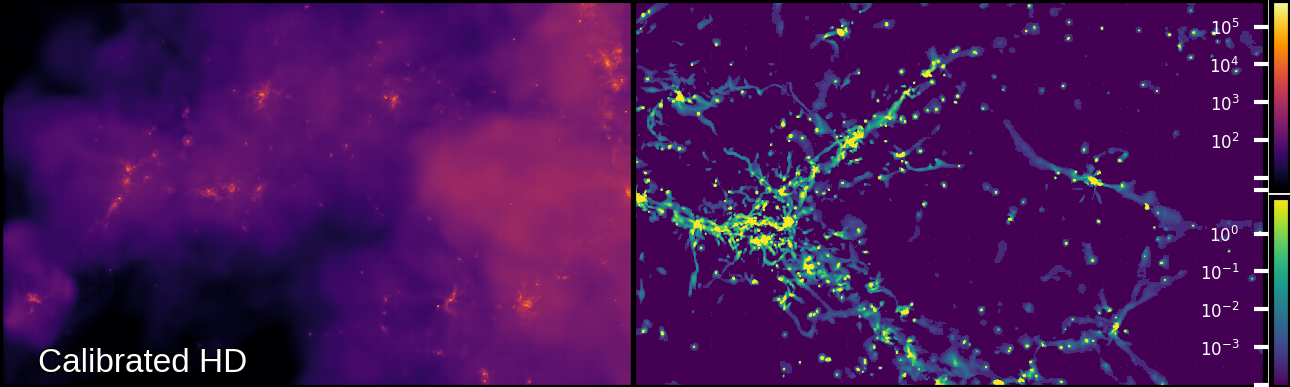}\\    
    \includegraphics[width=2\columnwidth]{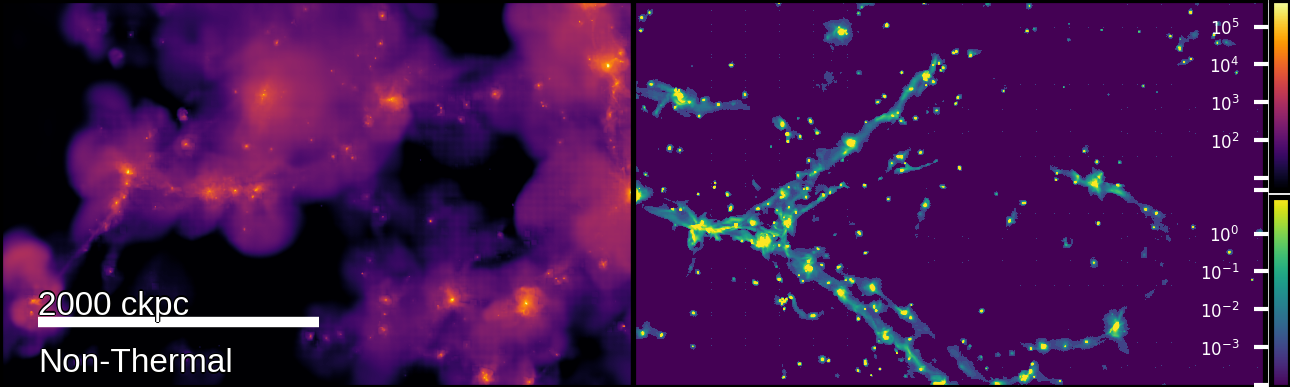}\\    
    \caption{Maps of the metal mass surface density (left panels, $\Sigma_{\rm Z}$) and Ly$\alpha$ optical depth measured over fixed line-of-sight scales (right panels, $\tau_{\rm Ly\alpha}$). Panels display the same cosmological region as in Fig.~\ref{fig:front_presentation}. 
    Different outflow strengths and metallicities imprint large variations in the CGM and IGM across our models. The \HD case (top) has confined enrichment, with metals concentrated in and around galaxies, as well as  highly opaque and coherent filaments. The \HDBoost model (center) shows an overly extended metal distribution, with fragmented filamentary structure, and reduced Ly$\alpha$ opacity in dense regions. Conversely, the \RTnsCRiMHDrun model (bottom) produces a structured cosmic web, with metals ejected to large distances, with cosmic filaments still being coherent but having a reduced opacity and smoother small-scale structure.}
    \label{fig:optical}
\end{figure*}

These differences in outflow strength and thermal state suggest three types of outflows: largely confined (\HD), overly extended (\HDBoost), and structured (\RTnsCRiMHD). Each of these will have a clear imprint on the ionization and metal distribution of the CGM and IGM. 
We show these in Fig.~\ref{fig:optical}, which displays the metal surface density maps (left panels) and Ly$\alpha$ optical depth (right panels) at $z = 3$ for the three simulations. The Ly$\alpha$ optical depth is computed using the neutral hydrogen fraction estimated with the self-shielding prescription of \citealt{Rahmati2013} (see Appendix~\ref{ap:Rahmati} for methods discussion and validation).

At the scale of galaxies, these differing outflows will affect how much gas and metals are removed from the densest regions of the cosmic web. The \HD model shows highly peaked Ly$\alpha$ absorption and overly concentrated metal distributions. The \HDBoost model significantly suppresses central Ly$\alpha$ absorption peaks, and has galaxies with significantly lower metallicities. Although the metal distribution is spatially smoother, galaxy metallicities remain high in the \RTnsCRiMHD model. A similar effect is seen in the Ly$\alpha$ absorption, which is less clumpy in these systems.

At the scale of cosmic filaments, the impact of the different models is most apparent in the Ly$\alpha$ absorption. In the \HD model, the filaments are narrow and coherent, and remain mostly unaffected by SN feedback. Feedback in the \HDBoost generates clear disruptions of filamentary structures, leading to fragmented Ly$\alpha$ absorption and a metal distribution that no longer traces the underlying density field. The feedback in the \RTnsCRiMHD model preserves filament connectivity, but reduces filament opacity and smooths their radial absorption profiles. 

At IGM scales, the differences in Ly$\alpha$ absorption between the models are dominated by variations in their small-scale structure. The absorption is smoother in the \RTnsCRiMHD model and patchier in the hydrodynamical cases. In contrast, metallicities have highly pronounced variations. In the \HD model, metals are confined to bubbles with steep radial profiles, tracing the cosmic web. Feedback in the \HDBoost simulation leads to highly extended and overlapping bubbles of enrichment, resulting in a quasi-uniform metallicity distribution. While the \RTnsCRiMHD model has a more extended metal enrichment than the \HD model, its metal distribution continues to trace the cosmic web. Metallicity profiles have significantly shallower declines in this model, leading to high enrichments on $\sim 1\,\cMpc$ scales, but maintaining extensive void pockets of pristine gas.

\begin{figure}[t!]
    \centering
    \includegraphics[width=\linewidth]{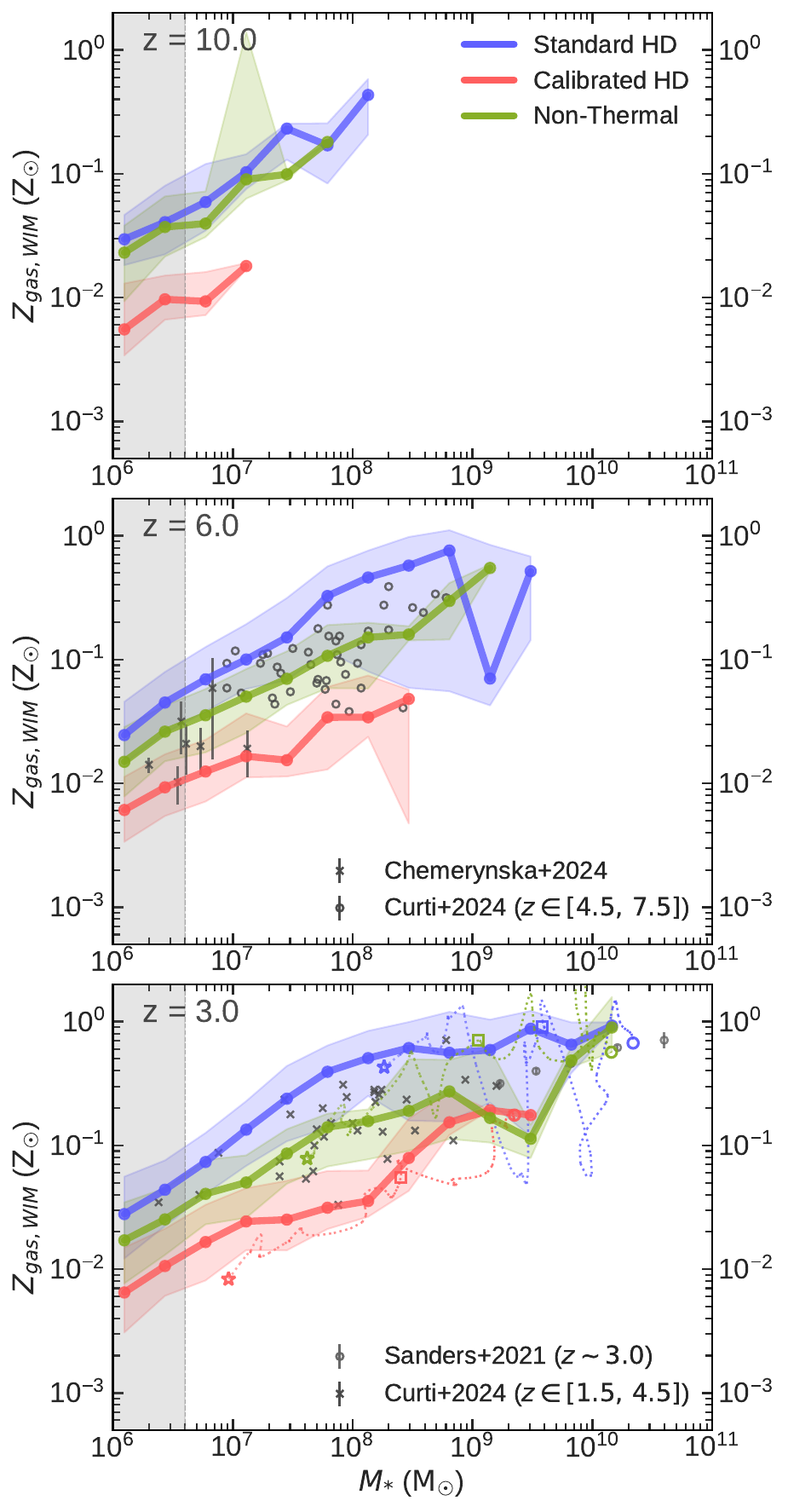}\\
    \caption{WIM gas mass-metallicity relation at $z = 10$ (top), $z = 6$ (center), and $z = 3$ (bottom), comparing the three simulations (\HD in blue, \HDBoostrun in red, and \RTnsCRiMHDrun in green) with observational data (grey). 
    Shaded bands correspond to the 1~$\sigma$ galaxy distribution scatter. Dotted lines in the bottom panel show the evolution of the most massive galaxy in each model.
    The \RTnsCRiMHDrun model is in best agreement with the observed MZR across all times, whereas the \HDBoostrun model underpredicts gas metallicities at $z = 6$ and $z = 3$, due to its over-ejective feedback self-regulation. This indicates that the observed MZR is a good discriminator of baryonic physics.}
    \label{fig:metallicity_mass_relation}
\end{figure}

The observed redistribution of metals across CGM and IGM is connected with metallicity variations within galaxies. These metallicities provide an accessible diagnostic of integrated star formation, feedback, and the internal chemical state of galaxies.
In Fig.~\ref{fig:metallicity_mass_relation}, we compare the mass-metallicity relation (MZR) of our models with recent observations at $z \gtrsim 3$ \citep{Sanders2021, Curti2024, Chemerynska2024}. Measurements combine all galaxies from snapshots within $\Delta z = 0.5$.
For a better comparison with emission-line metallicities, we show gas metallicities in the warm ionized medium (WIM) phase\footnote{We compute WIM gas density as $\rho_{\mathrm{WIM}} = \rho\, x_{\mathrm{HII}}$ for gas cells with temperatures $200\,\mathrm{K} < T < 10^{6}\,\mathrm{K}$.}. We review in Appendix~\ref{app:MZR_gas} the MZR of the total gas distribution, and show that the WIM MZR closely follows the trends of the total gas metallicity (subject to minor underestimation in the \HD model). 

The three models display a clear and systematic separation of their metallicities at all times. The \HD model consistently produces the highest enrichments, due to a combination of an overproduction of stars and less ejective feedback. Conversely, the \HDBoost model has the lowest metallicities, due to the opposite effects being at play. The \RTnsCRiMHD model lies between these two scenarios, reflecting a balance between regulated star formation, efficient but less disruptive feedback, and a reduced fraction of metals entrained in its outflows \citep{Martin-Alvarez2026}. This efficient but less disruptive feedback scenario and its consequences for the MZR are supported by new JWST observations \citep{Curti2024, Chemerynska2024}, as well as earlier results from \citet{Sanders2021}.
The MZR in our models does not evolve significantly across cosmic time,  undergoing only a slight increase in normalization. However, redshift evolution is more prominent for the total gas MZR (Appendix~\ref{app:MZR_gas}) and the stellar metallicities (not shown). While this, combined with the relatively narrow population scatter, suggests a more stable WIM phase enrichment, individual galaxy evolution trajectories (dotted colored lines) span a much broader range of metallicities. This highlights the dynamic nature of the WIM phase, as metals cycle in and out during accretion and outflow episodes.

\section{Conclusions}
\label{s:Conclusions}

In this work, we present the first results from our \azahar~cosmological simulation suite, which follows the formation of thousands of galaxies in a large zoom-in region at $\sim$20~pc resolution. Our simulations span ten galaxy formation models progressively incorporating non-thermal physical processes, to assess the impact of radiative transfer, cosmic rays, and magnetism on galaxy formation from the highest redshifts accessible by JWST to cosmic noon.

In this introductory paper we focus our study on three representative models: i) a non-calibrated hydrodynamical model (\HD); ii) a calibrated hydrodynamical model (\HDBoost; with supernova strength boosted by a factor of four); and iii) our non-calibrated, ``\textit{full-physics}'' model (\RTnsCRiMHD) which incorporates radiative transfer, cosmic rays, and magnetohydrodynamics on the fly. Our main results are:
\begin{enumerate}
\item The inclusion of non-thermal physics in the \RTnsCRiMHD model significantly changes the efficiency and timing of star formation. Early radiative feedback from massive stars and SN-injected cosmic rays lead to enhanced burstiness on short timescales ($\lesssim 10\,\Myr$), supported by the $\SFRHalpha / \SFRUV$ observational diagnostic. These physical processes help to prevent the overproduction of stars at very high redshifts, which has afflicted simulations similar to \HD in the literature for many years.   

\item Over cosmic time, successive SN explosion episodes accumulate cosmic rays in the ISM of galaxies. Consequently, the energy budget of the neutral ISM becomes progressively more dominated by the non-thermal component. The sustained cosmic-ray pressure buildup contributes to the suppression of in-situ star formation and drives outflows that transition from being ``ejective'' at very high redshifts to ``preventive'' at cosmic noon. These physical processes drive a distinct evolution of galaxy properties, which cannot be captured by simple calibrated hydrodynamical models such as \HDBoost, whose mass-loading factors are too high at low redshift.  

\item The \RTnsCRiMHD model simultaneously reproduces the UV luminosity function from $z \sim 14$ down to $z \sim 3$, the evolution of the cosmic star formation rate density, and the evolution of the galaxy stellar mass function, providing a unified theoretical framework to connect the highest redshift JWST galaxy observations with the wealth of data available at cosmic noon. 

\item While all three models are in good agreement with observations of the galaxy star-forming main sequence, their individual galaxies follow notably different evolutionary trajectories across this relation due to significant variations in the star formation histories of their underlying galaxy populations. This indicates that detailed studies of observed star formation histories for large galaxy samples may provide further means to constrain the galaxy formation physical processes at play. 

\item The gas mass-metallicity relation provides another key observational discriminant of galaxy formation physics. Overproduction of stars at very high redshift, such as in the \HD model, leads to gas metallicities that are too high. Conversely, overly ejective galactic outflows, characteristic of calibrated SN models, deplete the galaxies of metals and overpollute the CGM and IGM. The transition to more preventive but sustained outflows with cosmic time, facilitated by cosmic ray pressure support, provides a crucial avenue to retain a realistic proportion of metals in the ISM gas.        
\end{enumerate}

Overall, our results favor a scenario in which non-thermal physical processes, such as early radiative and cosmic ray feedback, are able to reshape the star formation cycle, its burstiness, and, consequently, outflow properties and metal transport into the CGM.
It is important to stress that the \RTnsCRiMHD model achieves this with a smaller thermal and kinetic energy budget per SN than the \HDrun model. In this simulation, additional regulation arises from energy redistribution and non-linear coupling across feedback channels that operate on different timescales and possess intrinsically different thermodynamical properties from those assumed in simple SN thermal feedback models. 
As we continue to detect the very first galaxies, while also characterizing large galaxy populations in terms of their morphologies, spatially resolved chemodynamics, and detailed star formation histories all the way to cosmic noon, simulation suites such as \azahar provide a unique means of physically interpreting this spectacular emergence of galaxies at cosmic dawn.

\begin{acknowledgments}
This research was supported by the Kavli Institute for Particle Astrophysics and Cosmology (KIPAC) and by NASA. S.M.A. is supported by a KIPAC Fellowship. D.S. acknowledges support from the Science and Technology Facilities Council (STFC) under grant ST/W000997/1.

We would like to thank Sandro Tacchella, Lily Whitler, Laura Keating, and the GFC group at Stanford for useful discussions. We are also grateful to Corey Pirie \citep{Pirie2025}, Andreas Faisst \citep{Faisst2019}, and Charlotte Simmonds \citep{Simmonds2025} for sharing their data.

This work used the DiRAC@Durham facility managed by the Institute for Computational Cosmology on behalf of the STFC DiRAC HPC Facility (\url{https://www.dirac.ac.uk}). The equipment was funded by BEIS capital funding via STFC capital grants ST/P002293/1, ST/R002371/1 and ST/S002502/1, Durham University and STFC operations grant ST/R000832/1. DiRAC is part of the National e-Infrastructure.

Some of the computing for this project was performed on the Sherlock cluster. We would like to thank Stanford University and the Stanford Research Computing Center for providing computational resources and support that contributed to these research results.
\end{acknowledgments}

\section*{Data Availability}
\label{s:availability}
The data employed in this manuscript will be shared upon reasonable request to the corresponding author. We plan to publicly release complete catalogs and provide access to the simulation snapshots in upcoming work, for the full \azahar simulation suite. Researchers interested in early access to the data are encouraged to contact the corresponding author.

\appendix
\section{Galaxy seeding, tracking and merging algorithm}
\label{ap:galaxy_tracker}
To identify galaxies in our simulations, we employ our new galaxy tracking algorithm in its post-processing mode. This OpenMP/MPI method identifies galaxies through a hybrid combination of halo finding and particle tracking. 

The tracker employs a dark matter halo catalog to `seed' galaxy tracker objects. We employ the \textsc{AdaptaHOP} mode \citep{Aubert2004} of the \textsc{halomaker} software \citep{Tweed2009} to generate our halo catalogs for each snapshot of the simulation. For those halos and subhalos from the catalog without an already existing associated galaxy tracker object, we seed a galaxy tracker in each of their centers, imposing a minimum halo mass threshold of $\Mhalo > 10^{7} \Msun$. New tracker objects are introduced at every snapshot and `accepted' only if they capture at least 20 stellar particles. 


The target number of tracked particles, $n_{\text{track}}$, for a given galaxy tracker is estimated from its measured stellar mass. Specifically, we require $40\%$ of its mass to be explicitly tracked with stellar particles, with $n_{\text{track}}$ bounded to a minimum of 50 and a maximum of 500 particles per system. At seeding time, we employ a rough estimate for the stellar mass to initialize the tracking. We use the virial mass of the (sub-)halo associated with the tracker, and set the stellar mass using the cosmic baryon fraction $\Omega_{\text{b}} / \Omega_{\text{m}}$, and a 3\% conversion efficiency. This naturally ensures that the linking radius and particle census of each tracker grow with galaxy masses. 

Each tracker stores the identities of its innermost particles as tracked particles, and employs them to recompute its new center of mass in each iteration, which serves as an initial approximation prior to the calculation of its true position. Galaxy tracker positions are refined with a shrinking-spheres procedure following \citet{Power2003}. This algorithm computes an initial mass-weighted centroid and iteratively contracts the enclosing radius by a factor of $0.975^2$ down to a lower limit of $10^{-8}$ of the starting radius (or an imposed distance cut), until the enclosed particle set converges. This produces stable centroids even in disturbed systems and guarantees continuity between snapshots. We set an effective search radius, $r_{\mathrm{max,use}}$ to be the previous tracked extent, or half of the current virial radius, whichever is shorter. The $r_{\mathrm{max,use}}$ value is bounded to a minimum of twice the minimum AMR cell size, and a maximum of $0.05$ of the full simulation box size.


To identify galaxy mergers, we enforce a single dominant tracker per main halo host. The code identifies the dominant system by stellar mass. Galaxy pairs are allowed to merge only when their separation falls below the size of 5 resolution elements ($5 \times \Delta x$), and the lower mass system is marked to merge into the dominant tracker. If a tracker was already marked to merge into the dominant system at the prior snapshot, then the galaxy is merged into the dominant system, and is archived as a defunct entry. At that point, the subordinate tracker transfers its accumulated metadata, allowing a direct and full reconstruction of its evolution, as well as the \textbf{per-galaxy} merger tree of the dominant galaxy. 

As part of our galaxy tracking, each system is assigned both a host halo and host sub-halo. The assignment of systems follows a stellar mass hierarchy, with each subhalo receiving a list of all the galaxies within $0.3$ of its virial radius, and then selecting its most massive (un-hosted) galaxy as its main occupant. This assignment proceeds in order of decreasing subhalo mass. All galaxies within a given halo are assigned the same halo as a host. In this paradigm, and for a given snapshot, all subhalos will have one galaxy associated with them, whether active, failed, or merged into a dominant system. However, a subset of galaxies is allowed to be orphaned if no viable subhalo host is found.

Prior to the measurement of any galaxy properties, we measure mass profiles built from 400 radial bins within a sphere of radius $r_{\text{prof}} = 0.2\,R_{\mathrm{vir}}$ (or the radius of the most distant tracked particle, $r_{\text{prof}} = r_{\text{max}}$, for orphaned galaxies), centered on the tracker. We use these to compute half-mass radii for stars, young stars with various age ranges, gas and dark matter; as well as angular momenta for each of these components. We also measure disk scale heights and disk scale lengths for each of these components within the same regions, informed by the computed angular momenta. To ensure that per-galaxy regions do not overlap, for galaxy $i$, we truncate $r_{\text{prof}}$ according to the distance to the nearest galaxy tracker $j$ as:
\begin{equation}
    r_{\text{prof}} = \min\left(r_{\text{prof}}, \frac{M_{*,\,i}^{1/3}}{M_{*,\,i}^{1/3} + M_{*,\,j}^{1/3}}\, r_{\text{nearest}}\right),
\label{eq:r_prof_truncation}
\end{equation}
where the $M_{*,\,i}^{1/3} / (M_{*,\,i}^{1/3} + M_{*,\,j}^{1/3})$ term ensures that the per-galaxy regions do not overlap, but is adjusted to ensure that no galaxy has $r_{\text{prof}}$ smaller than $10\,\Delta x$. Galaxies with overlapping positions and/or separations are tagged as overlapping, and excluded from our analysis.   

\section{High-Resolution Volume Estimation}
\label{ap:volume_estimation}
Some of our volumetric statistics (specifically the UV luminosity function and the galaxy stellar mass function) require the estimation of an effective volume over which number densities are computed. As our analysis is restricted to the high-resolution sub-volume of the cosmological domain, the determination of this volume requires careful consideration. Here we compare three different approaches for this estimation. Our three methods are:
\begin{enumerate}
    \item \textbf{Direct computational measurement}: By computing the volume occupied by high-resolution gas cells at each simulation output. Importantly, a large fraction of the extensive void volumes will be de-refined by the quasi-Lagrangian selection of the high-resolution region.
    
    \item \textbf{Convex hull method}: By constructing an expanded convex hull containing all dark matter halos contained in the high-resolution region.
    
    \item \textbf{Halo mass function matching}: By weighting the simulated halo population to match the expected cosmic halo mass function 
    \citep[following a similar approach to that of][]{Kannan2025}.
\end{enumerate}

\subsection{Direct computational measurement of the volume}
\label{ap:direct_volume}
The first method provides a direct estimate of the instantaneous high-resolution volume by integrating the volumes of all gas cells tagged as being high-resolution in a given snapshot. While this simple method precisely captures the instantaneous refined region, it does not account for the Lagrangian region from which galaxies are born. In a standard cosmological simulation with no high-resolution region, this approach would exclude regions such as cosmic voids, thus underestimating the volume. 

\subsection{Expanded convex hull estimate of the simulated volume}
\label{ap:ch_volume}
This second method aims to provide a better estimate of the approximate Lagrangian volume from which the studied galaxy population at a given time has evolved. 
We construct a convex hull containing all the identified dark matter halos in the high-resolution region (as defined by their virial radii), expanding the radius of each halo by the ratio accounting for mass conservation and the expansion of the Universe. Assuming spherical collapse, this ratio is $\sim$7 \citep{Bryan1998}, although it can range between $\sim$6 (computed with respect to the mean density; \citealt{PeacockBook}) and $\sim$8--9 (computed with respect to the critical density).  

\subsection{Halo Mass Function Weighting}
\label{ap:hmf_volume}
For our third method, which provides the volume we employ in our 
analysis, we follow the approach used by \citet{Kannan2025} to estimate the volume from the collection of zoom simulations of multiple halos that comprise the \textsc{Thesan}-zoom simulations. This relies on the approximate universality of the sampled halo mass function \citep[HMF;][]{Sheth2001, Tinker2008}, providing a robust method for volume estimation.
In this method, we estimate the effective volume by comparing the normalization of the halo mass function in our simulation with the theoretical Sheth-Tormen model \citep{Sheth2001}, assuming our cosmology. For each snapshot, we bin the simulated halos into logarithmically spaced mass intervals and fit them with a power law. We employ the ratio of the measured versus theoretical HMF to estimate the volume fraction $V_{\rm high-res}/V_{\rm box}$, from which we compute the effective comoving volume $V_{\rm high-res}$ of our sub-volume.

In our measurements, we employ the running average of this volume across our three simulations, and take the variance within the running window and across the three models as the volume uncertainty. Therefore, for a given redshift, our three simulations will employ the same numerical value for the volume and uncertainty of the high-resolution region.

\subsection{Volume estimates comparison}

\begin{figure}[t!]
    \centering
    \includegraphics[width=\columnwidth]{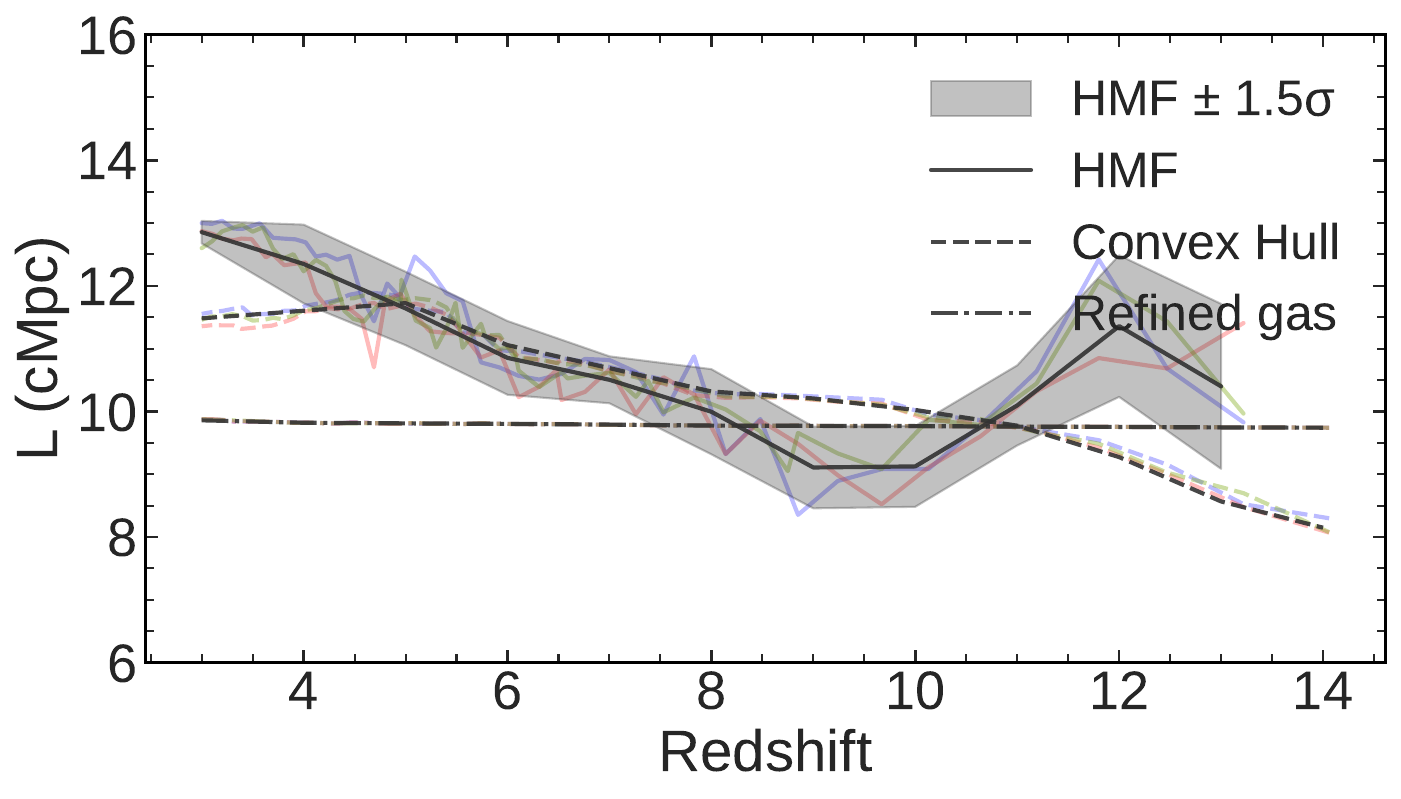}
    \caption{Comparison of the three volume estimation methods for the \HD (blue), \HDBoost (red), and \RTnsCRiMHD (green) simulations. Black lines show the combined estimate for each of the methods: HMF (solid lines), dark matter halos convex hull (dashed lines), and the volume spanned by refined gas cells (dot-dashed lines). The grey shaded band corresponds to the $\pm1.5\sigma$ variance of the combined HMF measure (rolling median and simulation variance).
    Once the population of dark matter halos builds up, the convex hull and HMF (employed in our main analysis) provide comparable results. At early times, the volume spanned by all the high-resolution cells provides a reasonable match to the HMF. The agreement between the Lagrangian-motivated convex hull and the statistical HMF approach validates our volume estimation across cosmic time.}
    \label{fig:volume_comparison}
\end{figure}

In Fig.~\ref{fig:volume_comparison} we show the resulting volumes as a function of redshift for each of our simulations, comparing the three methods described above.
At early times, while the halo population is still building up, the convex hull approach is affected by our limited resolution effects, and does not yet capture the Lagrangian region of the studied galaxy population. Conversely, the direct gas cell volume provides the most reliable estimate, as it directly traces the instantaneous refined region, which still remains close to the original Lagrangian region. 
As structure formation takes place, and our measured halo population becomes more complete, the convex hull and HMF methods converge to comparable values. These different strategies, based on a geometric construction and on matching the abundance of collapsed structures to theoretical measures, provide complementary estimates that highlight a reasonable robustness of the HMF method.  

\section{Star Formation Histories and Spectral Burstiness Computation}
\label{ap:burstiness_FFT}

\begin{figure}[t!]
    \centering
    \includegraphics[width=\columnwidth]{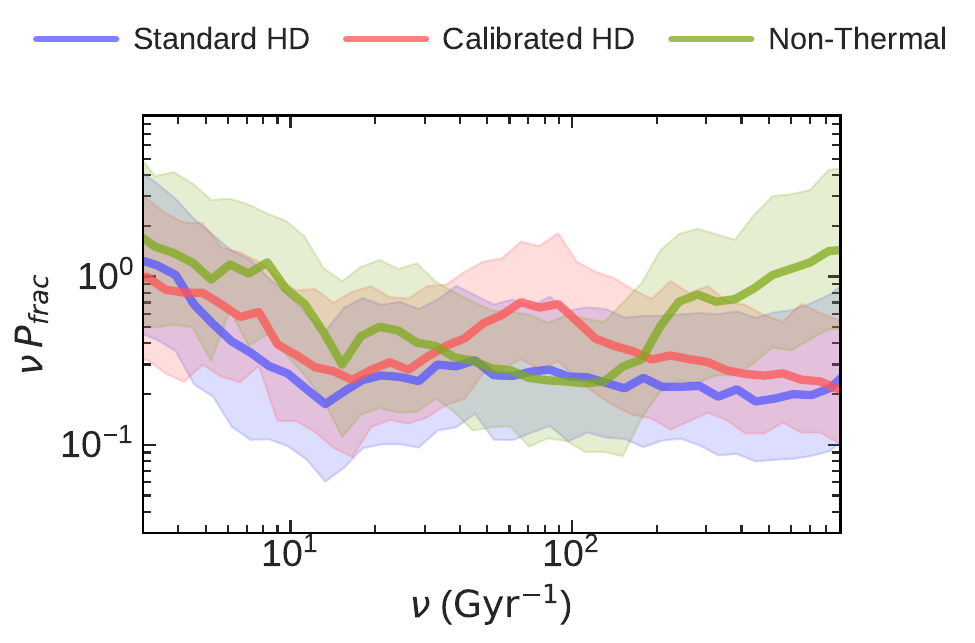}\\
    \caption{Power spectrum of the in-situ SFHs for the three simulations, showing the median variance per logarithmic frequency interval,     $\nu\,P_{\rm frac}$, as a function of frequency $\nu$, where $P_{\rm frac}$ denotes the PSD of the fractional SFH variability. The spectrum corresponds to galaxies with stellar mass $\Mstar(z = 3) > 5\times10^7\,\Msun$.}
    \label{fig:FFT_insitu_spectrum}
\end{figure}

To quantify the burstiness of star formation, we analyze in Section~\ref{ss:burstiness} the power spectral density (PSD) of the fractional variability of in-situ star formation histories (SFHs) for galaxies in our simulations. In this appendix, we describe in further detail how this burstiness is computed, and show the resulting spectra in Fig.~\ref{fig:FFT_insitu_spectrum}.

The burstiness diagnostic we investigate is the amount of power in the PSD of the SFHs for the galaxies, for each of our models.
To compute these spectra, we first compute for each galaxy its precise in-situ SFH. For this, we employ the galaxy tracking strategy described in Section~\ref{ss:trackers}, generating explicit, high-time-resolution in-situ SFHs, total SFHs, and in-situ and ex-situ $\Mstar$ histories for every galaxy in our simulations. We determine the SFH from stars formed in the galaxy between each pair of consecutive snapshots, including only stars within $2\,\rhalf$, and following the attribution criteria described in Section~\ref{ss:trackers}. For galaxies that eventually merge with other systems, we use the in-situ SFH up to the snapshot preceding the time at which they are: in a galaxy pair too close for stellar particles to be unequivocally assigned; or to the merging time, if the former is not captured in the snapshot sampling. We then compute the Fast Fourier Transform (FFT) of the fractional variability of the SFH ($f_{\rm var}$) for each galaxy, $i$
\begin{equation}
    f_{\rm var, i}(t) = \frac{\SFR(t)_{i} - \langle \SFR \rangle_{i}}{\langle \SFR \rangle_{i}}.
    \label{eq:frac_variability}
\end{equation}
We use these to compute the one-sided power spectral density, $P\,(\nu)$, with units $f_{\rm var}^2\,\text{Gyr}$. We focus our calculations on galaxies with well-resolved SFHs and, to reduce the impact of discrete sampling, additionally remove from each system the shot noise component computed from its average SFR and the $m_*$  resolution of the simulations. 
We select galaxies with stellar mass at $z = 3$ (or at their final merging time) above $\Mstar = 5 \times 10^7\,\Msun$.

\section{Selection of Star Formation Rate Timescales for Burstiness Measurement}
\label{ap:burst}

Through the comparison of star formation rates measured over different timescales, the burstiness parameter $\Bburst$ is used to estimate the short-term variability of star formation. Observationally, this is accessible through the ratio of emission-line diagnostics (frequently H$\alpha$), associated with very recent star formation, to UV continuum emission, which is used to trace longer timescales. 
The canonical timescales employed are often $\sim$10~Myr for H$\alpha$ and $\sim$100~Myr for UV continuum \citep{Kennicutt1998}. However, their characteristic timescales have different values, depending on the complex combination of stellar populations with varying metallicities and ages \citep{Calzetti2007, Haydon2020}.
Observations estimate star formation from the last 4--10~Myr through H$\alpha$, and back up to 100~Myr through UV emission \citep[e.g.,][]{Madau2014, Tacchella2022}, and the response curves are non-uniform (e.g., see Fig.~15 in \citealt{Iyer2024}; see also \citealt{Haydon2020}). This complicates a direct simulation-observation comparison. With H$\alpha$ emission dominated by O and B stars, its response peaks closer to $\lesssim$3--5~Myr. UV continuum response is instead concentrated within $\lesssim$30--50~Myr. Based on these considerations, and in the absence of a synthetic observation pipeline to measure this parameter, we adopt the simple ratio $\Bburst = \mathrm{SFR}_{3\,{\rm Myr}} / \mathrm{SFR}_{50\,{\rm Myr}}$ as our fiducial burstiness metric in Section~\ref{ss:burstiness}.

\begin{figure}
    \centering
    \includegraphics[width=\columnwidth]{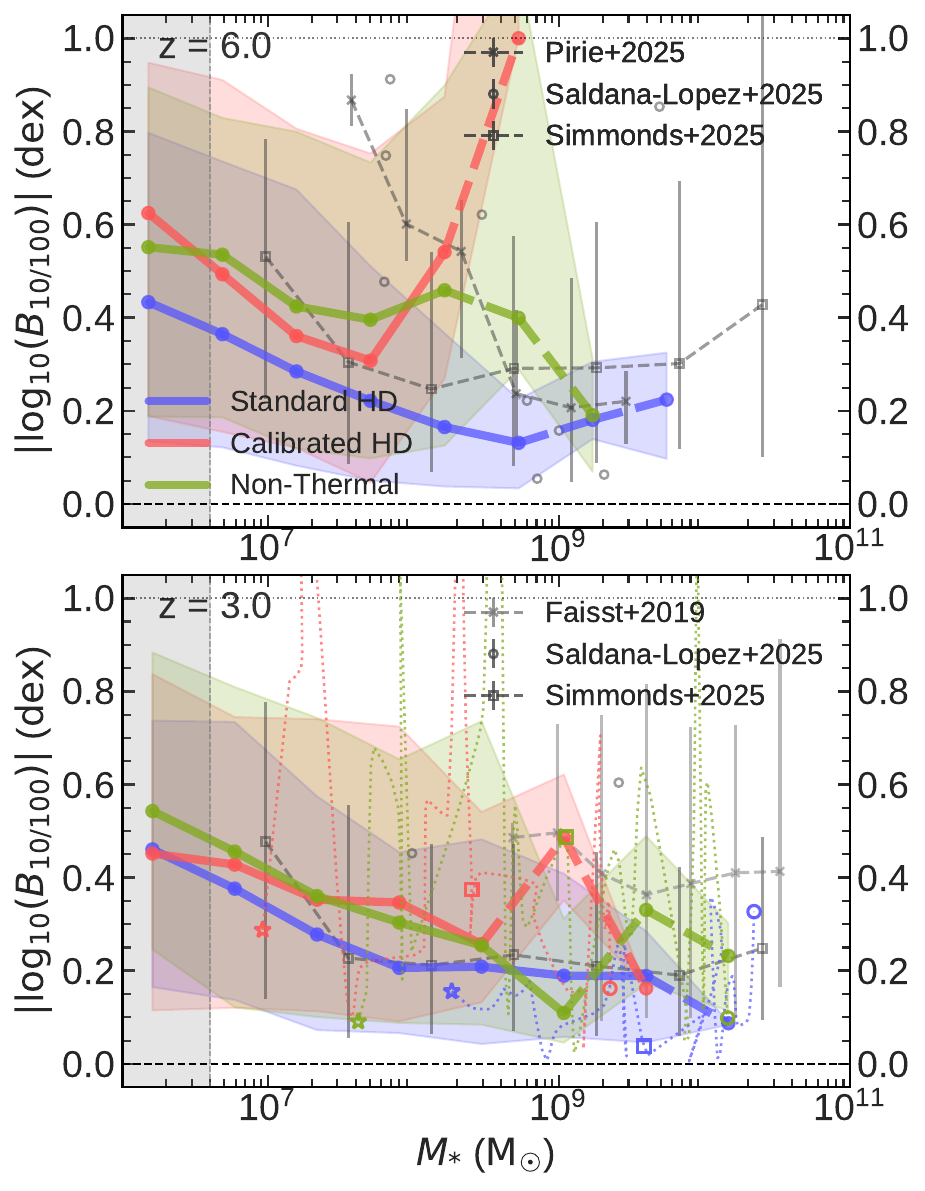}
    \caption{Comparison of burstiness measurements using different SFR timescales. Similar to Fig.~\ref{fig:burstiness_SF}, but now showing the ratio $\SFRten /\,\SFRhundred$ timescales instead of our fiducial $\SFRthree /\,\SFRfifty$. The results are very similar. The median, scatter, and relative ordering of our models 
    (\RTnsCRiMHD most bursty, \HDBoost slightly less bursty, and \HD being the lowest) are preserved at all redshifts. This reflects that our conclusions are robust with respect to the exact values selected for the approximate H$\alpha$ / UV star formation timescales.}
    \label{fig:burstiness_SF_comparison}
\end{figure}

Importantly, we show here how the choice between an $\SFRthree /\,\SFRfifty$ ratio and an $\SFRten /\,\SFRhundred$ ratio does not significantly affect our results, although the former better reproduces observational features across our fairly different galaxy formation models. Fig.~\ref{fig:burstiness_SF_comparison} shows the equivalent measurement to Fig.~\ref{fig:burstiness_SF}, now for the 10~Myr/100~Myr ratio ($B_{10/100}$). The 
results are quite comparable, with very minor differences in the median values and their scatter. Per-model trends are preserved, although the overall burstiness is lower when employing the longer timescales. The pronounced increase toward higher burstiness at lower masses is also reduced. Despite this, we conclude that burstiness estimation through the observational parameter is robust to the specific choice of timescales, although we recommend the use of shorter values when comparing simulations and observations directly (e.g., in the absence of a convolution with response curves).

\section{Energy budget of the neutral ISM across mass bins}
\label{ap:energy_budget}

\begin{figure}
    \centering
    \includegraphics[width=\columnwidth]{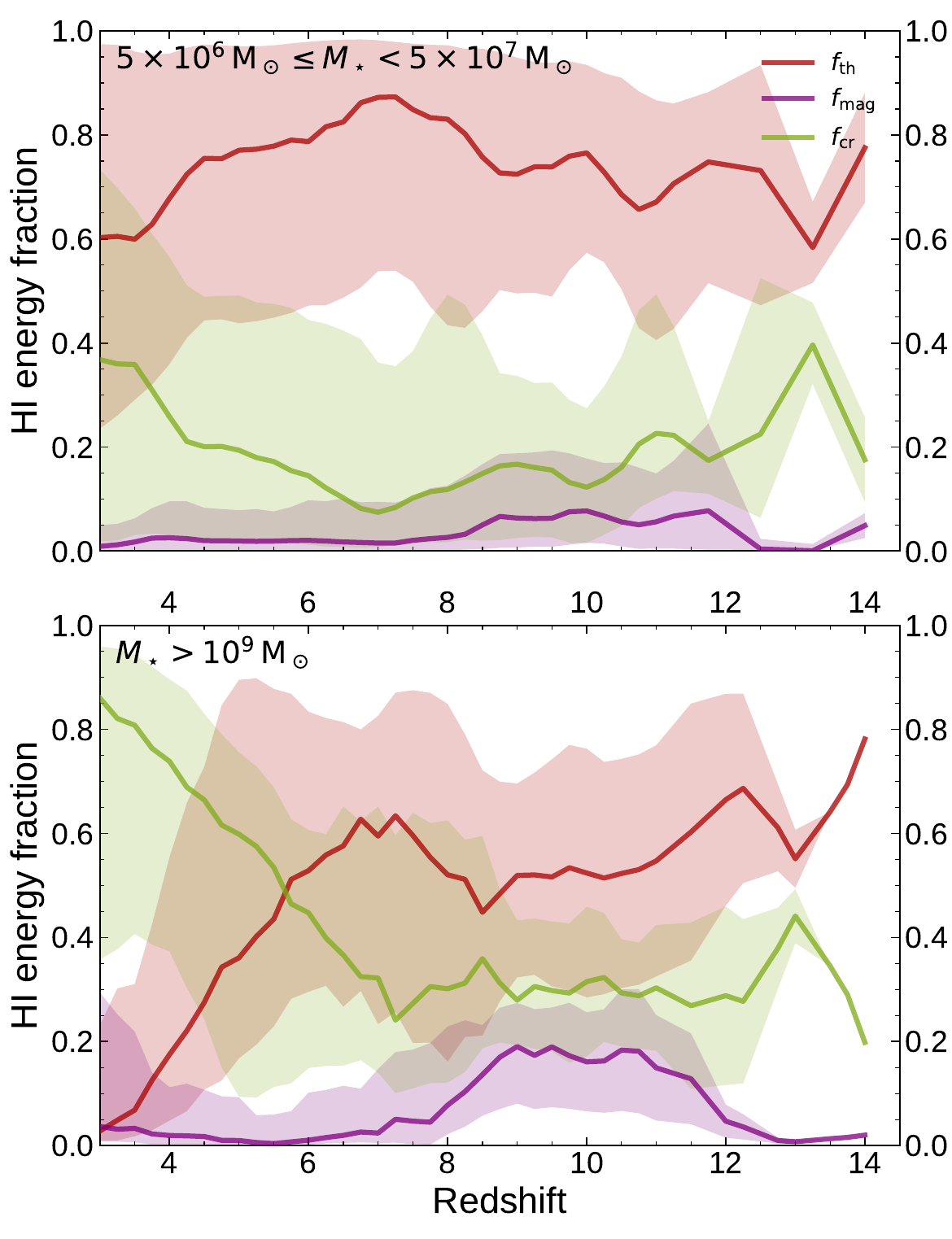}\\
    \caption{As Fig.~\ref{fig:energy_budget}, for the low ($5\times10^{6} \leq \Mstar < 5\times10^{7}\,\Msun$; top) and high ($\Mstar > 10^{9}\,\Msun$; bottom) mass bins of the \RTnsCRiMHDrun model. The increasing relative importance of cosmic ray energy with decreasing redshift is less pronounced toward lower stellar masses, and more rapid in the most massive galaxies.}
    \label{fig:energy_budget_appendix}
\end{figure}

To understand the relevance of the magnetic and cosmic ray energy contributions to the support of the neutral gas, in Fig.~\ref{fig:energy_budget} we showed their fractions for galaxies with intermediate masses across our sample. 
The transition from thermal domination to increasing cosmic-ray support is present across all stellar masses. However, this transition is more pronounced for more massive galaxies.
Despite this, our least massive galaxies ($5\times\,10^{6}\,\Msun \leq \Mstar < 5\times\,10^{7} \,\Msun$) preserve a dominant thermal component down to $z = 3$.

\section{Comparison of Lya optical depth estimates}
\label{ap:Rahmati}

In the main text, we analyzed the Ly$\alpha$ optical depth fields of our three simulations using the \citet{Rahmati2013} prescription to infer neutral fractions in both radiative-transfer and non-radiative-transfer runs.  
In this appendix, we outline the calculation method, and briefly describe the Rahmati approximation, comparing it against the map resulting from the ionization state in the simulation. 

Our Ly$\alpha$ optical depth maps are constructed by estimating the optical depth over segments of length $\Delta s = 50\,\kpc$, corresponding to a velocity scale comparable to the spectral resolution of upcoming Ly$\alpha$ forest observations such as those from the \Via Project.
Projected maps are computed along a line-of-sight depth of $\sim 1\,\Mpc$ ($\sim 4\,\cMpc$ comoving), corresponding to a velocity range of $\sim 400\,\kmps$.
To ensure a consistent treatment of $n_{\mathrm{HI}}$, we apply the \citet{Rahmati2013} prescription to all simulations. For the \RTnsCRiMHD model, we additionally compute maps directly from the simulated neutral hydrogen fraction. The resulting optical depth distributions are nearly identical, showing that the approximation is sufficiently accurate for the purposes of this work.

The \citet{Rahmati2013} approximation attenuates the ultraviolet background (UVB) photoionization rate in dense, self-shielded gas. The self-shielding density is given by
\begin{equation}
n_{\rm H,ss} \;=\; C\,\bar{\sigma}^{-2/3}\,T^{\,\alpha}\, \Gamma_{\rm UVB}^{\,2/3}\, f_g^{-1/3}.
\end{equation}
with effective photoionization rate
\iffalse
\begin{equation}
\Gamma_{\rm eff}(n_{\rm H}) \;=\; \Gamma_{\rm UVB}\!\left[A\,\left(1 + \left(\frac{n_{\rm H}}{n_{\rm H,ss}}\right)^{p}\right)^{-q} + B\,\left(1 + \frac{n_{\rm H}}{n_{\rm H,ss}}\right)^{-r} \right].
\end{equation}
\else
\begin{equation}
\Gamma_{\rm eff}(n_{\rm H})
= \Gamma_{\rm UVB}
\left[
A\left(1+x^{p}\right)^{-q}
+
B\left(1+x\right)^{-r}
\right],
\end{equation}
where $x \equiv {n_{\rm H}} / {n_{\rm H,ss}}$.
\fi
In these expressions, $\Gamma_{\rm UVB}$ corresponds to the unattenuated UV background photoionization rate, $\bar{\sigma}$ is the effective photoionization cross-section, and $f_g$ is the gas fraction. $A$, $p$, $q$, $r$, and $\alpha$ are dimensionless coefficients, and $C$ a normalization constant, tabulated in \citet{Rahmati2013}.

\begin{figure}
    \centering
    \includegraphics[width=\columnwidth]{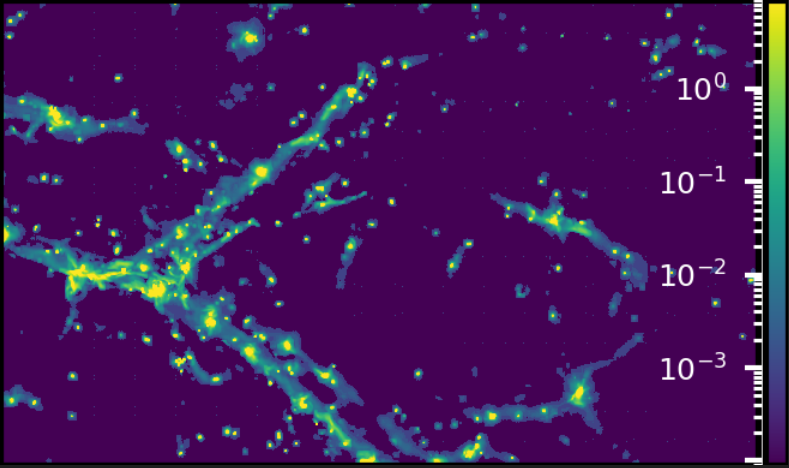}\\
    \caption{$\Lya$ optical depth per $\Delta s = 50\,\mathrm{kpc}$ for the \RTnsCRiMHDrun simulation at $z = 3$, as shown in Fig.~\ref{fig:optical}, now computed directly from the neutral hydrogen field in the simulation. Differences between maps are negligible, confirming the validity of the approximation.}
    \label{fig:optical_rahmati_comparison}
\end{figure}

Fig.~\ref{fig:optical_rahmati_comparison} shows the same map for the \RTnsCRiMHDrun model as that in Fig.~\ref{fig:optical}, except now computed through its direct ionization state. The resulting map is remarkably similar, confirming that the prescription is sufficiently accurate for our comparisons here and that variations between models are due to distinct physical feedback rather than numerical artifacts. 

\section{On the observed gas phase metallicities: warm-ionized vs total distribution}
\label{app:MZR_gas}

Observational measurements of metallicities in high-redshift galaxies frequently rely on optical emission lines such as [O~{\sc iii}], [N~{\sc ii}], and H$\alpha$ \citep[e.g.,][]{Maiolino2008, Curti2020, Sanders2021}. These lines arise from regions of the ISM with typical temperatures of $T \sim 10^4\,\K$. These regions, photoionized by young and massive stars, are some of the ISM regions categorized as the warm ionized medium. As a result, metallicities inferred from these lines only sample a subset of the WIM ISM, and do not necessarily reflect the full gas metallicity distribution within galaxies.

In contrast, when reviewing gas-phase metallicities in simulations, the distribution of gas across vastly different galactic environments has to be considered. To facilitate direct comparison with observations, our fiducial analysis employed WIM-phase gas metallicities. 

\begin{figure}
    \centering
    \includegraphics[width=\linewidth]{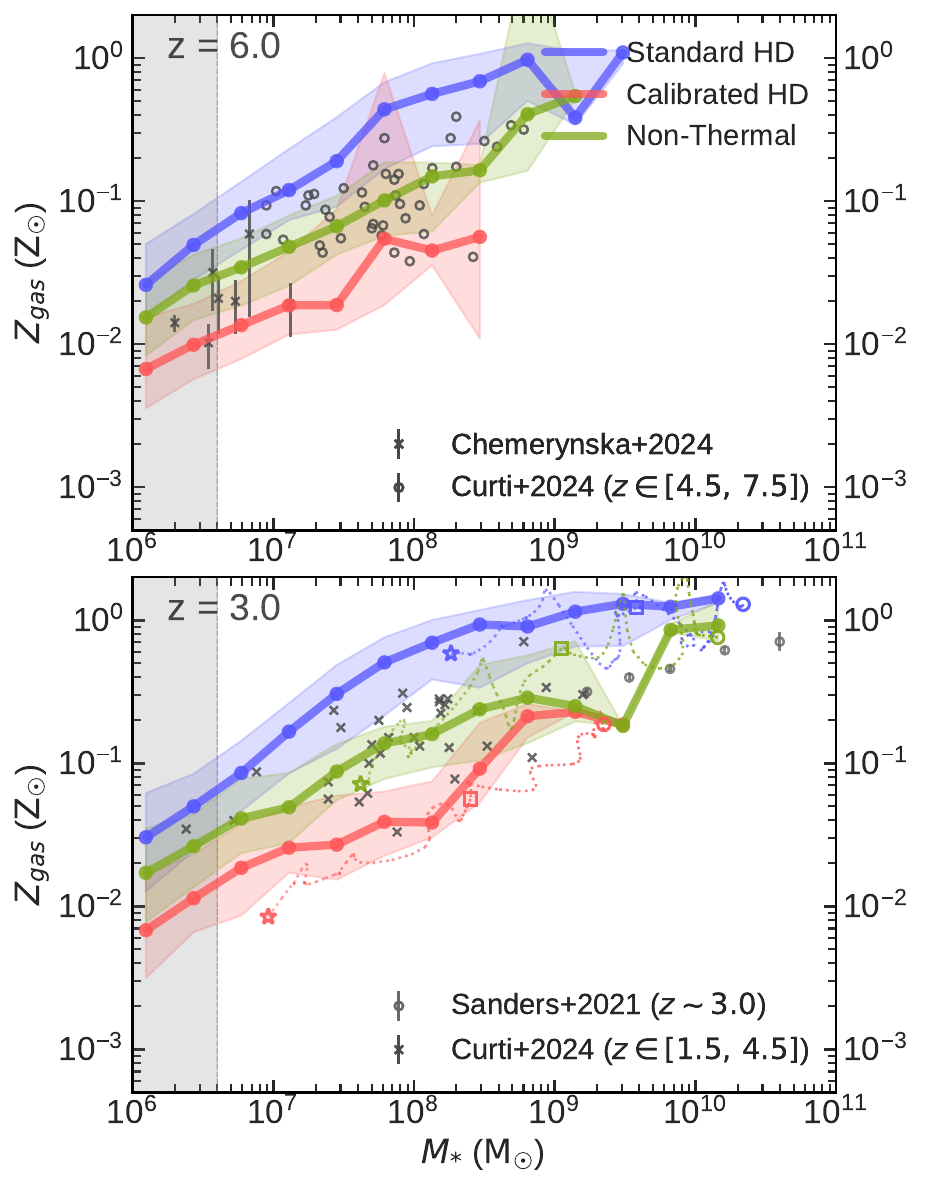}\\
    \caption{Comparison of the mass-metallicity relation shown in Fig.~\ref{fig:metallicity_mass_relation}, now for the total gas distribution instead of solely the WIM phase. 
    Overall, the same trends are preserved as for the WIM measurements, with slightly increased deviation toward higher (lower) values for the \HD (\HDBoost) model. Metallicity trajectories of individual galaxies (dotted lines, bottom panel) display more stable evolutions for the total gas of galaxies than for WIM metallicity.}
    \label{fig:MZR_comparison}
\end{figure}

To showcase the differences between the more standard and simple mass-metallicity relation for the full gas distribution and our WIM proxy, and to estimate how well observationally inferred metallicities reflect the full gas metallicity distribution, we show in Fig.~\ref{fig:MZR_comparison} the MZR for the total gas. 
Overall, we recover the same trends as in our fiducial analysis, with some minor differences. For our \RTnsCRiMHD model, WIM and total metallicities are comparable, reflecting a higher degree of ISM mixing in the presence of CRs \citep{Dashyan2020, Rathjen2023}. However, the \HD (\HDBoost) model shows slightly higher (lower) metallicity than its WIM counterpart.
Finally, while galaxy trajectories have highly stochastic evolutions across the WIM metallicity-stellar mass space, the total gas distribution leads to more stable evolutions, with the metallicity of galaxies strongly anchored by the neutral gas phases.

\bibliography{references,references_public}

@article{Harikane2023,
    title = {{A Comprehensive Study of Galaxies at z ∼ 9–16 Found in the Early JWST Data: Ultraviolet Luminosity Functions and Cosmic Star Formation History at the Pre-reionization Epoch}},
    year = {2023},
    journal = {The Astrophysical Journal Supplement Series},
    author = {Harikane, Yuichi and Ouchi, Masami and Oguri, Masamune and Ono, Yoshiaki and Nakajima, Kimihiko and Isobe, Yuki and Umeda, Hiroya and Mawatari, Ken and Zhang, Yechi},
    number = {1},
    month = {2},
    pages = {5},
    volume = {265},
    publisher = {IOP Publishing},
    url = {https://iopscience.iop.org/article/10.3847/1538-4365/acaaa9 https://iopscience.iop.org/article/10.3847/1538-4365/acaaa9/meta},
    doi = {10.3847/1538-4365/ACAAA9},
    issn = {0067-0049},
    arxivId = {2208.01612}
}

@article{Hegde2024,
    title = {{A hidden population of active galactic nuclei can explain the overabundance of luminous {\$}z>10{\$} objects observed by JWST}},
    year = {2024},
    journal = {Journal of Cosmology and Astroparticle Physics},
    author = {Hegde, Sahil and Wyatt, Michael M. and Furlanetto, Steven R.},
    number = {8},
    month = {5},
    volume = {2024},
    publisher = {Institute of Physics},
    url = {https://arxiv.org/pdf/2405.01629},
    doi = {10.1088/1475-7516/2024/08/025},
    issn = {14757516},
    arxivId = {2405.01629}
}

@article{Fromang2006,
    title = {{A high order Godunov scheme with constrained transport and adaptive mesh refinement for astrophysical magnetohydrodynamics}},
    year = {2006},
    journal = {Astronomy {\&} Astrophysics},
    author = {Fromang, S. and Hennebelle, P. and Teyssier, R.},
    number = {2},
    month = {10},
    pages = {371--384},
    volume = {457},
    publisher = {EDP Sciences},
    url = {http://www.aanda.org/10.1051/0004-6361:20065371},
    isbn = {0004-6361{\textbackslash}r1432-0746},
    doi = {10.1051/0004-6361:20065371},
    issn = {0004-6361},
    pmid = {9010224},
    arxivId = {astro-ph/0607230}
}

@article{Endsley2023,
    title = {{A JWST/NIRCam study of key contributors to reionization: the star-forming and ionizing properties of UV-faint z ∼7-8 galaxies}},
    year = {2023},
    journal = {Monthly Notices of the Royal Astronomical Society},
    author = {Endsley, Ryan and Stark, Daniel P. and Whitler, Lily and Topping, Michael W. and Chen, Zuyi and Plat, Adèle and Chisholm, John and Charlot, Stephane},
    number = {2},
    month = {7},
    pages = {2312--2330},
    volume = {524},
    publisher = {Oxford Academic},
    url = {https://dx.doi.org/10.1093/mnras/stad1919},
    doi = {10.1093/mnras/stad1919},
    issn = {13652966},
    arxivId = {2208.14999}
}

@article{Zhang2024,
    title = {{A magnetized Galactic halo from inner Galaxy outflows}},
    year = {2024},
    journal = {Nature Astronomy 2024 8:11},
    author = {Zhang, He Shou and Ponti, Gabriele and Carretti, Ettore and Liu, Ruo Yu and Morris, Mark R. and Haverkorn, Marijke and Locatelli, Nicola and Zheng, Xueying and Aharonian, Felix and Zhang, Hai Ming and Zhang, Yi and Stel, Giovanni and Strong, Andrew and Yeung, Michael C.H. and Merloni, Andrea},
    number = {11},
    month = {9},
    pages = {1416--1428},
    volume = {8},
    publisher = {Nature Publishing Group},
    url = {https://www.nature.com/articles/s41550-024-02362-0},
    doi = {10.1038/s41550-024-02362-0},
    issn = {2397-3366},
    arxivId = {2408.06312}
}

@article{Labbe2023,
    title = {{A population of red candidate massive galaxies {\~{}}600 Myr after the Big Bang}},
    year = {2023},
    journal = {Nature},
    author = {Labb{\'{e}}, Ivo and van Dokkum, Pieter and Nelson, Erica and Bezanson, Rachel and Suess, Katherine A. and Leja, Joel and Brammer, Gabriel and Whitaker, Katherine and Mathews, Elijah and Stefanon, Mauro and Wang, Bingjie},
    number = {7956},
    month = {2},
    pages = {266--269},
    volume = {616},
    publisher = {Nature Publishing Group},
    url = {https://www.nature.com/articles/s41586-023-05786-2},
    doi = {10.1038/s41586-023-05786-2},
    issn = {14764687},
    pmid = {36812940},
    arxivId = {2207.12446}
}

@article{Looser2024,
    title = {{A recently quenched galaxy 700 million years after the Big Bang}},
    year = {2024},
    journal = {Nature},
    author = {Looser, Tobias J. and D’Eugenio, Francesco and Maiolino, Roberto and Witstok, Joris and Sandles, Lester and Curtis-Lake, Emma and Chevallard, Jacopo and Tacchella, Sandro and Johnson, Benjamin D. and Baker, William M. and Suess, Katherine A. and Carniani, Stefano and Ferruit, Pierre and Arribas, Santiago and Bonaventura, Nina and Bunker, Andrew J. and Cameron, Alex J. and Charlot, Stephane and Curti, Mirko and de Graaff, Anna and Maseda, Michael V. and Rawle, Tim and Rix, Hans Walter and Del Pino, Bruno Rodríguez and Smit, Renske and {\"{U}}bler, Hannah and Willott, Chris and Alberts, Stacey and Egami, Eiichi and Eisenstein, Daniel J. and Endsley, Ryan and Hausen, Ryan and Rieke, Marcia and Robertson, Brant and Shivaei, Irene and Williams, Christina C. and Boyett, Kristan and Chen, Zuyi and Ji, Zhiyuan and Jones, Gareth C. and Kumari, Nimisha and Nelson, Erica and Perna, Michele and Saxena, Aayush and Scholtz, Jan},
    number = {8010},
    month = {3},
    pages = {53--57},
    volume = {629},
    publisher = {Nature Publishing Group},
    url = {https://www.nature.com/articles/s41586-024-07227-0},
    doi = {10.1038/s41586-024-07227-0},
    issn = {14764687},
    pmid = {38447669},
    arxivId = {2302.14155}
}

@article{Rosdahl2015a,
    title = {{A scheme for radiation pressure and photon diffusion with the M1 closure in RAMSES-RT}},
    year = {2015},
    journal = {Monthly Notices of the Royal Astronomical Society},
    author = {Rosdahl, J. and Teyssier, R.},
    number = {4},
    month = {6},
    pages = {4380--4403},
    volume = {449},
    publisher = {Oxford University Press},
    url = {http://academic.oup.com/mnras/article/449/4/4380/1194692/A-scheme-for-radiation-pressure-and-photon},
    doi = {10.1093/mnras/stv567},
    issn = {13652966}
}

@article{Maiolino2024,
    title = {{A small and vigorous black hole in the early Universe}},
    year = {2024},
    journal = {Nature},
    author = {Maiolino, Roberto and Scholtz, Jan and Witstok, Joris and Carniani, Stefano and D’Eugenio, Francesco and de Graaff, Anna and {\"{U}}bler, Hannah and Tacchella, Sandro and Curtis-Lake, Emma and Arribas, Santiago and Bunker, Andrew and Charlot, Stéphane and Chevallard, Jacopo and Curti, Mirko and Looser, Tobias J. and Maseda, Michael V. and Rawle, Timothy D. and Rodr{\'{i}}guez del Pino, Bruno and Willott, Chris J. and Egami, Eiichi and Eisenstein, Daniel J. and Hainline, Kevin N. and Robertson, Brant and Williams, Christina C. and Willmer, Christopher N.A. and Baker, William M. and Boyett, Kristan and DeCoursey, Christa and Fabian, Andrew C. and Helton, Jakob M. and Ji, Zhiyuan and Jones, Gareth C. and Kumari, Nimisha and Laporte, Nicolas and Nelson, Erica J. and Perna, Michele and Sandles, Lester and Shivaei, Irene and Sun, Fengwu},
    number = {8002},
    month = {1},
    pages = {59--63},
    volume = {627},
    publisher = {Nature Publishing Group},
    url = {https://www.nature.com/articles/s41586-024-07052-5},
    doi = {10.1038/s41586-024-07052-5},
    issn = {14764687},
    pmid = {38232944},
    arxivId = {2305.12492}
}

@article{Reddy2009,
    title = {{A STEEP FAINT-END SLOPE OF THE UV LUMINOSITY FUNCTION AT z ∼ 2–3: IMPLICATIONS FOR THE GLOBAL STELLAR MASS DENSITY AND STAR FORMATION IN LOW-MASS HALOS*}},
    year = {2009},
    journal = {The Astrophysical Journal},
    author = {Reddy, Naveen A. and Steidel, Charles C.},
    number = {1},
    month = {2},
    pages = {778},
    volume = {692},
    publisher = {IOP Publishing},
    url = {https://iopscience.iop.org/article/10.1088/0004-637X/692/1/778 https://iopscience.iop.org/article/10.1088/0004-637X/692/1/778/meta},
    doi = {10.1088/0004-637X/692/1/778},
    issn = {0004-637X},
    arxivId = {0810.2788}
}

@article{Geha2012,
    title = {{A stellar mass threshold for quenching of field galaxies}},
    year = {2012},
    journal = {Astrophysical Journal},
    author = {Geha, M. and Blanton, M. R. and Yan, R. and Tinker, J. L.},
    number = {1},
    month = {9},
    pages = {85},
    volume = {757},
    publisher = {IOP Publishing},
    url = {https://iopscience.iop.org/article/10.1088/0004-637X/757/1/85 https://iopscience.iop.org/article/10.1088/0004-637X/757/1/85/meta},
    doi = {10.1088/0004-637X/757/1/85},
    issn = {15384357},
    arxivId = {1206.3573}
}

@article{Pallottini2022,
    title = {{A survey of high-z galaxies: serra simulations}},
    year = {2022},
    journal = {Monthly Notices of the Royal Astronomical Society},
    author = {Pallottini, A. and Ferrara, A. and Gallerani, S. and Behrens, C. and Kohandel, M. and Carniani, S. and Vallini, L. and Salvadori, S. and Gelli, V. and Sommovigo, L. and D'Odorico, V. and Di Mascia, F. and Pizzati, E.},
    number = {4},
    month = {5},
    pages = {5621--5641},
    volume = {513},
    publisher = {Oxford Academic},
    url = {https://dx.doi.org/10.1093/mnras/stac1281},
    doi = {10.1093/mnras/stac1281},
    issn = {13652966},
    arxivId = {2201.02636}
}

@article{Brauer2025,
    title = {{Aeos: The Impact of Population III Initial Mass Function and Star-by-Star Models in Galaxy Simulations}},
    year = {2025},
    journal = {The Astrophysical Journal},
    author = {Brauer, Kaley and Mead, Jennifer and Wise, John H. and Bryan, Greg L. and Mac Low, Mordecai-Mark and Ji, Alexander P. and Emerick, Andrew and Andersson, Eric P. and Frebel, Anna and C{\^{o}}t{\'{e}}, Benoit},
    number = {1},
    month = {11},
    pages = {2},
    volume = {993},
    publisher = {American Astronomical Society},
    url = {http://arxiv.org/abs/2502.20433 http://dx.doi.org/10.3847/1538-4357/ae06a4},
    doi = {10.3847/1538-4357/ae06a4},
    arxivId = {2502.20433}
}

@article{Bouwens2016,
    title = {{ALMA SPECTROSCOPIC SURVEY IN THE HUBBLE ULTRA DEEP FIELD: THE INFRARED EXCESS OF UV-SELECTED z = 2–10 GALAXIES AS A FUNCTION OF UV-CONTINUUM SLOPE AND STELLAR MASS}},
    year = {2016},
    journal = {The Astrophysical Journal},
    author = {Bouwens, Rychard J. and Aravena, Manuel and Decarli, Roberto and Walter, Fabian and Cunha, Elisabete da and Labb{\'{e}}, Ivo and Bauer, Franz E. and Bertoldi, Frank and Carilli, Chris and Chapman, Scott and Daddi, Emanuele and Hodge, Jacqueline and Ivison, Rob J. and Karim, Alex and Fevre, Olivier Le and Magnelli, Benjamin and Ota, Kazuaki and Riechers, Dominik and Smail, Ian R. and Werf, Paul van der and Weiss, Axel and Cox, Pierre and Elbaz, David and Gonzalez-Lopez, Jorge and Infante, Leopoldo and Oesch, Pascal and Wagg, Jeff and Wilkins, Steve},
    number = {1},
    month = {12},
    pages = {72},
    volume = {833},
    publisher = {IOP Publishing},
    url = {https://iopscience.iop.org/article/10.3847/1538-4357/833/1/72 https://iopscience.iop.org/article/10.3847/1538-4357/833/1/72/meta},
    doi = {10.3847/1538-4357/833/1/72},
    issn = {0004-637X},
    arxivId = {1606.05280}
}

@article{Maiolino2008,
    title = {{AMAZE - I. The evolution of the mass–metallicity relation at z > 3}},
    year = {2008},
    journal = {Astronomy {\&} Astrophysics},
    author = {Maiolino, R. and Nagao, T. and Grazian, A. and Cocchia, F. and Marconi, A. and Mannucci, F. and Cimatti, A. and Pipino, A. and Ballero, S. and Calura, F. and Chiappini, C. and Fontana, A. and Granato, L. and Matteucci, F. and Pastorini, G. and Pentericci, L. and Risaliti, G. and Salvati, M. and Silva, L.},
    number = {2},
    month = {9},
    pages = {463--479},
    volume = {488},
    publisher = {EDP Sciences},
    url = {https://www.aanda.org/articles/aa/abs/2008/35/aa09678-08/aa09678-08.html},
    doi = {10.1051/0004-6361:200809678},
    issn = {0004-6361},
    arxivId = {0806.2410}
}

@article{Dubois2016,
    title = {{An implicit scheme for solving the anisotropic diffusion of heat and cosmic rays in the RAMSES code}},
    year = {2016},
    journal = {Astronomy {\&} Astrophysics},
    author = {Dubois, Yohan and Commer{\c{c}}on, Benoît},
    month = {1},
    pages = {A138},
    volume = {585},
    publisher = {EDP Sciences},
    url = {http://www.aanda.org/10.1051/0004-6361/201527126 http://arxiv.org/abs/1509.07037%0Ahttp://dx.doi.org/10.1051/0004-6361/201527126},
    doi = {10.1051/0004-6361/201527126},
    issn = {0004-6361},
    arxivId = {1509.07037}
}

@article{Haydon2020,
    title = {{An uncertainty principle for star formation – III. The characteristic emission time-scales of star formation rate tracers}},
    year = {2020},
    journal = {Monthly Notices of the Royal Astronomical Society},
    author = {Haydon, Daniel T. and Kruijssen, J. M.Diederik and Chevance, Mélanie and Hygate, Alexander P.S. and Krumholz, Mark R. and Schruba, Andreas and Longmore, Steven N.},
    number = {1},
    month = {9},
    pages = {235--257},
    volume = {498},
    publisher = {Oxford Academic},
    url = {https://dx.doi.org/10.1093/mnras/staa2430},
    doi = {10.1093/MNRAS/STAA2430},
    issn = {0035-8711},
    arxivId = {1810.10897}
}

@article{MaxRey2025,
    title = {{ARCHITECTS I: impact of subgrid physics on the simulated properties of the circumgalactic medium}},
    year = {2025},
    journal = {Monthly Notices of the Royal Astronomical Society},
    author = {Rey, Maxime and Blaizot, Jérémy and Kimm, Taysun and Rosdahl, Joakim and Michel-Dansac, Léo},
    number = {1},
    month = {9},
    pages = {12--27},
    volume = {543},
    publisher = {Oxford Academic},
    url = {https://dx.doi.org/10.1093/mnras/staf1372},
    doi = {10.1093/mnras/staf1372},
    issn = {13652966}
}

@article{Smith2023,
    title = {{Arkenstone – I. A novel method for robustly capturing high specific energy outflows in cosmological simulations}},
    year = {2023},
    journal = {Monthly Notices of the Royal Astronomical Society},
    author = {Smith, Matthew C. and Fielding, Drummond B. and Bryan, Greg L. and Kim, Chang Goo and Ostriker, Eve C. and Somerville, Rachel S. and Stern, Jonathan and Su, Kung Yi and Weinberger, Rainer and Hu, Chia Yu and Forbes, John C. and Hernquist, Lars and Burkhart, Blakesley and Li, Yuan},
    number = {1},
    month = {10},
    pages = {1216--1243},
    volume = {527},
    publisher = {Oxford Academic},
    url = {https://dx.doi.org/10.1093/mnras/stad3168},
    doi = {10.1093/mnras/stad3168},
    issn = {13652966},
    arxivId = {2301.07116}
}

@article{Concas2022,
    title = {{Being KLEVER at cosmic noon: Ionized gas outflows are inconspicuous in low-mass star-forming galaxies but prominent in massive AGN hosts}},
    year = {2022},
    journal = {Monthly Notices of the Royal Astronomical Society},
    author = {Concas, Alice and Maiolino, Roberto and Curti, Mirko and Hayden-Pawson, Connor and Cirasuolo, Michele and Jones, Gareth C. and Mercurio, Amata and Belfiore, Francesco and Cresci, Giovanni and Cullen, Fergus and Mannucci, Filippo and Marconi, Alessandro and Cappellari, Michele and Cicone, Claudia and Peng, Yingjie and Troncoso, Paulina},
    number = {2},
    month = {5},
    pages = {2535--2562},
    volume = {513},
    publisher = {Oxford Academic},
    url = {https://dx.doi.org/10.1093/mnras/stac1026},
    doi = {10.1093/mnras/stac1026},
    issn = {13652966},
    arxivId = {2203.11958}
}

@article{Ubler2025,
    title = {{BlackTHUNDER: evidence for three massive black holes in a z{\~{}}5 galaxy}},
    year = {2025},
    journal = {arXiv},
    author = {{\"{U}}bler, Hannah and Mazzolari, Giovanni and Maiolino, Roberto and D'Eugenio, Francesco and Davari, Nazanin and Juod{\v{z}}balis, Ignas and Schneider, Raffaella and Valiante, Rosa and Arribas, Santiago and Bertola, Elena and Bunker, Andrew J. and Bromm, Volker and Carniani, Stefano and Charlot, Stéphane and Cresci, Giovanni and Curti, Mirko and Davies, Richard and Eisenhauer, Frank and Fabian, Andrew and Schreiber, Natascha M. Förster and Genzel, Reinhard and Inayoshi, Kohei and Ivey, Lucy R. and Jones, Gareth C. and Liu, Boyuan and Lutz, Dieter and Mackenzie, Ruari and Matthee, Jorryt and Parlanti, Eleonora and Perna, Michele and Robertson, Brant and del Pino, Bruno Rodríguez and Shimizu, T. Taro and Sijacki, Debora and Sturm, Eckhard and Tacchella, Sandro and Tacconi, Linda and Tozzi, Giulia and Trinca, Alessandro and Venturi, Giacomo and Volonteri, Marta and Willot, Chris and Zhang, Saiyang},
    month = {9},
    url = {https://arxiv.org/pdf/2509.21575},
    arxivId = {2509.21575}
}

@article{Tweed2009,
    title = {{Building Merger Trees from Cosmological N-body Simulations}},
    year = {2009},
    journal = {Astronomy {\&} Astrophysics},
    author = {Tweed, D. and Devriendt, J. and Blaizot, J. and Colombi, S. and Slyz, A.},
    number = {2},
    month = {11},
    pages = {647--660},
    volume = {506},
    publisher = {EDP Sciences},
    url = {http://www.aanda.org/10.1051/0004-6361/200911787 http://arxiv.org/abs/0902.0679%0Ahttp://dx.doi.org/10.1051/0004-6361/200911787},
    doi = {10.1051/0004-6361/200911787},
    issn = {0004-6361},
    arxivId = {0902.0679}
}

@article{Governato2010,
    title = {{Bulgeless dwarf galaxies and dark matter cores from supernova-driven outflows}},
    year = {2010},
    journal = {Nature 2010 463:7278},
    author = {Governato, F. and Brook, C. and Mayer, L. and Brooks, A. and Rhee, G. and Wadsley, J. and Jonsson, P. and Willman, B. and Stinson, G. and Quinn, T. and Madau, P.},
    number = {7278},
    month = {1},
    pages = {203--206},
    volume = {463},
    publisher = {Nature Publishing Group},
    url = {https://www.nature.com/articles/nature08640},
    doi = {10.1038/nature08640},
    issn = {1476-4687}
}

@article{Simmonds2025,
    title = {{Bursting at the seams: the star-forming main sequence and its scatter at z = 3–9 using NIRCam photometry from JADES}},
    year = {2025},
    journal = {Monthly Notices of the Royal Astronomical Society},
    author = {Simmonds, C and Tacchella, S and McClymont, W and Curtis-Lake, E and D’Eugenio, F and Hainline, K and Johnson, B D and Kravtsov, A and Pusk{\'{a}}s, D and Robertson, B and Stoffers, A and Willott, C and Baker, W M and Belokurov, V A and Bhatawdekar, R and Bunker, A J and Carniani, S and Chevallard, J and Curti, M and Duan, Q and Helton, J M and Ji, Z and Looser, T J and Maiolino, R and Maseda, M V and Shivaei, I and Williams, C C},
    number = {4},
    month = {11},
    pages = {4551--4575},
    volume = {544},
    publisher = {Oxford Academic},
    url = {https://dx.doi.org/10.1093/mnras/staf1950},
    doi = {10.1093/MNRAS/STAF1950},
    issn = {0035-8711},
    arxivId = {2508.04410}
}

@article{Asada2023,
    title = {{Bursty star formation and galaxy–galaxy interactions in low-mass galaxies 1 Gyr after the Big Bang}},
    year = {2023},
    journal = {Monthly Notices of the Royal Astronomical Society},
    author = {Asada, Yoshihisa and Sawicki, Marcin and Abraham, Roberto and Brada{\v{c}}, Maruša and Brammer, Gabriel and Desprez, Guillaume and Estrada-Carpenter, Vince and Iyer, Kartheik and Martis, Nicholas and Matharu, Jasleen and Mowla, Lamiya and Muzzin, Adam and Noirot, Gaël and Sarrouh, Ghassan T.E. and Strait, Victoria and Willott, Chris J. and Harshan, Anishya},
    number = {4},
    month = {12},
    pages = {11372--11392},
    volume = {527},
    publisher = {Oxford Academic},
    url = {https://dx.doi.org/10.1093/mnras/stad3902},
    doi = {10.1093/MNRAS/STAD3902},
    issn = {0035-8711},
    arxivId = {2310.02314}
}

@article{Curro2026,
    title = {{CALIMA: On-the-fly dust and PAH evolution for radiation-hydrodynamics galaxy formation simulations}},
    year = {2026},
    author = {Rodr{\'{i}}guez Montero, Francisco and Dubois, Yohan and Katz, Harley and Slyz, Adrianne and Devriendt, Julien},
    month = {2},
    url = {http://arxiv.org/abs/2602.21790},
    arxivId = {2602.21790}
}

@article{Kokorev2025,
    title = {{CAPERS Observations of Two UV-Bright Galaxies at z>10. More Evidence for Bursting Star Formation in the Early Universe}},
    year = {2025},
    journal = {The Astrophysical Journal Letters},
    author = {Kokorev, Vasily and Ch{\'{a}}vez Ortiz, Oscar A and Taylor, Anthony J and Finkelstein, Steven L and Arrabal Haro, Pablo and Dickinson, Mark and Chisholm, John and Fujimoto, Seiji and Mu{\~{n}}oz, Julian B and Endsley, Ryan and Hu, Weida and Napolitano, Lorenzo and Wilkins, Stephen M and Akins, Hollis B and Amori{\'{i}}n, Ricardo and Casey, Caitlin M and Cheng, Yingjie and Cleri, Nikko J and Cole, Justin and Cullen, Fergus and Daddi, Emanuele and Davis, Kelcey and Donnan, Callum T and Dunlop, James S and Fern{\'{a}}ndez, Vital and Giavalisco, Mauro and Grogin, Norman A and Hathi, Nimish and Hirschmann, Michaela and Kartaltepe, Jeyhan S and Koekemoer, Anton M and Leung, Ho-Hin and Lucas, Ray A and McLeod, Derek and Papovich, Casey and Pentericci, Laura and P{\'{e}}rez-Gonz{\'{a}}lez, Pablo G and Somerville, Rachel S and Wang, Xin and Aaron Yung, L Y and Zavala, Jorge A and Woods, Cynthia},
    number = {1},
    month = {4},
    pages = {L10},
    volume = {988},
    publisher = {American Astronomical Society},
    url = {https://arxiv.org/pdf/2504.12504},
    doi = {10.3847/2041-8213/ade8f5},
    issn = {2041-8205},
    arxivId = {2504.12504}
}

@article{Finkelstein2023,
    title = {{CEERS Key Paper. I. An Early Look into the First 500 Myr of Galaxy Formation with JWST}},
    year = {2023},
    journal = {The Astrophysical Journal Letters},
    author = {Finkelstein, Steven L and Bagley, Micaela B and Ferguson, Henry C and Wilkins, Stephen M and Kartaltepe, Jeyhan S and Papovich, Casey and Yung, L. Y. Aaron and Arrabal Haro, Pablo and Behroozi, Peter and Dickinson, Mark and Kocevski, Dale D and Koekemoer, Anton M and Larson, Rebecca L and Le Bail, Aurélien and Morales, Alexa M and P{\'{e}}rez-Gonz{\'{a}}lez, Pablo G and Burgarella, Denis and Dav{\'{e}}, Romeel and Hirschmann, Michaela and Somerville, Rachel S and Wuyts, Stijn and Bromm, Volker and Casey, Caitlin M and Fontana, Adriano and Fujimoto, Seiji and Gardner, Jonathan P and Giavalisco, Mauro and Grazian, Andrea and Grogin, Norman A and Hathi, Nimish P and Hutchison, Taylor A and Jha, Saurabh W and Jogee, Shardha and Kewley, Lisa J and Kirkpatrick, Allison and Long, Arianna S and Lotz, Jennifer M and Pentericci, Laura and Pierel, Justin D. R. and Pirzkal, Nor and Ravindranath, Swara and Ryan, Russell E and Trump, Jonathan R and Yang, Guang and Bhatawdekar, Rachana and Bisigello, Laura and Buat, Véronique and Calabr{\`{o}}, Antonello and Castellano, Marco and Cleri, Nikko J and Cooper, M C and Croton, Darren and Daddi, Emanuele and Dekel, Avishai and Elbaz, David and Franco, Maximilien and Gawiser, Eric and Holwerda, Benne W and Huertas-Company, Marc and Jaskot, Anne E and Leung, Gene C. K. and Lucas, Ray A and Mobasher, Bahram and Pandya, Viraj and Tacchella, Sandro and Weiner, Benjamin J and Zavala, Jorge A},
    number = {1},
    pages = {L13},
    volume = {946},
    url = {https://doi.org/10.3847/2041-8213/acade4},
    doi = {10.3847/2041-8213/acade4},
    issn = {2041-8205},
    arxivId = {2211.05792}
}

@article{Cole2025,
    title = {{CEERS: Increasing Scatter along the Star-forming Main Sequence Indicates Early Galaxies Form in Bursts}},
    year = {2025},
    journal = {The Astrophysical Journal},
    author = {Cole, Justin W. and Papovich, Casey and Finkelstein, Steven L. and Bagley, Micaela B. and Dickinson, Mark and Iyer, Kartheik G. and Yung, L. Y. Aaron and Ciesla, Laure and Amor{\'{i}}n, Ricardo O. and Haro, Pablo Arrabal and Bhatawdekar, Rachana and Calabr{\`{o}}, Antonello and Cleri, Nikko J. and Vega, Alexander de la and Dekel, Avishai and Endsley, Ryan and Gawiser, Eric and Giavalisco, Mauro and Hathi, Nimish P. and Hirschmann, Michaela and Holwerda, Benne W. and Kartaltepe, Jeyhan S. and Koekemoer, Anton M. and Lucas, Ray A. and Mascia, Sara and Mobasher, Bahram and P{\'{e}}rez-Gonz{\'{a}}lez, Pablo G. and Rodighiero, Giulia and Ronayne, Kaila and Tacchella, Sandro and Weiner, Benjamin J. and Wilkins, Stephen M.},
    number = {2},
    month = {1},
    pages = {193},
    volume = {979},
    publisher = {IOP Publishing},
    url = {https://iopscience.iop.org/article/10.3847/1538-4357/ad9a6a https://iopscience.iop.org/article/10.3847/1538-4357/ad9a6a/meta},
    doi = {10.3847/1538-4357/AD9A6A},
    issn = {0004-637X},
    arxivId = {2312.10152}
}

@article{Roca-Fabrega2019,
    title = {{CGM properties in VELA and NIHAO simulations; the OVI ionization mechanism: dependence on redshift, halo mass, and radius}},
    year = {2019},
    journal = {Monthly Notices of the Royal Astronomical Society},
    author = {Roca-F{\`{a}}brega, S. and Dekel, A. and Faerman, Y. and Gnat, O. and Strawn, C. and Ceverino, D. and Primack, J. and Macci{\`{o}}, A. V. and Dutton, A. A. and Prochaska, J. X. and Stern, J.},
    number = {3},
    month = {4},
    pages = {3625--3645},
    volume = {484},
    publisher = {Oxford Academic},
    url = {https://dx.doi.org/10.1093/mnras/stz063},
    doi = {10.1093/MNRAS/STZ063},
    issn = {0035-8711},
    arxivId = {1808.09973}
}

@article{Koprowski2024,
    title = {{Charting the main sequence of star-forming galaxies out to redshifts z ≲ 5.7}},
    year = {2024},
    journal = {Astronomy {\&} Astrophysics},
    author = {Koprowski, M. P. and Wijesekera, J. V. and Dunlop, J. S. and Mcleod, D. J. and Michalowski, M. J. and Lisiecki, K. and Mclure, R. J.},
    month = {11},
    pages = {A164},
    volume = {691},
    publisher = {EDP Sciences},
    url = {https://www.aanda.org/articles/aa/full_html/2024/11/aa49948-24/aa49948-24.html https://www.aanda.org/articles/aa/abs/2024/11/aa49948-24/aa49948-24.html},
    doi = {10.1051/0004-6361/202449948},
    issn = {0004-6361},
    arxivId = {2403.06575}
}

@article{Ferland1998,
    title = {{CLOUDY 90: Numerical Simulation of Plasmas and Their Spectra}},
    year = {1998},
    journal = {Publications of the Astronomical Society of the Pacific},
    author = {Ferland, G. J. and Korista, K. T. and Verner, D. A. and Ferguson, J. W. and Kingdon, J. B. and Verner, E. M.},
    number = {749},
    month = {7},
    pages = {761--778},
    volume = {110},
    publisher = {IOP Publishing},
    url = {https://iopscience.iop.org/article/10.1086/316190 https://iopscience.iop.org/article/10.1086/316190/meta},
    doi = {10.1086/316190},
    issn = {0004-6280}
}

@article{Diemer2018,
    title = {{COLOSSUS: A Python Toolkit for Cosmology, Large-scale Structure, and Dark Matter Halos}},
    year = {2018},
    journal = {The Astrophysical Journal Supplement Series},
    author = {Diemer, Benedikt},
    number = {2},
    month = {12},
    pages = {35},
    volume = {239},
    publisher = {IOP Publishing},
    url = {https://iopscience.iop.org/article/10.3847/1538-4365/aaee8c https://iopscience.iop.org/article/10.3847/1538-4365/aaee8c/meta},
    doi = {10.3847/1538-4365/AAEE8C},
    issn = {0067-0049},
    arxivId = {1712.04512}
}

@article{Fichtner2024,
    title = {{Connecting stellar and galactic scales: Energetic feedback from stellar wind bubbles to supernova remnants}},
    year = {2024},
    journal = {Astronomy {\&} Astrophysics},
    author = {Fichtner, Yvonne A. and Mackey, Jonathan and Grassitelli, Luca and Romano-D{\'{i}}az, Emilio and Porciani, Cristiano},
    month = {10},
    pages = {A72},
    volume = {690},
    publisher = {EDP Sciences},
    url = {https://www.aanda.org/articles/aa/full_html/2024/10/aa49638-24/aa49638-24.html https://www.aanda.org/articles/aa/abs/2024/10/aa49638-24/aa49638-24.html},
    doi = {10.1051/0004-6361/202449638},
    issn = {0004-6361},
    arxivId = {2402.11008}
}

@article{Trotta2011,
    title = {{Constraints on cosmic-ray propagation models from a global bayesian analysis}},
    year = {2011},
    journal = {Astrophysical Journal},
    author = {Trotta, R. and J{\'{o}}hannesson, G. and Moskalenko, I. V. and Porter, T. A. and Ruiz De Austri, R. and Strong, A. W.},
    number = {2},
    month = {3},
    pages = {106},
    volume = {729},
    publisher = {Institute of Physics Publishing},
    url = {http://galprop.stanford.edu/webrun},
    doi = {10.1088/0004-637X/729/2/106},
    issn = {15384357},
    arxivId = {1011.0037}
}

@article{Navarro-Carrera2024,
    title = {{Constraints on the Faint End of the Galaxy Stellar Mass Function at z ≃ 4–8 from Deep JWST Data}},
    year = {2024},
    journal = {The Astrophysical Journal},
    author = {Navarro-Carrera, Rafael and Rinaldi, Pierluigi and Caputi, Karina I. and Iani, Edoardo and Kokorev, Vasily and Mierlo, Sophie E. van},
    number = {2},
    month = {1},
    pages = {207},
    volume = {961},
    publisher = {IOP Publishing},
    url = {https://iopscience.iop.org/article/10.3847/1538-4357/ad0df6 https://iopscience.iop.org/article/10.3847/1538-4357/ad0df6/meta},
    doi = {10.3847/1538-4357/AD0DF6},
    issn = {0004-637X},
    arxivId = {2305.16141}
}

@article{Girichidis2018,
    title = {{Cooler and smoother - the impact of cosmic rays on the phase structure of galactic outflows}},
    year = {2018},
    journal = {Monthly Notices of the Royal Astronomical Society},
    author = {Girichidis, Philipp and Naab, Thorsten and Hanasz, Michał and Walch, Stefanie},
    number = {3},
    pages = {3042--3067},
    volume = {479},
    url = {http://flash.uchicago.edu/site/},
    doi = {10.1093/mnras/sty1653},
    issn = {13652966},
    arxivId = {1805.09333}
}

@article{Hopkins2021a,
    title = {{Cosmic ray driven outflows to Mpc scales from L*galaxies}},
    year = {2021},
    journal = {Monthly Notices of the Royal Astronomical Society},
    author = {Hopkins, Philip F and Chan, T K and Ji, Suoqing and Hummels, Cameron B and Kere{\v{s}}, Dušan and Quataert, Eliot and Faucher-Gigu{\`{e}}re, Claude André},
    number = {3},
    pages = {3640--3662},
    volume = {501},
    url = {http://fire.northwestern.edu},
    doi = {10.1093/mnras/staa3690},
    issn = {13652966},
    arxivId = {2002.02462}
}

@article{Dashyan2020,
    title = {{Cosmic ray feedback from supernovae in dwarf galaxies}},
    year = {2020},
    journal = {Astronomy and Astrophysics},
    author = {Dashyan, Gohar and Dubois, Yohan},
    month = {6},
    pages = {A123},
    volume = {638},
    publisher = {EDP Sciences},
    url = {https://www.aanda.org/10.1051/0004-6361/201936339},
    doi = {10.1051/0004-6361/201936339},
    issn = {14320746},
    arxivId = {2003.09900}
}

@article{Ruszkowski2023,
    title = {{Cosmic ray feedback in galaxies and galaxy clusters: A pedagogical introduction and a topical review of the acceleration, transport, observables, and dynamical impact of cosmic rays}},
    year = {2023},
    journal = {Astronomy and Astrophysics Review},
    author = {Ruszkowski, Mateusz and Pfrommer, Christoph},
    number = {1},
    month = {6},
    volume = {31},
    url = {http://arxiv.org/abs/2306.03141},
    doi = {10.1007/s00159-023-00149-2},
    issn = {09354956},
    arxivId = {2306.03141}
}

@article{Wiener2017,
    title = {{Cosmic ray-driven galactic winds: streaming or diffusion?}},
    year = {2017},
    journal = {Monthly Notices of the Royal Astronomical Society},
    author = {Ji, Suoqing and Peng Oh, S. and McCourt, Michael},
    number = {1},
    month = {1},
    pages = {906},
    volume = {467},
    publisher = {Narnia},
    url = {https://academic.oup.com/mnras/article-lookup/doi/10.1093/mnras/stx127},
    doi = {10.1093/mnras/stx127},
    issn = {0035-8711}
}

@article{Hanasz2013,
    title = {{Cosmic rays can drive strong outflows from gas-rich high-redshift disk galaxies}},
    year = {2013},
    journal = {Astrophysical Journal Letters},
    author = {Hanasz, M and Lesch, H and Naab, T and Gawryszczak, A and Kowalik, K and W{\'{o}}lta{\'{n}}ski, D.},
    number = {2},
    pages = {L38},
    volume = {777},
    doi = {10.1088/2041-8205/777/2/L38},
    issn = {20418205},
    arxivId = {1310.3273}
}

@article{Madau2014,
    title = {{Cosmic Star Formation History}},
    year = {2014},
    journal = {Annual Review of Astronomy and Astrophysics},
    author = {Madau, Piero and Dickinson, Mark},
    number = {1},
    month = {8},
    pages = {415--486},
    volume = {52},
    publisher = {Annual Reviews},
    url = {http://www.annualreviews.org/doi/10.1146/annurev-astro-081811-125615 http://arxiv.org/abs/1403.0007%0Ahttp://dx.doi.org/10.1146/annurev-astro-081811-125615},
    isbn = {1539-3704 (Electronic){\textbackslash}r0003-4819 (Linking)},
    doi = {10.1146/annurev-astro-081811-125615},
    issn = {0066-4146},
    pmid = {23856689},
    arxivId = {1403.0007}
}

@article{Armillotta2021,
    title = {{Cosmic-Ray Transport in Simulations of Star-forming Galactic Disks}},
    year = {2021},
    journal = {The Astrophysical Journal},
    author = {Armillotta, Lucia and Ostriker, Eve C. and Jiang, Yan-Fei},
    number = {1},
    month = {11},
    pages = {11},
    volume = {922},
    publisher = {IOP Publishing},
    url = {https://iopscience.iop.org/article/10.3847/1538-4357/ac1db2 https://iopscience.iop.org/article/10.3847/1538-4357/ac1db2/meta},
    doi = {10.3847/1538-4357/AC1DB2},
    issn = {0004-637X},
    arxivId = {2108.09356}
}

@article{Sike2025,
    title = {{Cosmic-Ray-driven Galactic Winds with Resolved Interstellar Medium and Ion-neutral Damping}},
    year = {2025},
    journal = {The Astrophysical Journal},
    author = {Sike, Brandon and Thomas, Timon and Ruszkowski, Mateusz and Pfrommer, Christoph and Weber, Matthias},
    number = {2},
    month = {7},
    pages = {204},
    volume = {987},
    publisher = {IOP Publishing},
    url = {https://iopscience.iop.org/article/10.3847/1538-4357/adda3d https://iopscience.iop.org/article/10.3847/1538-4357/adda3d/meta},
    doi = {10.3847/1538-4357/adda3d},
    issn = {0004-637X}
}

@article{Teyssier2002,
    title = {{Cosmological Hydrodynamics with Adaptive Mesh Refinement: a new high resolution code called RAMSES}},
    year = {2002},
    journal = {Astronomy {\&} Astrophysics},
    author = {Teyssier, Romain},
    number = {1},
    month = {4},
    pages = {337--364},
    volume = {385},
    publisher = {EDP Sciences},
    url = {http://www.aanda.org/10.1051/0004-6361:20011817 http://arxiv.org/abs/astro-ph/0111367%0Ahttp://dx.doi.org/10.1051/0004-6361:20011817},
    isbn = {0902009192},
    doi = {10.1051/0004-6361:20011817},
    issn = {14052059},
    pmid = {9010224},
    arxivId = {astro-ph/0111367}
}

@article{Attia2021,
    title = {{Cosmological magnetogenesis: The Biermann battery during the Epoch of reionization}},
    year = {2021},
    journal = {Monthly Notices of the Royal Astronomical Society},
    author = {Attia, Omar and Teyssier, Romain and Katz, Harley and Kimm, Taysun and Martin-Alvarez, Sergio and Ocvirk, Pierre and Rosdahl, Joakim},
    number = {2},
    month = {4},
    pages = {2346--2359},
    volume = {504},
    publisher = {Oxford Academic},
    url = {https://academic.oup.com/mnras/article/504/2/2346/6226645},
    doi = {10.1093/mnras/stab1030},
    issn = {13652966},
    arxivId = {2102.09535}
}

@article{Smith2019,
    title = {{Cosmological simulations of dwarfs: the need for ISM physics beyond SN feedback alone}},
    year = {2019},
    journal = {Monthly Notices of the Royal Astronomical Society},
    author = {Smith, Matthew C and Sijacki, Debora and Shen, Sijing},
    number = {3},
    month = {5},
    pages = {3317--3333},
    volume = {485},
    publisher = {Narnia},
    url = {https://academic.oup.com/mnras/article/485/3/3317/5368355 http://arxiv.org/abs/1807.04288%0Ahttp://dx.doi.org/10.1093/mnras/stz599},
    doi = {10.1093/mnras/stz599},
    issn = {0035-8711},
    arxivId = {1807.04288}
}

@article{Vogelsberger2020,
    title = {{Cosmological simulations of galaxy formation}},
    year = {2020},
    journal = {Nature Reviews Physics},
    author = {Vogelsberger, Mark and Marinacci, Federico and Torrey, Paul and Puchwein, Ewald},
    number = {1},
    month = {1},
    pages = {42--66},
    volume = {2},
    publisher = {Nature Publishing Group},
    url = {https://www.nature.com/articles/s42254-019-0127-2},
    doi = {10.1038/s42254-019-0127-2},
    issn = {25225820},
    arxivId = {1909.07976}
}

@article{Springel2003a,
    title = {{Cosmological SPH simulations: A hybrid multi-phase model for star formation}},
    year = {2003},
    journal = {Monthly Notices of the Royal Astronomical Society},
    author = {Springel, Volker and Hernquist, Lars},
    number = {2},
    month = {2},
    pages = {289--311},
    volume = {339},
    publisher = {Narnia},
    url = {https://academic.oup.com/mnras/article-lookup/doi/10.1046/j.1365-8711.2003.06206.x http://arxiv.org/abs/astro-ph/0206393%0Ahttp://dx.doi.org/10.1046/j.1365-8711.2003.06206.x},
    doi = {10.1046/j.1365-8711.2003.06206.x},
    issn = {0035-8711},
    arxivId = {astro-ph/0206393}
}

@article{Shuntov2025,
    title = {{COSMOS-Web: Stellar mass assembly in relation to dark matter halos across 0.2 < z < 12 of cosmic history}},
    year = {2025},
    journal = {Astronomy {\&} Astrophysics},
    author = {Shuntov, M. and Ilbert, O. and Toft, S. and Arango-Toro, R. C. and Akins, H. B. and Casey, C. M. and Franco, M. and Harish, S. and Kartaltepe, J. S. and Koekemoer, A. M. and McCracken, H. J. and Paquereau, L. and Laigle, C. and Bethermin, M. and Dubois, Y. and Drakos, N. E. and Faisst, A. and Gozaliasl, G. and Gillman, S. and Hayward, C. C. and Hirschmann, M. and Huertas-Company, M. and Jespersen, C. K. and Jin, S. and Kokorev, V. and Lambrides, E. and Le Borgne, D. and Liu, D. and Magdis, G. and Massey, R. and McPartland, C. J.R. and Mercier, W. and McCleary, J. E. and McKinney, J. and Oesch, P. A. and Renzini, A. and Rhodes, J. D. and Rich, R. M. and Robertson, B. E. and Sanders, D. and Trebitsch, M. and Tresse, L. and Valentino, F. and Vijayan, A. P. and Weaver, J. R. and Weibel, A. and Wilkins, S. M. and Yang, L.},
    month = {3},
    pages = {A20},
    volume = {695},
    publisher = {EDP Sciences},
    url = {https://www.aanda.org/articles/aa/full_html/2025/03/aa52570-24/aa52570-24.html https://www.aanda.org/articles/aa/abs/2025/03/aa52570-24/aa52570-24.html},
    doi = {10.1051/0004-6361/202452570},
    issn = {0004-6361},
    arxivId = {2410.08290}
}

@article{Dubois2014,
    title = {{Dancing in the dark: Galactic properties trace spin swings along the cosmic web}},
    year = {2014},
    journal = {Monthly Notices of the Royal Astronomical Society},
    author = {Dubois, Y. and Pichon, C. and Welker, C. and Le Borgne, D. and Devriendt, J. and Laigle, C. and Codis, S. and Pogosyan, D. and Arnouts, S. and Benabed, K. and Bertin, E. and Blaizot, J. and Bouchet, F. and Cardoso, J. F. and Colombi, S. and De Lapparent, V. and Desjacques, V. and Gavazzi, R. and Kassin, S. and Kimm, T. and McCracken, H. and Milliard, B. and Peirani, S. and Prunet, S. and Rouberol, S. and Silk, J. and Slyz, A. and Sousbie, T. and Teyssier, R. and Tresse, L. and Treyer, M. and Vibert, D. and Volonteri, M.},
    number = {2},
    month = {10},
    pages = {1453--1468},
    volume = {444},
    publisher = {Oxford University Press},
    url = {https://academic.oup.com/mnras/article/444/2/1453/990507},
    doi = {10.1093/mnras/stu1227},
    issn = {13652966},
    arxivId = {1402.1165}
}

@article{Sanders2024,
    title = {{Direct T e-based Metallicities of z = 2–9 Galaxies with JWST/NIRSpec: Empirical Metallicity Calibrations Applicable from Reionization to Cosmic Noon}},
    year = {2024},
    journal = {The Astrophysical Journal},
    author = {Sanders, Ryan L. and Shapley, Alice E. and Topping, Michael W. and Reddy, Naveen A. and Brammer, Gabriel B.},
    number = {1},
    month = {2},
    pages = {24},
    volume = {962},
    publisher = {IOP Publishing},
    url = {https://iopscience.iop.org/article/10.3847/1538-4357/ad15fc https://iopscience.iop.org/article/10.3847/1538-4357/ad15fc/meta},
    doi = {10.3847/1538-4357/AD15FC},
    issn = {0004-637X},
    arxivId = {2303.08149}
}

@article{Meurer1999,
    title = {{Dust Absorption and the Ultraviolet Luminosity Density at z ≈ 3 as Calibrated by Local Starburst Galaxies*}},
    year = {1999},
    journal = {The Astrophysical Journal},
    author = {Meurer, Gerhardt R. and Heckman, Timothy M. and Calzetti, Daniela},
    number = {1},
    month = {8},
    pages = {64},
    volume = {521},
    publisher = {IOP Publishing},
    url = {https://iopscience.iop.org/article/10.1086/307523 https://iopscience.iop.org/article/10.1086/307523/meta},
    doi = {10.1086/307523},
    issn = {0004-637X},
    arxivId = {astro-ph/9903054}
}

@article{Sanati2024,
    title = {{Dwarf galaxies as a probe of a primordially magnetized Universe}},
    year = {2024},
    journal = {Astronomy {\&} Astrophysics},
    author = {Sanati, Mahsa and Martin-Alvarez, Sergio and Schober, Jennifer and Revaz, Yves and Slyz, Adrianne and Devriendt, Julien},
    month = {10},
    pages = {A59},
    volume = {690},
    publisher = {EDP Sciences},
    url = {https://www.aanda.org/articles/aa/full_html/2024/10/aa49822-24/aa49822-24.html https://www.aanda.org/articles/aa/abs/2024/10/aa49822-24/aa49822-24.html},
    doi = {10.1051/0004-6361/202449822},
    issn = {0004-6361},
    arxivId = {2403.05672}
}

@article{Robertson2024,
    title = {{Earliest Galaxies in the JADES Origins Field: Luminosity Function and Cosmic Star Formation Rate Density 300 Myr after the Big Bang}},
    year = {2024},
    journal = {The Astrophysical Journal},
    author = {Robertson, Brant and Johnson, Benjamin D. and Tacchella, Sandro and Eisenstein, Daniel J. and Hainline, Kevin and Arribas, Santiago and Baker, William M. and Bunker, Andrew J. and Carniani, Stefano and Cargile, Phillip A. and Carreira, Courtney and Charlot, Stephane and Chevallard, Jacopo and Curti, Mirko and Curtis-Lake, Emma and D’Eugenio, Francesco and Egami, Eiichi and Hausen, Ryan and Helton, Jakob M. and Jakobsen, Peter and Ji, Zhiyuan and Jones, Gareth C. and Maiolino, Roberto and Maseda, Michael V. and Nelson, Erica and P{\'{e}}rez-Gonz{\'{a}}lez, Pablo G. and Pusk{\'{a}}s, Dávid and Rieke, Marcia and Smit, Renske and Sun, Fengwu and {\"{U}}bler, Hannah and Whitler, Lily and Williams, Christina C. and Willmer, Christopher N. A. and Willott, Chris and Witstok, Joris},
    number = {1},
    month = {7},
    pages = {31},
    volume = {970},
    publisher = {IOP Publishing},
    url = {https://iopscience.iop.org/article/10.3847/1538-4357/ad463d https://iopscience.iop.org/article/10.3847/1538-4357/ad463d/meta},
    doi = {10.3847/1538-4357/AD463D},
    issn = {0004-637X},
    arxivId = {2312.10033}
}

@article{Dekel2023,
    title = {{Efficient formation of massive galaxies at cosmic dawn by feedback-free starbursts}},
    year = {2023},
    journal = {Monthly Notices of the Royal Astronomical Society},
    author = {Dekel, Avishai and Sarkar, Kartick C. and Birnboim, Yuval and Mandelker, Nir and Li, Zhaozhou},
    number = {3},
    month = {6},
    pages = {3201--3218},
    volume = {523},
    publisher = {Oxford Academic},
    url = {https://dx.doi.org/10.1093/mnras/stad1557},
    doi = {10.1093/MNRAS/STAD1557},
    issn = {0035-8711},
    arxivId = {2303.04827}
}

@article{Sheth2001,
    title = {{Ellipsoidal collapse and an improved model for the number and spatial distribution of dark matter haloes}},
    year = {2001},
    journal = {Monthly Notices of the Royal Astronomical Society},
    author = {Sheth, Ravi K. and Mo, H. J. and Tormen, Giuseppe},
    number = {1},
    month = {5},
    pages = {1--12},
    volume = {323},
    publisher = {Oxford Academic},
    url = {https://dx.doi.org/10.1046/j.1365-8711.2001.04006.x},
    doi = {10.1046/J.1365-8711.2001.04006.X},
    issn = {0035-8711},
    arxivId = {astro-ph/9907024}
}

@article{Harvey2025,
    title = {{EPOCHS. IV. SED Modeling Assumptions and Their Impact on the Stellar Mass Function at 6.5 ≤ z ≤ 13.5 Using PEARLS and Public JWST Observations}},
    year = {2025},
    journal = {The Astrophysical Journal},
    author = {Harvey, Thomas and Conselice, Christopher J. and Adams, Nathan J. and Austin, Duncan and Juod{\v{z}}balis, Ignas and Trussler, James and Li, Qiong and Ormerod, Katherine and Ferreira, Leonardo and Lovell, Christopher C. and Duan, Qiao and Westcott, Lewi and Harris, Honor and Bhatawdekar, Rachana and Coe, Dan and Cohen, Seth H. and Caruana, Joseph and Cheng, Cheng and Driver, Simon P. and Frye, Brenda and Furtak, Lukas J. and Grogin, Norman A. and Hathi, Nimish P. and Holwerda, Benne W. and Jansen, Rolf A. and Koekemoer, Anton M. and Marshall, Madeline A. and Nonino, Mario and Vijayan, Aswin P. and Wilkins, Stephen M. and Windhorst, Rogier and Willmer, Christopher N. A. and Yan, Haojing and Zitrin, Adi},
    number = {1},
    month = {12},
    pages = {89},
    volume = {978},
    publisher = {IOP Publishing},
    url = {https://iopscience.iop.org/article/10.3847/1538-4357/ad8c29 https://iopscience.iop.org/article/10.3847/1538-4357/ad8c29/meta},
    doi = {10.3847/1538-4357/AD8C29},
    issn = {0004-637X},
    arxivId = {2403.03908}
}

@article{Kimm2014,
    title = {{Escape fraction of ionizing photons during reionization: Effects due to supernova feedback and runaway ob stars}},
    year = {2014},
    journal = {Astrophysical Journal},
    author = {Kimm, Taysun and Cen, Renyue},
    number = {2},
    month = {5},
    pages = {121},
    volume = {788},
    publisher = {IOP Publishing},
    url = {http://stacks.iop.org/0004-637X/788/i=2/a=121?key=crossref.0f0973026711d57f08aaf5f5ed24ac18},
    doi = {10.1088/0004-637X/788/2/121},
    issn = {15384357},
    arxivId = {1405.0552}
}

@article{Taziaux2025,
    title = {{Exploring magnetised galactic outflows in starburst dwarf galaxies NGC 3125 and IC 4662}},
    year = {2025},
    journal = {Astronomy {\&} Astrophysics},
    author = {Taziaux, Sam and M{\"{u}}ller, Ancla and Adebahr, Björn and Basu, Aritra and Pfrommer, Christoph and Stein, Michael and Chy{\.{z}}y, Krysztof T. and Bomans, Dominik J. and En{\ss}lin, Torsten and Heesen, Volker and Kamphuis, Peter and Soida, Marian and Wezgowiec, Marek and Dettmar, Ralf Jürgen and Das, Samata and Tjus, Julia},
    month = {4},
    pages = {A226},
    volume = {696},
    publisher = {EDP Sciences},
    url = {https://www.aanda.org/articles/aa/full_html/2025/04/aa53311-24/aa53311-24.html https://www.aanda.org/articles/aa/abs/2025/04/aa53311-24/aa53311-24.html},
    doi = {10.1051/0004-6361/202453311},
    issn = {0004-6361}
}

@article{vanDaalen2020,
    title = {{Exploring the effects of galaxy formation on matter clustering through a library of simulation power spectra}},
    year = {2020},
    journal = {Monthly Notices of the Royal Astronomical Society},
    author = {van Daalen, Marcel P. and McCarthy, Ian G. and Schaye, Joop},
    number = {2},
    month = {1},
    pages = {2424--2446},
    volume = {491},
    publisher = {Oxford Academic},
    url = {https://dx.doi.org/10.1093/mnras/stz3199},
    doi = {10.1093/mnras/stz3199},
    issn = {13652966},
    arxivId = {1906.00968}
}

@article{Yuan2025,
    title = {{Extended red wings and the visibility of reionization-epoch Lyman-{$\alpha$} emitters}},
    year = {2025},
    journal = {Monthly Notices of the Royal Astronomical Society},
    author = {Yuan, Yuxuan and Martin-Alvarez, Sergio and Haehnelt, Martin G and Garel, Thibault and Keating, Laura and Witstok, Joris and Sijacki, Debora},
    number = {2},
    month = {8},
    pages = {762--789},
    volume = {542},
    publisher = {Oxford Academic},
    url = {https://dx.doi.org/10.1093/mnras/staf1252},
    doi = {10.1093/MNRAS/STAF1252},
    issn = {0035-8711},
    arxivId = {2412.07970}
}

@article{Martin-Alvarez2024,
    title = {{Extragalactic Magnetism with SOFIA (SALSA Legacy Program). VII. A tomographic view of far infrared and radio polarimetric observations through MHD simulations of galaxies}},
    year = {2024},
    journal = {The Astrophysical Journal},
    author = {Martin-Alvarez, Sergio and Lopez-Rodriguez, Enrique and Dacunha, Tara and Clark, Susan E. and Borlaff, Alejandro S. and Beck, Rainer and Montero, Francisco Rodríguez and Jung, Seoyoung Lyla and Devriendt, Julien and Slyz, Adrianne and Roman-Duval, Julia and Ntormousi, Evangelia and Tahani, Mehrnoosh and Subramanian, Kandaswamy and Dale, Daniel A. and Marcum, Pamela M. and Tassis, Konstantinos and del Moral-Castro, Ignacio and Tram, Le Ngoc and Jarvis, Matt J.},
    number = {1},
    month = {4},
    pages = {43},
    volume = {966},
    publisher = {IOP Publishing},
    url = {https://iopscience.iop.org/article/10.3847/1538-4357/ad2e9e https://iopscience.iop.org/article/10.3847/1538-4357/ad2e9e/meta http://arxiv.org/abs/2311.06356},
    doi = {10.3847/1538-4357/ad2e9e},
    issn = {15384357},
    arxivId = {2311.06356}
}

@article{Saldana-Lopez2025,
    title = {{Feedback and dynamical masses in high-z galaxies: the advent of high-resolution NIRSpec spectroscopy}},
    year = {2025},
    journal = {Monthly Notices of the Royal Astronomical Society},
    author = {Saldana-Lopez, A. and Chisholm, J. and Gazagnes, S. and Endsley, R. and Hayes, M. J. and Berg, D. A. and Finkelstein, S. L. and Flury, S. R. and Guseva, N. G. and Henry, A. and Izotov, Y. I. and Lambrides, E. and Marques-Chaves, R. and Richardson, C. T.},
    number = {1},
    month = {10},
    pages = {132--151},
    volume = {544},
    publisher = {Oxford Academic},
    url = {https://dx.doi.org/10.1093/mnras/staf1680},
    doi = {10.1093/MNRAS/STAF1680},
    issn = {0035-8711},
    arxivId = {2501.17145}
}

@article{Oppenheimer2010,
    title = {{Feedback and recycled wind accretion: assembling the z= 0 galaxy mass function}},
    year = {2010},
    journal = {Monthly Notices of the Royal Astronomical Society},
    author = {Oppenheimer, Benjamin D. and Dav{\'{e}}, Romeel and Kere{\v{s}}, Dušan and Fardal, Mark and Katz, Neal and Kollmeier, Juna A. and Weinberg, David H.},
    number = {4},
    month = {8},
    pages = {2325--2338},
    volume = {406},
    publisher = {Oxford Academic},
    url = {https://dx.doi.org/10.1111/j.1365-2966.2010.16872.x},
    doi = {10.1111/J.1365-2966.2010.16872.X},
    issn = {0035-8711},
    arxivId = {0912.0519}
}

@article{Guo2008,
    title = {{Feedback heating by cosmic rays in clusters of galaxies}},
    year = {2008},
    journal = {Monthly Notices of the Royal Astronomical Society},
    author = {Guo, Fulai and Oh, S. Peng},
    number = {1},
    month = {2},
    pages = {251--266},
    volume = {384},
    publisher = {Oxford Academic},
    url = {https://academic.oup.com/mnras/article-lookup/doi/10.1111/j.1365-2966.2007.12692.x},
    doi = {10.1111/j.1365-2966.2007.12692.x},
    issn = {00358711}
}

@article{Kimm2017,
    title = {{Feedback-regulated star formation and escape of LyC photons from mini-haloes during reionization}},
    year = {2017},
    journal = {Monthly Notices of the Royal Astronomical Society},
    author = {Kimm, Taysun and Katz, Harley and Haehnelt, Martin and Rosdahl, Joakim and Devriendt, Julien and Slyz, Adrianne},
    number = {4},
    month = {5},
    pages = {4826--4846},
    volume = {466},
    publisher = {Oxford Academic},
    url = {https://dx.doi.org/10.1093/mnras/stx052},
    doi = {10.1093/MNRAS/STX052},
    issn = {0035-8711},
    arxivId = {1608.04762}
}

@article{Faucher-Giguere2013,
    title = {{Feedback-regulated star formation in molecular clouds and galactic discs}},
    year = {2013},
    journal = {Monthly Notices of the Royal Astronomical Society},
    author = {Faucher-Gigu{\`{e}}re, Claude André and Quataert, Eliot and Hopkins, Philip F.},
    number = {3},
    month = {8},
    pages = {1970--1990},
    volume = {433},
    publisher = {Oxford Academic},
    url = {https://dx.doi.org/10.1093/mnras/stt866},
    doi = {10.1093/MNRAS/STT866},
    issn = {0035-8711},
    arxivId = {1301.3905}
}

@article{Hopkins2022a,
    title = {{First predicted cosmic ray spectra, primary-to-secondary ratios, and ionization rates from MHD galaxy formation simulations}},
    year = {2022},
    journal = {Monthly Notices of the Royal Astronomical Society},
    author = {Hopkins, Philip F. and Butsky, Iryna S. and Panopoulou, Georgia V. and Ji, Suoqing and Quataert, Eliot and Faucher-Gigu{\`{e}}re, Claude André and Kere{\v{s}}, Dušan},
    number = {3},
    month = {9},
    pages = {3470--3514},
    volume = {516},
    publisher = {Oxford Academic},
    url = {https://dx.doi.org/10.1093/mnras/stac1791},
    doi = {10.1093/MNRAS/STAC1791},
    issn = {0035-8711},
    arxivId = {2109.09762}
}

@article{Pillepich2018b,
    title = {{First results from the illustristng simulations: The stellar mass content of groups and clusters of galaxies}},
    year = {2018},
    journal = {Monthly Notices of the Royal Astronomical Society},
    author = {Pillepich, Annalisa and Nelson, Dylan and Hernquist, Lars and Springe, Volker and {R{\"{u}}diger Pakmor} and Torrey, Paul and Weinberger, Rainer and Gene, Shy and Naiman, Jill P and Marinacci, Federico and Vogelsberger, Mark},
    number = {1},
    month = {3},
    pages = {648--675},
    volume = {475},
    publisher = {Narnia},
    url = {https://academic.oup.com/mnras/article/475/1/648/4683271},
    doi = {10.1093/mnras/stx3112},
    issn = {13652966}
}

@article{Kugel2023,
    title = {{FLAMINGO: calibrating large cosmological hydrodynamical simulations with machine learning}},
    year = {2023},
    journal = {Monthly Notices of the Royal Astronomical Society},
    author = {Kugel, Roi and Schaye, Joop and Schaller, Matthieu and Helly, John C. and Braspenning, Joey and Elbers, Willem and Frenk, Carlos S. and McCarthy, Ian G. and Kwan, Juliana and Salcido, Jaime and van Daalen, Marcel P. and Vandenbroucke, Bert and Bah{\'{e}}, Yannick M. and Borrow, Josh and Chaikin, Evgenii and Hu{\v{s}}ko, Filip and Jenkins, Adrian and Lacey, Cedric G. and Nobels, Folkert S.J. and Vernon, Ian},
    number = {4},
    month = {10},
    pages = {6103--6127},
    volume = {526},
    publisher = {Oxford Academic},
    url = {https://dx.doi.org/10.1093/mnras/stad2540},
    doi = {10.1093/MNRAS/STAD2540},
    issn = {0035-8711},
    pmid = {37900898},
    arxivId = {2306.05492}
}

@article{Semenov2025,
    title = {{From UV-bright Galaxies to Early Disks: The Importance of Turbulent Star Formation in the Early Universe}},
    year = {2025},
    journal = {The Astrophysical Journal},
    author = {Semenov, Vadim A and Conroy, Charlie and Hernquist, Lars},
    number = {2},
    month = {8},
    pages = {219},
    volume = {989},
    publisher = {IOP Publishing},
    url = {https://iopscience.iop.org/article/10.3847/1538-4357/ade22d https://iopscience.iop.org/article/10.3847/1538-4357/ade22d/meta},
    doi = {10.3847/1538-4357/ADE22D},
    issn = {0004-637X},
    arxivId = {2410.09205}
}

@article{Ishigaki2018,
    title = {{Full-data Results of Hubble Frontier Fields: UV Luminosity Functions at z ∼ 6–10 and a Consistent Picture of Cosmic Reionization}},
    year = {2018},
    journal = {The Astrophysical Journal},
    author = {Ishigaki, Masafumi and Kawamata, Ryota and Ouchi, Masami and Oguri, Masamune and Shimasaku, Kazuhiro and Ono, Yoshiaki},
    number = {1},
    month = {2},
    pages = {73},
    volume = {854},
    publisher = {IOP Publishing},
    url = {https://iopscience.iop.org/article/10.3847/1538-4357/aaa544 https://iopscience.iop.org/article/10.3847/1538-4357/aaa544/meta},
    doi = {10.3847/1538-4357/AAA544},
    issn = {0004-637X},
    arxivId = {1702.04867}
}

@article{Cummings2016,
    title = {{GALACTIC COSMIC RAYS IN THE LOCAL INTERSTELLAR MEDIUM: VOYAGER 1 OBSERVATIONS AND MODEL RESULTS}},
    year = {2016},
    journal = {The Astrophysical Journal},
    author = {Cummings, A C and Stone, E C and Heikkila, B C and Lal, N and Webber, W R and J{\'{o}}hannesson, G and Moskalenko, I V and Orlando, E and Porter, T A},
    number = {1},
    pages = {18},
    volume = {831},
    doi = {10.3847/0004-637x/831/1/18}
}

@misc{McQuinn2019,
    title = {{Galactic Winds in Low-Mass Galaxies}},
    year = {2019},
    booktitle = {arXiv},
    author = {McQuinn, Kristen B.W. and van Zee, Liese and Skillman, Evan D},
    url = {https://doi.org/10.3847/1538-4357/ab4c37},
    doi = {10.1017/s1743921319000085},
    issn = {23318422},
    arxivId = {1910.04167}
}

@article{Hopkins2014,
    title = {{Galaxies on FIRE (Feedback In Realistic Environments): Stellar feedback explains cosmologically inefficient star formation}},
    year = {2014},
    journal = {Monthly Notices of the Royal Astronomical Society},
    author = {Hopkins, Philip F. and Kere{\v{s}}, Dušan and O{\~{n}}orbe, José and Faucher-Gigu{\`{e}}re, Claude André and Quataert, Eliot and Murray, Norman and Bullock, James S.},
    number = {1},
    month = {11},
    pages = {581--603},
    volume = {445},
    publisher = {Narnia},
    url = {http://academic.oup.com/mnras/article/445/1/581/988797/Galaxies-on-FIRE-Feedback-In-Realistic},
    doi = {10.1093/mnras/stu1738},
    issn = {13652966}
}

@article{Rosdahl2015b,
    title = {{Galaxies that shine: Radiation-hydrodynamical simulations of disc galaxies}},
    year = {2015},
    journal = {Monthly Notices of the Royal Astronomical Society},
    author = {Rosdahl, Joakim and Schaye, Joop and Teyssier, Romain and Agertz, Oscar},
    number = {1},
    month = {5},
    pages = {34--58},
    volume = {451},
    publisher = {Oxford University Press},
    url = {https://academic.oup.com/mnras/article/451/1/34/1362527},
    doi = {10.1093/mnras/stv937},
    issn = {13652966},
    arxivId = {1501.04632}
}

@article{Dubois2024,
    title = {{Galaxies with grains: unraveling dust evolution and extinction curves with hydrodynamical simulations}},
    year = {2024},
    journal = {Astronomy and Astrophysics},
    author = {Dubois, Yohan and Rodr{\'{i}}guez Montero, Francisco and Guerra, Corentin and Trebitsch, Maxime and Han, San and Beckmann, Ricarda and Yi, Sukyoung K. and Lewis, Joseph and Jang, J. K.},
    month = {7},
    pages = {A240},
    volume = {687},
    url = {https://www.aanda.org/10.1051/0004-6361/202449784},
    doi = {10.1051/0004-6361/202449784},
    issn = {14320746},
    arxivId = {2402.18515}
}

@article{Weibel2024,
    title = {{Galaxy build-up in the first 1.5 Gyr of cosmic history: insights from the stellar mass function at z {\~{}} 4–9 from JWST NIRCam observations}},
    year = {2024},
    journal = {Monthly Notices of the Royal Astronomical Society},
    author = {Weibel, Andrea and Oesch, Pascal A. and Barrufet, Laia and Gottumukkala, Rashmi and Ellis, Richard S. and Santini, Paola and Weaver, John R. and Allen, Natalie and Bouwens, Rychard and Bowler, Rebecca A.A. and Brammer, Gabe and Carnall, Adam C. and Cullen, Fergus and Dayal, Pratika and Dickinson, Mark and Donnan, Callum T. and Dunlop, James S. and Giavalisco, Mauro and Grogin, Norman A. and Illingworth, Garth D. and Koekemoer, Anton M. and Labbe, Ivo and Marchesini, Danilo and McLeod, Derek J. and McLure, Ross J. and Naidu, Rohan P. and P{\'{e}}rez-Gonz{\'{a}}lez, Pablo G. and Shuntov, Marko and Stefanon, Mauro and Toft, Sune and Xiao, Mengyuan},
    number = {2},
    month = {8},
    pages = {1808--1838},
    volume = {533},
    publisher = {Oxford Academic},
    url = {https://dx.doi.org/10.1093/mnras/stae1891},
    doi = {10.1093/MNRAS/STAE1891},
    issn = {0035-8711},
    arxivId = {2403.08872}
}

@article{Stefanon2021,
    title = {{Galaxy Stellar Mass Functions from z ∼ 10 to z ∼ 6 using the Deepest Spitzer/Infrared Array Camera Data: No Significant Evolution in the Stellar-to-halo Mass Ratio of Galaxies in the First Gigayear of Cosmic Time}},
    year = {2021},
    journal = {The Astrophysical Journal},
    author = {Stefanon, Mauro and Bouwens, Rychard J. and Labb{\'{e}}, Ivo and Illingworth, Garth D. and Gonzalez, Valentino and Oesch, Pascal A.},
    number = {1},
    month = {11},
    pages = {29},
    volume = {922},
    publisher = {IOP Publishing},
    url = {https://iopscience.iop.org/article/10.3847/1538-4357/ac1bb6 https://iopscience.iop.org/article/10.3847/1538-4357/ac1bb6/meta},
    doi = {10.3847/1538-4357/AC1BB6},
    issn = {0004-637X}
}

@article{Ackermann2012,
    title = {{GeV observations of star-forming galaxies with the fermi large area telescope}},
    year = {2012},
    journal = {Astrophysical Journal},
    author = {Ackermann, M. and Ajello, M. and Allafort, A. and Baldini, L. and Ballet, J. and Bastieri, D. and Bechtol, K. and Bellazzini, R. and Berenji, B. and Bloom, E. D. and Bonamente, E. and Borgland, A. W. and Bouvier, A. and Bregeon, J. and Brigida, M. and Bruel, P. and Buehler, R. and Buson, S. and Caliandro, G. A. and Cameron, R. A. and Caraveo, P. A. and Casandjian, J. M. and Cecchi, C. and Charles, E. and Chekhtman, A. and Cheung, C. C. and Chiang, J. and Cillis, A. N. and Ciprini, S. and Claus, R. and Cohen-Tanugi, J. and Conrad, J. and Cutini, S. and De Palma, F. and Dermer, C. D. and Digel, S. W. and Do Couto E Silva, E. and Drell, P. S. and Drlica-Wagner, A. and Favuzzi, C. and Fegan, S. J. and Fortin, P. and Fukazawa, Y. and Funk, S. and Fusco, P. and Gargano, F. and Gasparrini, D. and Germani, S. and Giglietto, N. and Giordano, F. and Glanzman, T. and Godfrey, G. and Grenier, I. A. and Guiriec, S. and Gustafsson, M. and Hadasch, D. and Hayashida, M. and Hays, E. and Hughes, R. E. and J{\'{o}}hannesson, G. and Johnson, A. S. and Kamae, T. and Katagiri, H. and Kataoka, J. and Kn{\"{o}}dlseder, J. and Kuss, M. and Lande, J. and Longo, F. and Loparco, F. and Lott, B. and Lovellette, M. N. and Lubrano, P. and Madejski, G. M. and Martin, P. and Mazziotta, M. N. and McEnery, J. E. and Michelson, P. F. and Mizuno, T. and Monte, C. and Monzani, M. E. and Morselli, A. and Moskalenko, I. V. and Murgia, S. and Nishino, S. and Norris, J. P. and Nuss, E. and Ohno, M. and Ohsugi, T. and Okumura, A. and Omodei, N. and Orlando, E. and Ozaki, M. and Parent, D. and Persic, M. and Pesce-Rollins, M. and Petrosian, V. and Pierbattista, M. and Piron, F. and Pivato, G. and Porter, T. A. and Rain{\`{o}}, S. and Rando, R. and Razzano, M. and Reimer, A. and Reimer, O. and Ritz, S. and Roth, M. and Sbarra, C. and Sgr{\`{o}}, C. and Siskind, E. J. and Spandre, G. and Spinelli, P. and Stawarz, Łukasz and Strong, A. W. and Takahashi, H. and Tanaka, T. and Thayer, J. B. and Tibaldo, L. and Tinivella, M. and Torres, D. F. and Tosti, G. and Troja, E. and Uchiyama, Y. and Vandenbroucke, J. and Vianello, G. and Vitale, V. and Waite, A. P. and Wood, M. and Yang, Z.},
    number = {2},
    month = {8},
    pages = {164},
    volume = {755},
    publisher = {Institute of Physics Publishing},
    url = {https://iopscience.iop.org/article/10.1088/0004-637X/755/2/164 https://iopscience.iop.org/article/10.1088/0004-637X/755/2/164/meta},
    doi = {10.1088/0004-637X/755/2/164},
    issn = {15384357}
}

@article{Rosen1995,
    title = {{Global Models of the Interstellar Medium in Disk Galaxies}},
    year = {1995},
    journal = {The Astrophysical Journal},
    author = {Rosen, Alexander and Bregman, Joel N.},
    month = {2},
    pages = {634},
    volume = {440},
    url = {http://adsabs.harvard.edu/doi/10.1086/175303},
    doi = {10.1086/175303},
    issn = {0004-637X}
}

@article{Muratov2015,
    title = {{Gusty, gaseous flows of FIRE: Galactic winds in cosmological simulations with explicit stellar feedback}},
    year = {2015},
    journal = {Monthly Notices of the Royal Astronomical Society},
    author = {Muratov, Alexander L. and Kere{\v{s}}, Dušan and Faucher-Gigu{\`{e}}re, Claude André and Hopkins, Philip F. and Quataert, Eliot and Murray, Norman},
    number = {3},
    month = {12},
    pages = {2691--2713},
    volume = {454},
    publisher = {Narnia},
    url = {https://academic.oup.com/mnras/article-lookup/doi/10.1093/mnras/stv2126},
    doi = {10.1093/mnras/stv2126},
    issn = {13652966}
}

@article{Tacchella2022,
    title = {{H {$\alpha$} emission in local galaxies: star formation, time variability, and the diffuse ionized gas}},
    year = {2022},
    journal = {Monthly Notices of the Royal Astronomical Society},
    author = {Tacchella, Sandro and Smith, Aaron and Kannan, Rahul and Marinacci, Federico and Hernquist, Lars and Vogelsberger, Mark and Torrey, Paul and Sales, Laura and Li, Hui},
    number = {2},
    month = {5},
    pages = {2904--2929},
    volume = {513},
    publisher = {Oxford Academic},
    url = {https://dx.doi.org/10.1093/mnras/stac818},
    doi = {10.1093/MNRAS/STAC818},
    issn = {0035-8711},
    arxivId = {2112.00027}
}

@article{Saxena2026,
    title = {{Hitting the slopes: a spectroscopic view of UV continuum slopes of galaxies reveals a reddening at z > 9.5}},
    year = {2026},
    journal = {Monthly Notices of the Royal Astronomical Society},
    author = {Saxena, Aayush and Cameron, Alex J. and Katz, Harley and Bunker, Andrew J. and Chevallard, Jacopo and D'Eugenio, Francesco and Arribas, Santiago and Bhatawdekar, Rachana and Boyett, Kristan and Cargile, Phillip A. and Carniani, Stefano and Charlot, Stephane and Curti, Mirko and Curtis-Lake, Emma and Hainline, Kevin and Ji, Zhiyuan and Johnson, Benjamin D. and Jones, Gareth C. and Kumari, Nimisha and Laseter, Isaac and Maseda, Michael V. and Robertson, Brant and Simmonds, Charlotte and Tacchella, Sandro and Ubler, Hannah and Williams, Christina C. and Willott, Chris and Witstok, Joris and Zhu, Yongda},
    number = {4},
    month = {5},
    volume = {548},
    publisher = {Oxford Academic},
    url = {https://dx.doi.org/10.1093/mnras/stag808},
    doi = {10.1093/MNRAS/STAG808},
    issn = {0035-8711},
    arxivId = {2411.14532}
}

@article{Martin-Alvarez2020,
    title = {{How primordial magnetic fields shrink galaxies}},
    year = {2020},
    journal = {Monthly Notices of the Royal Astronomical Society},
    author = {Martin-Alvarez, Sergio and Slyz, Adrianne and Devriendt, Julien and G{\'{o}}mez-Guijarro, Carlos},
    pages = {4475--4495},
    volume = {495},
    url = {https://academic.oup.com/mnras/article-abstract/495/4/4475/5843277},
    doi = {10.1093/mnras/staa1438},
    issn = {0035-8711},
    arxivId = {2005.10269}
}

@article{Hayward2017,
    title = {{How stellar feedback simultaneously regulates star formation and drives outflows}},
    year = {2017},
    journal = {Monthly Notices of the Royal Astronomical Society},
    author = {Hayward, Christopher C. and Hopkins, Philip F.},
    number = {2},
    month = {2},
    pages = {1682--1698},
    volume = {465},
    publisher = {Narnia},
    url = {https://academic.oup.com/mnras/article-lookup/doi/10.1093/mnras/stw2888 http://arxiv.org/abs/1510.05650%0Ahttp://dx.doi.org/10.1093/mnras/stw2888},
    doi = {10.1093/mnras/stw2888},
    issn = {0035-8711},
    arxivId = {1510.05650}
}

@article{Katz2020,
    title = {{How to quench a dwarf galaxy: The impact of inhomogeneous reionization on dwarf galaxies and cosmic filaments}},
    year = {2020},
    journal = {Monthly Notices of the Royal Astronomical Society},
    author = {Katz, Harley and Ramsoy, Marius and Rosdahl, Joakim and Kimm, Taysun and Blaizot, Jérémy and Haehnelt, Martin G and Michel-Dansac, Léo and Garel, Thibault and Laigle, Clotilde and Devriendt, Julien and Slyz, Adrianne},
    number = {2},
    month = {5},
    pages = {2200--2220},
    volume = {494},
    publisher = {Oxford University Press (OUP)},
    url = {https://academic.oup.com/mnras/article/494/2/2200/5780248},
    doi = {10.1093/mnras/staa639},
    issn = {0035-8711},
    arxivId = {1905.11414}
}

@article{Dome2025,
    title = {{Increased burstiness at high redshift in multiphysics models combining supernova feedback, radiative transfer, and cosmic rays}},
    year = {2025},
    journal = {Monthly Notices of the Royal Astronomical Society},
    author = {Dome, Tibor and Martin-Alvarez, Sergio and Tacchella, Sandro and Yuan, Yuxuan and Sijacki, Debora},
    number = {2},
    month = {1},
    pages = {629--639},
    volume = {537},
    publisher = {Oxford Academic},
    url = {https://dx.doi.org/10.1093/mnras/staf006},
    doi = {10.1093/MNRAS/STAF006},
    issn = {0035-8711}
}

@article{Dubois2021,
    title = {{Introducing the NEWHORIZON simulation: Galaxy properties with resolved internal dynamics across cosmic time}},
    year = {2021},
    journal = {Astronomy {\&} Astrophysics},
    author = {Dubois, Yohan and Beckmann, Ricarda and Bournaud, Frédéric and Choi, Hoseung and Devriendt, Julien and Jackson, Ryan and Kaviraj, Sugata and Kimm, Taysun and Kraljic, Katarina and Laigle, Clotilde and Martin, Garreth and Park, Min Jung and Peirani, Sébastien and Pichon, Christophe and Volonteri, Marta and Yi, Sukyoung K.},
    month = {7},
    pages = {A109},
    volume = {651},
    publisher = {EDP Sciences},
    url = {https://www.aanda.org/articles/aa/full_html/2021/07/aa39429-20/aa39429-20.html https://www.aanda.org/articles/aa/abs/2021/07/aa39429-20/aa39429-20.html},
    doi = {10.1051/0004-6361/202039429},
    issn = {0004-6361},
    arxivId = {2009.10578}
}

@article{Kannan2021,
    title = {{Introducing the thesan project: radiation-magneto-hydrodynamic simulations of the epoch of reionization}},
    year = {2021},
    journal = {Monthly Notices of the Royal Astronomical Society},
    author = {Kannan, R and Garaldi, E and Smith, A and Pakmor, R and Springel, V and Vogelsberger, M and Hernquist, L},
    month = {12},
    url = {https://academic.oup.com/mnras/advance-article/doi/10.1093/mnras/stab3710/6484814},
    doi = {10.1093/MNRAS/STAB3710},
    issn = {0035-8711}
}

@article{Kannan2025,
    title = {{Introducing the THESAN-ZOOM project: radiation-hydrodynamic simulations of high-redshift galaxies with a multi-phase interstellar medium}},
    year = {2025},
    journal = {The Open Journal of Astrophysics},
    author = {Kannan, Rahul and Puchwein, Ewald and Smith, Aaron and Borrow, Josh and Garaldi, Enrico and Keating, Laura and Vogelsberger, Mark and Zier, Oliver and McClymont, William and Shen, Xuejian and Popovic, Filip and Tacchella, Sandro and Hernquist, Lars and Springel, Volker},
    month = {10},
    volume = {8},
    publisher = {Maynooth Academic Publishing},
    doi = {10.33232/001C.145804},
    arxivId = {2502.20437}
}

@article{Kim2023Tigress,
    title = {{Introducing TIGRESS-NCR. I. Coregulation of the Multiphase Interstellar Medium and Star Formation Rates}},
    year = {2023},
    journal = {The Astrophysical Journal},
    author = {Kim, Chang-Goo and Kim, Jeong-Gyu and Gong, Munan and Ostriker, Eve C.},
    number = {1},
    month = {3},
    pages = {3},
    volume = {946},
    publisher = {IOP Publishing},
    url = {https://iopscience.iop.org/article/10.3847/1538-4357/acbd3a https://iopscience.iop.org/article/10.3847/1538-4357/acbd3a/meta},
    doi = {10.3847/1538-4357/acbd3a},
    issn = {0004-637X},
    arxivId = {2211.13293}
}

@article{Llerena2023,
    title = {{Ionized gas kinematics and chemical abundances of low-mass star-forming galaxies at z  ∼  3}},
    year = {2023},
    journal = {Astronomy {\&} Astrophysics},
    author = {Llerena, M. and Amor{\'{i}}n, R. and Pentericci, L. and Calabr{\`{o}}, A. and Shapley, A. E. and Boutsia, K. and P{\'{e}}rez-Montero, E. and V{\'{i}}lchez, J. M. and Nakajima, K.},
    month = {8},
    pages = {A53},
    volume = {676},
    publisher = {EDP Sciences},
    url = {https://www.aanda.org/articles/aa/full_html/2023/08/aa46232-23/aa46232-23.html https://www.aanda.org/articles/aa/abs/2023/08/aa46232-23/aa46232-23.html},
    doi = {10.1051/0004-6361/202346232},
    issn = {0004-6361},
    arxivId = {2303.01536}
}

@article{Curti2024,
    title = {{JADES: Insights into the low-mass end of the mass–metallicity–SFR relation at 3 < z < 10 from deep JWST/NIRSpec spectroscopy}},
    year = {2024},
    journal = {Astronomy {\&} Astrophysics},
    author = {Curti, Mirko and Maiolino, Roberto and Curtis-Lake, Emma and Chevallard, Jacopo and Carniani, Stefano and D'Eugenio, Francesco and Looser, Tobias J. and Scholtz, Jan and Charlot, Stephane and Cameron, Alex and {\"{U}}bler, Hannah and Witstok, Joris and Boyett, Kristian and Laseter, Isaac and Sandles, Lester and Arribas, Santiago and Bunker, Andrew and Giardino, Giovanna and Maseda, Michael V. and Rawle, Tim and Del Pino, Bruno Rodríguez and Smit, Renske and Willott, Chris J. and Eisenstein, Daniel J. and Hausen, Ryan and Johnson, Benjamin and Rieke, Marcia and Robertson, Brant and Tacchella, Sandro and Williams, Christina C. and Willmer, Christopher and Baker, William M. and Bhatawdekar, Rachana and Egami, Eiichi and Helton, Jakob M. and Ji, Zhiyuan and Kumari, Nimisha and Perna, Michele and Shivaei, Irene and Sun, Fengwu},
    month = {4},
    pages = {A75},
    volume = {684},
    publisher = {EDP Sciences},
    url = {https://www.aanda.org/articles/aa/full_html/2024/04/aa46698-23/aa46698-23.html https://www.aanda.org/articles/aa/abs/2024/04/aa46698-23/aa46698-23.html},
    doi = {10.1051/0004-6361/202346698},
    issn = {0004-6361}
}

@article{Carniani2024,
    title = {{JADES: The incidence rate and properties of galactic outflows in low-mass galaxies across 3 < z < 9}},
    year = {2024},
    journal = {Astronomy {\&} Astrophysics},
    author = {Carniani, Stefano and Venturi, Giacomo and Parlanti, Eleonora and de Graaff, Anna and Maiolino, Roberto and Arribas, Santiago and Bonaventura, Nina and Boyett, Kristan and Bunker, Andrew J. and Cameron, Alex J. and Charlot, Stephane and Chevallard, Jacopo and Curti, Mirko and Curtis-Lake, Emma and Eisenstein, Daniel J. and Giardino, Giovanna and Hausen, Ryan and Kumari, Nimisha and Maseda, Michael V. and Nelson, Erica and Perna, Michele and Rix, Hans Walter and Robertson, Brant and Rodr{\'{i}}guez Del Pino, Bruno and Sandles, Lester and Scholtz, Jan and Simmonds, Charlotte and Smit, Renske and Tacchella, Sandro and {\"{U}}bler, Hannah and Williams, Christina C. and Willott, Chris and Witstok, Joris},
    month = {5},
    pages = {A99},
    volume = {685},
    publisher = {EDP Sciences},
    url = {https://www.aanda.org/articles/aa/full_html/2024/05/aa47230-23/aa47230-23.html https://www.aanda.org/articles/aa/abs/2024/05/aa47230-23/aa47230-23.html},
    doi = {10.1051/0004-6361/202347230},
    issn = {0004-6361},
    arxivId = {2306.11801}
}

@article{Donnan2024,
    title = {{JWST PRIMER: a new multifield determination of the evolving galaxy UV luminosity function at redshifts z ≃ 9 – 15}},
    year = {2024},
    journal = {Monthly Notices of the Royal Astronomical Society},
    author = {Donnan, C. T. and McLure, R. J. and Dunlop, J. S. and McLeod, D. J. and Magee, D. and Arellano-C{\'{o}}rdova, K. Z. and Barrufet, L. and Begley, R. and Bowler, R. A.A. and Carnall, A. C. and Cullen, F. and Ellis, R. S. and Fontana, A. and Illingworth, G. D. and Grogin, N. A. and Hamadouche, M. L. and Koekemoer, A. M. and Liu, F. Y. and Mason, C. and Santini, P. and Stanton, T. M.},
    number = {3},
    month = {8},
    pages = {3222--3237},
    volume = {533},
    publisher = {Oxford Academic},
    url = {https://dx.doi.org/10.1093/mnras/stae2037},
    doi = {10.1093/MNRAS/STAE2037},
    issn = {0035-8711}
}

@article{Nelson2023,
    title = {{JWST Reveals a Population of Ultrared, Flattened Galaxies at 2 ≲ z ≲ 6 Previously Missed by HST}},
    year = {2023},
    journal = {The Astrophysical Journal Letters},
    author = {Nelson, Erica J. and Suess, Katherine A. and Bezanson, Rachel and Price, Sedona H. and Dokkum, Pieter van and Leja, Joel and Wang, Bingjie and 王, 冰洁 and Whitaker, Katherine E. and Labb{\'{e}}, Ivo and Barrufet, Laia and Brammer, Gabriel and Eisenstein, Daniel J. and Gibson, Justus and Hartley, Abigail I. and Johnson, Benjamin D. and Heintz, Kasper E. and Mathews, Elijah and Miller, Tim B. and Oesch, Pascal A. and Sandles, Lester and Setton, David J. and Speagle, Joshua S. and 沈, 佳士 and Tacchella, Sandro and Tadaki, Ken-ichi and {\"{U}}bler, Hannah and Weaver, John. R.},
    number = {2},
    month = {5},
    pages = {L18},
    volume = {948},
    publisher = {IOP Publishing},
    url = {https://iopscience.iop.org/article/10.3847/2041-8213/acc1e1 https://iopscience.iop.org/article/10.3847/2041-8213/acc1e1/meta},
    doi = {10.3847/2041-8213/ACC1E1},
    issn = {2041-8205},
    arxivId = {2208.01630}
}

@article{Harikane2025,
    title = {{JWST, ALMA, and Keck Spectroscopic Constraints on the UV Luminosity Functions at z ∼ 7–14: Clumpiness and Compactness of the Brightest Galaxies in the Early Universe}},
    year = {2025},
    journal = {The Astrophysical Journal},
    author = {Harikane, Yuichi and Inoue, Akio K. and Ellis, Richard S. and Ouchi, Masami and Nakazato, Yurina and Yoshida, Naoki and Ono, Yoshiaki and Sun, Fengwu and Sato, Riku A. and Ferrami, Giovanni and Fujimoto, Seiji and Kashikawa, Nobunari and McLeod, Derek J. and P{\'{e}}rez-Gonz{\'{a}}lez, Pablo G. and Sawicki, Marcin and Sugahara, Yuma and Xu, Yi and Yamanaka, Satoshi and Carnall, Adam C. and Cullen, Fergus and Dunlop, James S. and Egami, Eiichi and Grogin, Norman and Isobe, Yuki and Koekemoer, Anton M. and Laporte, Nicolas and Lee, Chien-Hsiu and Magee, Dan and Matsuo, Hiroshi and Matsuoka, Yoshiki and Mawatari, Ken and Nakajima, Kimihiko and Nakane, Minami and Tamura, Yoichi and Umeda, Hiroya and Yanagisawa, Hiroto},
    number = {1},
    month = {2},
    pages = {138},
    volume = {980},
    publisher = {IOP Publishing},
    url = {https://iopscience.iop.org/article/10.3847/1538-4357/ad9b2c https://iopscience.iop.org/article/10.3847/1538-4357/ad9b2c/meta},
    doi = {10.3847/1538-4357/AD9B2C},
    issn = {0004-637X},
    arxivId = {2406.18352}
}

@article{Teyssier2006,
    title = {{Kinematic dynamos using constrained transport with high order Godunov schemes and adaptive mesh refinement}},
    year = {2006},
    journal = {Journal of Computational Physics},
    author = {Teyssier, Romain and Fromang, Sébastien and Dormy, Emmanuel},
    number = {1},
    month = {10},
    pages = {44--67},
    volume = {218},
    publisher = {Academic Press},
    url = {https://www.sciencedirect.com/science/article/pii/S0021999106000593?via%3Dihub},
    doi = {10.1016/j.jcp.2006.01.042},
    issn = {00219991},
    arxivId = {astro-ph/0601715}
}

@article{Yuan2024,
    title = {{Ly{$\alpha$} emission as a sensitive probe of feedback-regulated LyC escape from dwarf galaxies}},
    year = {2024},
    journal = {Monthly Notices of the Royal Astronomical Society},
    author = {Yuan, Yuxuan and Martin-Alvarez, Sergio and Haehnelt, Martin G. and Garel, Thibault and Sijacki, Debora},
    number = {4},
    month = {7},
    pages = {3643--3668},
    volume = {532},
    publisher = {Oxford Academic},
    url = {https://dx.doi.org/10.1093/mnras/stae1606},
    doi = {10.1093/MNRAS/STAE1606},
    issn = {0035-8711},
    arxivId = {2401.02572}
}

@techreport{Shukurov2018,
    title = {{Magnetic field effects on the ISM structure and galactic outflows}},
    year = {2018},
    booktitle = {XXIXth IAU General Assembly},
    author = {Shukurov, A. and Evirgen, C. C. and Fletcher, A. and Bushby, P. J. and Gent, F. A.},
    month = {10},
    volume = {1},
    url = {http://arxiv.org/abs/1810.01202 https://arxiv.org/pdf/1810.01202.pdf},
    institution = {International Astronomical Union},
    arxivId = {arXiv:1810.01202v1}
}

@article{Katz2025,
    title = {{MEGATRON: Reproducing the Diversity of High-Redshift Galaxy Spectra with Cosmological Radiation Hydrodynamics Simulations}},
    year = {2025},
    journal = {arXiv},
    author = {Katz, Harley and Rey, Martin P. and Cadiou, Corentin and Agertz, Oscar and Blaizot, Jeremy and Cameron, Alex J. and Choustikov, Nicholas and Devriendt, Julien and Hauk, Uliana and Jones, Gareth C. and Kimm, Taysun and Laseter, Isaac and Martin-Alvarez, Sergio and Matsumoto, Kosei and Pearce, Autumn and Montero, Francisco Rodríguez and Rosdahl, Joki and Sanati, Mahsa and Saxena, Aayush and Slyz, Adrianne and Stiskalek, Richard and Storck, Anatole and Veenema, Oscar and Yee, Wonjae},
    month = {10},
    url = {https://arxiv.org/pdf/2510.05201},
    arxivId = {2510.05201}
}

@article{Cadiou2025,
    title = {{MEGATRON: the impact of non-equilibrium effects and local radiation fields on the circumgalactic medium at cosmic noon}},
    year = {2025},
    journal = {arXiv},
    author = {Cadiou, Corentin and Katz, Harley and Rey, Martin P and Agertz, Oscar and Blaizot, Jeremy and Cameron, Alex J and Choustikov, Nicholas and Devriendt, Julien and Hauk, Uliana and Jones, Gareth C and Kimm, Taysun and Laseter, Isaac and Mart{\'{i}}n, Sergio and Mart{\'{i}}n´alvarez, Martín´ and Matsumoto, Kosei and Nyhagen, Camilla T and Pearce, Autumn and Montero, Francisco Rodríguez and Rosdahl, Joki and Rufo Pastor, Víctor and Sanati, Mahsa and Saxena, Aayush and Slyz, Adrianne and Stiskalek, Richard and Storck, Anatole and Yee, Wonjae},
    month = {10},
    url = {https://arxiv.org/pdf/2510.05667},
    arxivId = {2510.05667}
}

@article{Chisholm2018,
    title = {{Metal-enriched galactic outflows shape the mass–metallicity relationship}},
    year = {2018},
    journal = {Monthly Notices of the Royal Astronomical Society},
    author = {Chisholm, J. and Tremonti, C. and Leitherer, C.},
    number = {2},
    month = {12},
    pages = {1690--1706},
    volume = {481},
    publisher = {Oxford Academic},
    url = {https://dx.doi.org/10.1093/mnras/sty2380},
    doi = {10.1093/MNRAS/STY2380},
    issn = {0035-8711},
    arxivId = {1808.10453}
}

@article{Chisari2019,
    title = {{Modelling baryonic feedback for survey cosmology}},
    year = {2019},
    journal = {The Open Journal of Astrophysics},
    author = {Chisari, Nora Elisa and Mead, Alexander J. and Joudaki, Shahab and Ferreira, Pedro and Schneider, Aurel and Mohr, Joseph and Tr{\"{o}}ster, Tilman and Alonso, David and McCarthy, Ian G. and Martin-Alvarez, Sergio and Devriendt, Julien and Slyz, Adrianne and van Daalen, Marcel P.},
    number = {1},
    month = {6},
    pages = {9452},
    volume = {2},
    publisher = {Maynooth Academic Publishing},
    url = {https://astro.theoj.org/article/9452-modelling-baryonic-feedback-for-survey-cosmology http://arxiv.org/abs/1905.06082},
    doi = {10.21105/astro.1905.06082},
    arxivId = {1905.06082}
}

@article{Trayford2026,
    title = {{Modelling the evolution and influence of dust in cosmological simulations that include the cold phase of the interstellar medium}},
    year = {2026},
    journal = {Monthly Notices of the Royal Astronomical Society},
    author = {Trayford, James W. and Schaye, Joop and Correa, Camila and Ploeckinger, Sylvia and Richings, Alexander J. and Chaikin, Evgenii and Schaller, Matthieu and Ben{\'{i}}tez-Llambay, Alejandro and Frenk, Carlos and Hu{\v{s}}ko, Filip},
    number = {4},
    month = {1},
    volume = {545},
    publisher = {Oxford Academic},
    url = {https://dx.doi.org/10.1093/mnras/staf2040},
    doi = {10.1093/mnras/staf2040},
    issn = {13652966},
    arxivId = {2505.13056}
}

@article{Hahn2011,
    title = {{Multi-scale initial conditions for cosmological simulations}},
    year = {2011},
    journal = {Monthly Notices of the Royal Astronomical Society},
    author = {Hahn, Oliver and Abel, Tom},
    number = {3},
    month = {8},
    pages = {2101--2121},
    volume = {415},
    publisher = {Oxford University Press},
    url = {https://academic.oup.com/mnras/article-lookup/doi/10.1111/j.1365-2966.2011.18820.x},
    doi = {10.1111/j.1365-2966.2011.18820.x},
    issn = {00358711}
}

@article{Bouwens2021,
    title = {{New Determinations of the UV Luminosity Functions from z ∼ 9 to 2 Show a Remarkable Consistency with Halo Growth and a Constant Star Formation Efficiency}},
    year = {2021},
    journal = {The Astronomical Journal},
    author = {Bouwens, R. J. and Oesch, P. A. and Stefanon, M. and Illingworth, G. and Labb{\'{e}}, I. and Reddy, N. and Atek, H. and Montes, M. and Naidu, R. and Nanayakkara, T. and Nelson, E. and Wilkins, S.},
    number = {2},
    month = {7},
    pages = {47},
    volume = {162},
    publisher = {IOP Publishing},
    url = {https://iopscience.iop.org/article/10.3847/1538-3881/abf83e https://iopscience.iop.org/article/10.3847/1538-3881/abf83e/meta},
    doi = {10.3847/1538-3881/ABF83E},
    issn = {1538-3881},
    arxivId = {2102.07775}
}

@article{Leung2023,
    title = {{NGDEEP Epoch 1: The Faint End of the Luminosity Function at z ∼ 9–12 from Ultradeep JWST Imaging}},
    year = {2023},
    journal = {The Astrophysical Journal Letters},
    author = {Leung, Gene C. K. and Bagley, Micaela B. and Finkelstein, Steven L. and Ferguson, Henry C. and Koekemoer, Anton M. and P{\'{e}}rez-Gonz{\'{a}}lez, Pablo G. and Morales, Alexa and Kocevski, Dale D. and Yang, Guang and 杨, 光 and Somerville, Rachel S. and Wilkins, Stephen M. and Yung, L. Y. Aaron and Fujimoto, Seiji and Larson, Rebecca L. and Papovich, Casey and Pirzkal, Nor and Berg, Danielle A. and Lotz, Jennifer M. and Castellano, Marco and Ortiz, Óscar A. Chávez and Cheng, Yingjie and Dickinson, Mark and Giavalisco, Mauro and Hathi, Nimish P. and Hutchison, Taylor A. and Jung, Intae and Kartaltepe, Jeyhan S. and Natarajan, Priyamvada and Rothberg, Barry},
    number = {2},
    month = {9},
    pages = {L46},
    volume = {954},
    publisher = {IOP Publishing},
    url = {https://iopscience.iop.org/article/10.3847/2041-8213/acf365 https://iopscience.iop.org/article/10.3847/2041-8213/acf365/meta},
    doi = {10.3847/2041-8213/ACF365},
    issn = {2041-8205},
    arxivId = {2306.06244}
}

@article{Parizot2006,
    title = {{Observational constraints on energetic particle diffusion in young supernovae remnants: Amplified magnetic field and maximum energy}},
    year = {2006},
    journal = {Astronomy and Astrophysics},
    author = {Parizot, E and Marcowith, A and Ballet, J and Gallant, Y A},
    number = {2},
    pages = {387--395},
    volume = {453},
    url = {http://dx.doi.org/10.1051/0004-6361:20064985},
    doi = {10.1051/0004-6361:20064985},
    issn = {14320746}
}

@article{Rahmati2013,
    title = {{On the evolution of the H i column density distribution in cosmological simulations}},
    year = {2013},
    journal = {Monthly Notices of the Royal Astronomical Society},
    author = {Rahmati, Alireza and Pawlik, Andreas H. and Raicevic, Milan and Schaye, Joop},
    number = {3},
    month = {4},
    pages = {2427--2445},
    volume = {430},
    publisher = {Oxford Academic},
    url = {https://dx.doi.org/10.1093/mnras/stt066},
    doi = {10.1093/MNRAS/STT066},
    issn = {0035-8711},
    arxivId = {1210.7808}
}

@article{Ferrara2023,
    title = {{On the stunning abundance of super-early, luminous galaxies revealed by JWST}},
    year = {2023},
    journal = {Monthly Notices of the Royal Astronomical Society},
    author = {Ferrara, Andrea and Pallottini, Andrea and Dayal, Pratika},
    number = {3},
    month = {5},
    pages = {3986--3991},
    volume = {522},
    publisher = {Oxford Academic},
    url = {https://dx.doi.org/10.1093/mnras/stad1095},
    doi = {10.1093/MNRAS/STAD1095},
    issn = {0035-8711}
}

@article{Kroupa2001,
    title = {{On the variation of the initial mass function}},
    year = {2001},
    journal = {Monthly Notices of the Royal Astronomical Society},
    author = {Kroupa, Pavel},
    number = {2},
    month = {4},
    pages = {231--246},
    volume = {322},
    publisher = {Oxford University Press},
    url = {https://academic.oup.com/mnras/article-lookup/doi/10.1046/j.1365-8711.2001.04022.x},
    doi = {10.1046/j.1365-8711.2001.04022.x},
    issn = {00358711}
}

@article{Geen2015b,
    title = {{Photoionization feedback in a self-gravitating, magnetized, turbulent cloud}},
    year = {2015},
    journal = {Monthly Notices of the Royal Astronomical Society},
    author = {Geen, Sam and Hennebelle, Patrick and Tremblin, Pascal and Rosdahl, Joakim},
    number = {4},
    month = {12},
    pages = {4484--4502},
    volume = {454},
    publisher = {Oxford University Press},
    url = {https://academic.oup.com/mnras/article/454/4/4484/1001540},
    doi = {10.1093/mnras/stv2272},
    issn = {13652966},
    arxivId = {1507.02981}
}

@article{Sartorio2021,
    title = {{Photoionization feedback in turbulent molecular clouds}},
    year = {2020},
    journal = {Monthly Notices of the Royal Astronomical Society},
    author = {Sartorio, Nina S. and Vandenbroucke, Bert and Falceta-Goncalves, Diego and Wood, Kenneth},
    number = {2},
    month = {12},
    pages = {1833--1843},
    volume = {500},
    publisher = {Oxford Academic},
    url = {https://dx.doi.org/10.1093/mnras/staa3380},
    doi = {10.1093/MNRAS/STAA3380},
    issn = {0035-8711},
    arxivId = {2011.00020}
}

@article{Somerville2015,
    title = {{Physical models of galaxy formation in a cosmological framework}},
    year = {2015},
    journal = {Annual Review of Astronomy and Astrophysics},
    author = {Somerville, Rachel S. and Dav{\'{e}}, Romeel},
    number = {1},
    month = {8},
    pages = {51--113},
    volume = {53},
    publisher = {Annual Reviews Inc.},
    url = {https://www.annualreviews.org/content/journals/10.1146/annurev-astro-082812-140951},
    doi = {10.1146/ANNUREV-ASTRO-082812-140951/CITE/REFWORKS},
    issn = {00664146},
    arxivId = {1412.2712}
}

@article{PlanckCollaboration2016cosmo,
    title = {{Planck 2015 results: XIII. Cosmological parameters}},
    year = {2016},
    journal = {Astronomy and Astrophysics},
    author = {{Planck Collaboration} and Ade, P. A.R. and Aghanim, N. and Arnaud, M. and Ashdown, M. and Aumont, J. and Baccigalupi, C. and Banday, A. J. and Barreiro, R. B. and Bartlett, J. G. and Bartolo, N. and Battaner, E. and Battye, R. and Benabed, K. and Beno{\^{i}}t, A. and Benoit-L{\'{e}}vy, A. and Bernard, J. P. and Bersanelli, M. and Bielewicz, P. and Bock, J. J. and Bonaldi, A. and Bonavera, L. and Bond, J. R. and Borrill, J. and Bouchet, F. R. and Boulanger, F. and Bucher, M. and Burigana, C. and Butler, R. C. and Calabrese, E. and Cardoso, J. F. and Catalano, A. and Challinor, A. and Chamballu, A. and Chary, R. R. and Chiang, H. C. and Chluba, J. and Christensen, P. R. and Church, S. and Clements, D. L. and Colombi, S. and Colombo, L. P.L. and Combet, C. and Coulais, A. and Crill, B. P. and Curto, A. and Cuttaia, F. and Danese, L. and Davies, R. D. and Davis, R. J. and De Bernardis, P. and De Rosa, A. and De Zotti, G. and Delabrouille, J. and D{\'{e}}sert, F. X. and Di Valentino, E. and Dickinson, C. and Diego, J. M. and Dolag, K. and Dole, H. and Donzelli, S. and Dor{\'{e}}, O. and Douspis, M. and Ducout, A. and Dunkley, J. and Dupac, X. and Efstathiou, G. and Elsner, F. and En{\ss}lin, T. A. and Eriksen, H. K. and Farhang, M. and Fergusson, J. and Finelli, F. and Forni, O. and Frailis, M. and Fraisse, A. A. and Franceschi, E. and Frejsel, A. and Galeotta, S. and Galli, S. and Ganga, K. and Gauthier, C. and Gerbino, M. and Ghosh, T. and Giard, M. and Giraud-H{\'{e}}raud, Y. and Giusarma, E. and Gjerl{\o}w, E. and Gonz{\'{a}}lez-Nuevo, J. and G{\'{o}}rski, K. M. and Gratton, S. and Gregorio, A. and Gruppuso, A. and Gudmundsson, J. E. and Hamann, J. and Hansen, F. K. and Hanson, D. and Harrison, D. L. and Helou, G. and Henrot-Versill{\'{e}}, S. and Hern{\'{a}}ndez-Monteagudo, C. and Herranz, D. and Hildebrandt, S. R. and Hivon, E. and Hobson, M. and Holmes, W. A. and Hornstrup, A. and Hovest, W. and Huang, Z. and Huffenberger, K. M. and Hurier, G. and Jaffe, A. H. and Jaffe, T. R. and Jones, W. C. and Juvela, M. and Keih{\"{a}}nen, E. and Keskitalo, R. and Kisner, T. S. and Kneissl, R. and Knoche, J. and Knox, L. and Kunz, M. and Kurki-Suonio, H. and Lagache, G. and L{\"{a}}hteenm{\"{a}}ki, A. and Lamarre, J. M. and Lasenby, A. and Lattanzi, M. and Lawrence, C. R. and Leahy, J. P. and Leonardi, R. and Lesgourgues, J. and Levrier, F. and Lewis, A. and Liguori, M. and Lilje, P. B. and Linden-V{\o}rnle, M. and L{\'{o}}pez-Caniego, M. and Lubin, P. M. and Maci{\'{a}}s-P{\'{e}}rez, J. F. and Maggio, G. and Maino, D. and Mandolesi, N. and Mangilli, A. and Marchini, A. and Maris, M. and Martin, P. G. and Martinelli, M. and Mart{\'{i}}nez-Gonz{\'{a}}lez, E. and Masi, S. and Matarrese, S. and Mcgehee, P. and Meinhold, P. R. and Melchiorri, A. and Melin, J. B. and Mendes, L. and Mennella, A. and Migliaccio, M. and Millea, M. and Mitra, S. and Miville-Desch{\^{e}}nes, M. A. and Moneti, A. and Montier, L. and Morgante, G. and Mortlock, D. and Moss, A. and Munshi, D. and Murphy, J. A. and Naselsky, P. and Nati, F. and Natoli, P. and Netterfield, C. B. and N{\o}rgaard-Nielsen, H. U. and Noviello, F. and Novikov, D. and Novikov, I. and Oxborrow, C. A. and Paci, F. and Pagano, L. and Pajot, F. and Paladini, R. and Paoletti, D. and Partridge, B. and Pasian, F. and Patanchon, G. and Pearson, T. J. and Perdereau, O. and Perotto, L. and Perrotta, F. and Pettorino, V. and Piacentini, F. and Piat, M. and Pierpaoli, E. and Pietrobon, D. and Plaszczynski, S. and Pointecouteau, E. and Polenta, G. and Popa, L. and Pratt, G. W. and Pr{\'{e}}zeau, G. and Prunet, S. and Puget, J. L. and Rachen, J. P. and Reach, W. T. and Rebolo, R. and Reinecke, M. and Remazeilles, M. and Renault, C. and Renzi, A. and Ristorcelli, I. and Rocha, G. and Rosset, C. and Rossetti, M. and Roudier, G. and Rouill{\'{e}} D'orfeuil, B. and Rowan-Robinson, M. and Rubinõ-Mart{\'{i}}n, J. A. and Rusholme, B. and Said, N. and Salvatelli, V. and Salvati, L. and Sandri, M. and Santos, D. and Savelainen, M. and Savini, G. and Scott, D. and Seiffert, M. D. and Serra, P. and Shellard, E. P.S. and Spencer, L. D. and Spinelli, M. and Stolyarov, V. and Stompor, R. and Sudiwala, R. and Sunyaev, R. and Sutton, D. and Suur-Uski, A. S. and Sygnet, J. F. and Tauber, J. A. and Terenzi, L. and Toffolatti, L. and Tomasi, M. and Tristram, M. and Trombetti, T. and Tucci, M. and Tuovinen, J. and T{\"{u}}rler, M. and Umana, G. and Valenziano, L. and Valiviita, J. and Van Tent, F. and Vielva, P. and Villa, F. and Wade, L. A. and Wandelt, B. D. and Wehus, I. K. and White, M. and White, S. D.M. and Wilkinson, A. and Yvon, D. and Zacchei, A. and Zonca, A.},
    month = {10},
    pages = {A13},
    volume = {594},
    publisher = {EDP Sciences},
    url = {http://www.esa.int/Planck},
    doi = {10.1051/0004-6361/201525830},
    issn = {14320746},
    arxivId = {1502.01589}
}

@article{Andersson2024,
    title = {{Pre-supernova feedback sets the star cluster mass function to a power law and reduces the cluster formation efficiency}},
    year = {2024},
    journal = {Astronomy {\&} Astrophysics},
    author = {Andersson, Eric P. and Mac Low, Mordecai Mark and Agertz, Oscar and Renaud, Florent and Li, Hui},
    month = {1},
    pages = {A28},
    volume = {681},
    publisher = {EDP Sciences},
    url = {https://www.aanda.org/articles/aa/full_html/2024/01/aa47792-23/aa47792-23.html https://www.aanda.org/articles/aa/abs/2024/01/aa47792-23/aa47792-23.html},
    doi = {10.1051/0004-6361/202347792},
    issn = {14320746}
}

@article{Oesch2013,
    title = {{PROBING THE DAWN OF GALAXIES AT z ∼ 9–12: NEW CONSTRAINTS FROM HUDF12/XDF AND CANDELS DATA*}},
    year = {2013},
    journal = {The Astrophysical Journal},
    author = {Oesch, P. A. and Bouwens, R. J. and Illingworth, G. D. and Labb{\'{e}}, I. and Franx, M. and Van Dokkum, P. G. and Trenti, M. and Stiavelli, M. and Gonzalez, V. and Magee, D.},
    number = {1},
    month = {7},
    pages = {75},
    volume = {773},
    publisher = {IOP Publishing},
    url = {https://iopscience.iop.org/article/10.1088/0004-637X/773/1/75 https://iopscience.iop.org/article/10.1088/0004-637X/773/1/75/meta},
    doi = {10.1088/0004-637X/773/1/75},
    issn = {0004-637X},
    arxivId = {1301.6162}
}

@article{Helder2013,
    title = {{Proper motions of H{$\alpha$} filaments in the supernova remnant RCW 86}},
    year = {2013},
    journal = {Monthly Notices of the Royal Astronomical Society},
    author = {Helder, E. A. and Vink, J. and Bamba, A. and Bleeker, J. A.M. and Burrows, D. N. and Ghavamian, P. and Yamazaki, R.},
    number = {2},
    month = {10},
    pages = {910--916},
    volume = {435},
    publisher = {Oxford Academic},
    url = {https://academic.oup.com/mnras/article/435/2/910/1047123},
    doi = {10.1093/mnras/stt993},
    issn = {00358711}
}

@article{Harikane2024,
    title = {{Pure Spectroscopic Constraints on UV Luminosity Functions and Cosmic Star Formation History from 25 Galaxies at z spec = 8.61–13.20 Confirmed with JWST/NIRSpec}},
    year = {2024},
    journal = {The Astrophysical Journal},
    author = {Harikane, Yuichi and Nakajima, Kimihiko and Ouchi, Masami and Umeda, Hiroya and Isobe, Yuki and Ono, Yoshiaki and Xu, Yi and Zhang, Yechi},
    number = {1},
    month = {12},
    pages = {56},
    volume = {960},
    publisher = {IOP Publishing},
    url = {https://iopscience.iop.org/article/10.3847/1538-4357/ad0b7e https://iopscience.iop.org/article/10.3847/1538-4357/ad0b7e/meta},
    doi = {10.3847/1538-4357/AD0B7E},
    issn = {0004-637X},
    arxivId = {2304.06658}
}

@article{Farcy2022,
    title = {{Radiation-magnetohydrodynamics simulations of cosmic ray feedback in disc galaxies}},
    year = {2022},
    journal = {Monthly Notices of the Royal Astronomical Society},
    author = {Farcy, Marion and Rosdahl, Joakim and Dubois, Yohan and Blaizot, Jérémy and Martin-Alvarez, Sergio},
    number = {4},
    month = {2},
    pages = {5000--5019},
    volume = {513},
    url = {http://arxiv.org/abs/2202.01245},
    isbn = {2202.01245v1},
    doi = {10.1093/mnras/stac1196},
    issn = {0035-8711},
    arxivId = {2202.01245}
}

@article{Haardt1996,
    title = {{Radiative Transfer in a Clumpy Universe. II. The Ultraviolet Extragalactic Background}},
    year = {1996},
    journal = {The Astrophysical Journal},
    author = {Haardt, Francesco and Madau, Piero},
    month = {4},
    pages = {20},
    volume = {461},
    url = {http://adsabs.harvard.edu/doi/10.1086/177035},
    doi = {10.1086/177035},
    issn = {0004-637X},
    arxivId = {astro-ph/9509093}
}

@article{Rosdahl2013,
    title = {{Ramses-rt: Radiation hydrodynamics in the cosmological context}},
    year = {2013},
    journal = {Monthly Notices of the Royal Astronomical Society},
    author = {Rosdahl, J. and Blaizot, J. and Aubert, D. and Stranex, T. and Teyssier, R.},
    number = {3},
    month = {12},
    pages = {2188--2231},
    volume = {436},
    publisher = {Oxford Academic},
    url = {https://academic.oup.com/mnras/article/436/3/2188/1247446},
    doi = {10.1093/mnras/stt1722},
    issn = {00358711}
}

@article{Stanway2018,
    title = {{Re-evaluating old stellar populations}},
    year = {2018},
    journal = {Monthly Notices of the Royal Astronomical Society},
    author = {Stanway, E. R. and Eldridge, J. J.},
    number = {1},
    month = {9},
    pages = {75--93},
    volume = {479},
    publisher = {Oxford Academic},
    url = {https://dx.doi.org/10.1093/mnras/sty1353},
    doi = {10.1093/MNRAS/STY1353},
    issn = {0035-8711},
    arxivId = {1805.08784}
}

@article{Robinson2024,
    title = {{Regulating star formation in a magnetized disc galaxy}},
    year = {2024},
    journal = {Monthly Notices of the Royal Astronomical Society},
    author = {Robinson, Hector and Wadsley, James},
    number = {2},
    month = {9},
    pages = {1420--1432},
    volume = {534},
    publisher = {Oxford Academic},
    url = {https://dx.doi.org/10.1093/mnras/stae2132},
    doi = {10.1093/MNRAS/STAE2132},
    issn = {0035-8711},
    arxivId = {2310.15244}
}

@article{Deng2024,
    title = {{RIGEL: Simulating dwarf galaxies at solar mass resolution with radiative transfer and feedback from individual massive stars}},
    year = {2024},
    journal = {Astronomy {\&} Astrophysics},
    author = {Deng, Yunwei and Li, Hui and Liu, Boyuan and Kannan, Rahul and Smith, Aaron and Bryan, Greg L.},
    month = {11},
    pages = {A231},
    volume = {691},
    publisher = {EDP Sciences},
    url = {https://www.aanda.org/articles/aa/full_html/2024/11/aa50699-24/aa50699-24.html https://www.aanda.org/articles/aa/abs/2024/11/aa50699-24/aa50699-24.html},
    doi = {10.1051/0004-6361/202450699},
    issn = {14320746},
    arxivId = {2405.08869}
}

@article{Salem2016,
    title = {{Role of cosmic rays in the circumgalactic medium}},
    year = {2016},
    journal = {Monthly Notices of the Royal Astronomical Society},
    author = {Salem, Munier and Bryan, Greg L. and Corlies, Lauren},
    number = {1},
    month = {2},
    pages = {582--601},
    volume = {456},
    publisher = {Oxford University Press},
    url = {https://academic.oup.com/mnras/article/456/1/582/1066363},
    doi = {10.1093/mnras/stv2641},
    issn = {13652966},
    arxivId = {1511.05144}
}

@article{Weibel2025,
    title = {{RUBIES Reveals a Massive Quiescent Galaxy at z = 7.3}},
    year = {2025},
    journal = {The Astrophysical Journal},
    author = {Weibel, Andrea and Graaff, Anna de and Setton, David J. and Miller, Tim B. and Oesch, Pascal A. and Brammer, Gabriel and Lagos, Claudia D. P. and Whitaker, Katherine E. and Williams, Christina C. and Baggen, Josephine F.W. and Bezanson, Rachel and Boogaard, Leindert A. and Cleri, Nikko J. and Greene, Jenny E. and Hirschmann, Michaela and Hviding, Raphael E. and Kuruvanthodi, Adarsh and Labb{\'{e}}, Ivo and Leja, Joel and Maseda, Michael V. and Matthee, Jorryt and McConachie, Ian and Naidu, Rohan P. and Roberts-Borsani, Guido and Schaerer, Daniel and Suess, Katherine A. and Valentino, Francesco and Dokkum, Pieter van and Wang, Bingjie and 王, 冰洁},
    number = {1},
    month = {4},
    pages = {11},
    volume = {983},
    publisher = {IOP Publishing},
    url = {https://iopscience.iop.org/article/10.3847/1538-4357/adab7a https://iopscience.iop.org/article/10.3847/1538-4357/adab7a/meta},
    doi = {10.3847/1538-4357/adab7a},
    issn = {0004-637X},
    arxivId = {2409.03829}
}

@article{Kado-Fong2024,
    title = {{SAGAbg. I. A Near-unity Mass-loading Factor in Low-mass Galaxies via Their Low-redshift Evolution in Stellar Mass, Oxygen Abundance, and Star Formation Rate}},
    year = {2024},
    journal = {The Astrophysical Journal},
    author = {Kado-Fong, Erin and Geha, Marla and Mao, Yao-Yuan and Reyes, Mithi A. C. de los and Wechsler, Risa H. and Asali, Yasmeen and Kallivayalil, Nitya and Nadler, Ethan O. and Tollerud, Erik J. and Weiner, Benjamin},
    number = {1},
    month = {5},
    pages = {129},
    volume = {966},
    publisher = {IOP Publishing},
    url = {https://iopscience.iop.org/article/10.3847/1538-4357/ad3042 https://iopscience.iop.org/article/10.3847/1538-4357/ad3042/meta},
    doi = {10.3847/1538-4357/AD3042},
    issn = {0004-637X},
    arxivId = {2401.16469}
}

@article{Sun2023a,
    title = {{Seen and unseen: bursty star formation and its implications for observations of high-redshift galaxies with JWST}},
    year = {2023},
    journal = {Monthly Notices of the Royal Astronomical Society},
    author = {Sun, Guochao and Faucher-Gigu{\`{e}}re, Claude André and Hayward, Christopher C. and Shen, Xuejian},
    number = {2},
    month = {9},
    pages = {2665--2672},
    volume = {526},
    publisher = {Oxford Academic},
    url = {https://dx.doi.org/10.1093/mnras/stad2902},
    doi = {10.1093/MNRAS/STAD2902},
    issn = {0035-8711},
    arxivId = {2305.02713}
}

@article{Marasco2023,
    title = {{Shaken, but not expelled: Gentle baryonic feedback from nearby starburst dwarf galaxies}},
    year = {2023},
    journal = {Astronomy {\&} Astrophysics},
    author = {Marasco, A. and Belfiore, F. and Cresci, G. and Lelli, F. and Venturi, G. and Hunt, L. K. and Concas, A. and Marconi, A. and Mannucci, F. and Mingozzi, M. and McLeod, A. F. and Kumari, N. and Carniani, S. and Vanzi, L. and Ginolfi, M.},
    month = {2},
    pages = {A92},
    volume = {670},
    publisher = {EDP Sciences},
    url = {https://www.aanda.org/articles/aa/full_html/2023/02/aa44895-22/aa44895-22.html https://www.aanda.org/articles/aa/abs/2023/02/aa44895-22/aa44895-22.html},
    doi = {10.1051/0004-6361/202244895},
    issn = {0004-6361}
}

@article{Dubois2019,
    title = {{Shock-accelerated cosmic rays and streaming instability in the adaptive mesh refinement code Ramses}},
    year = {2019},
    journal = {Astronomy and Astrophysics},
    author = {Dubois, Yohan and Commer{\c{c}}on, Benoît and Marcowith, Alexandre and Brahimi, Loann},
    month = {11},
    pages = {A121},
    volume = {631},
    url = {https://www.aanda.org/10.1051/0004-6361/201936275},
    doi = {10.1051/0004-6361/201936275},
    issn = {14320746},
    arxivId = {1907.04300}
}

@article{Rathjen2023,
    title = {{SILCC – VII. Gas kinematics and multiphase outflows of the simulated ISM at high gas surface densities}},
    year = {2023},
    journal = {Monthly Notices of the Royal Astronomical Society},
    author = {Rathjen, Tim Eric and Naab, Thorsten and Walch, Stefanie and Seifried, Daniel and Girichidis, Philipp and W{\"{u}}nsch, Richard},
    number = {2},
    month = {4},
    pages = {1843--1862},
    volume = {522},
    publisher = {Oxford Academic},
    url = {https://dx.doi.org/10.1093/mnras/stad1104},
    doi = {10.1093/MNRAS/STAD1104},
    issn = {0035-8711},
    arxivId = {2211.15419}
}

@article{Dave2019,
    title = {{simba: Cosmological simulations with black hole growth and feedback}},
    year = {2019},
    journal = {Monthly Notices of the Royal Astronomical Society},
    author = {Dav{\'{e}}, Romeel and Angl{\'{e}}s-Alc{\'{a}}zar, Daniel and Narayanan, Desika and Li, Qi and Rafieferantsoa, Mika H. and Appleby, Sarah},
    number = {2},
    month = {6},
    pages = {2827--2849},
    volume = {486},
    publisher = {Oxford Academic},
    url = {https://dx.doi.org/10.1093/mnras/stz937},
    doi = {10.1093/MNRAS/STZ937},
    issn = {0035-8711},
    arxivId = {1901.10203}
}

@article{Pfrommer2017b,
    title = {{Simulating cosmic ray physics on a moving mesh}},
    year = {2017},
    journal = {Monthly Notices of the Royal Astronomical Society},
    author = {Pfrommer, C. and Pakmor, R. and Schaal, K. and Simpson, C. M. and Springel, V.},
    number = {4},
    month = {3},
    pages = {4500--4529},
    volume = {465},
    publisher = {Oxford University Press},
    url = {https://academic.oup.com/mnras/article-lookup/doi/10.1093/mnras/stw2941},
    doi = {10.1093/mnras/stw2941},
    issn = {0035-8711}
}

@article{Pfrommer2007,
    title = {{Simulating cosmic rays in clusters of galaxies - I. Effects on the Sunyaev-Zel'dovich effect and the X-ray emission}},
    year = {2007},
    journal = {Monthly Notices of the Royal Astronomical Society},
    author = {Pfrommer, C. and En{\ss}lin, T. A. and Springel, V. and Jubelgas, M. and Dolag, K.},
    number = {2},
    month = {6},
    pages = {385--408},
    volume = {378},
    publisher = {Blackwell Publishing Ltd},
    url = {https://dx.doi.org/10.1111/j.1365-2966.2007.11732.x},
    doi = {10.1111/J.1365-2966.2007.11732.X/2/M{\_}MNRAS0378-0385-M13.GIF},
    issn = {13652966}
}

@article{Weinberger2017,
    title = {{Simulating galaxy formation with black hole driven thermal and kinetic feedback}},
    year = {2017},
    journal = {Monthly Notices of the Royal Astronomical Society},
    author = {Weinberger, Rainer and Springel, Volker and Hernquist, Lars and Pillepich, Annalisa and Marinacci, Federico and Pakmor, Rüdiger and Nelson, Dylan and Genel, Shy and Vogelsberger, Mark and Naiman, Jill and Torrey, Paul},
    number = {3},
    month = {3},
    pages = {3291--3308},
    volume = {465},
    publisher = {Oxford Academic},
    url = {https://dx.doi.org/10.1093/mnras/stw2944},
    doi = {10.1093/mnras/stw2944},
    issn = {13652966},
    arxivId = {1607.03486}
}

@article{Pfrommer2017a,
    title = {{Simulating Gamma-Ray Emission in Star-forming Galaxies}},
    year = {2017},
    journal = {The Astrophysical Journal},
    author = {Pfrommer, Christoph and Pakmor, Rüdiger and Simpson, Christine M. and Springel, Volker},
    number = {2},
    month = {9},
    pages = {L13},
    volume = {847},
    publisher = {American Astronomical Society},
    url = {https://doi.org/10.3847/2041-8213/aa8bb1},
    doi = {10.3847/2041-8213/aa8bb1},
    issn = {2041-8213},
    arxivId = {1709.05343}
}

@article{Marinacci2019,
    title = {{Simulating the interstellar medium and stellar feedback on a moving mesh: implementation and isolated galaxies}},
    year = {2019},
    journal = {Monthly Notices of the Royal Astronomical Society},
    author = {Marinacci, Federico and Sales, Laura V. and Vogelsberger, Mark and Torrey, Paul and Springel, Volker},
    number = {3},
    month = {11},
    pages = {4233--4260},
    volume = {489},
    publisher = {Oxford Academic},
    url = {https://dx.doi.org/10.1093/mnras/stz2391},
    doi = {10.1093/mnras/stz2391},
    issn = {13652966},
    arxivId = {1905.08806}
}

@article{Girichidis2022,
    title = {{Spectrally resolved cosmic rays – II. Momentum-dependent cosmic ray diffusion drives powerful galactic winds}},
    year = {2022},
    journal = {Monthly Notices of the Royal Astronomical Society},
    author = {Girichidis, Philipp and Pfrommer, Christoph and Pakmor, Rüdiger and Springel, Volker},
    number = {3},
    month = {1},
    pages = {3917--3938},
    volume = {510},
    publisher = {Oxford Academic},
    url = {https://academic.oup.com/mnras/article/510/3/3917/6446005},
    doi = {10.1093/mnras/stab3462},
    issn = {0035-8711},
    arxivId = {2109.13250}
}

@article{Bryan1998,
    title = {{Statistical Properties of X‐Ray Clusters: Analytic and Numerical Comparisons}},
    year = {1998},
    journal = {The Astrophysical Journal},
    author = {Bryan, Greg L. and Norman, Michael L.},
    number = {1},
    month = {3},
    pages = {80--99},
    volume = {495},
    publisher = {IOP Publishing},
    url = {http://stacks.iop.org/0004-637X/495/i=1/a=80},
    doi = {10.1086/305262},
    issn = {0004-637X}
}

@article{Hopkins2012a,
    title = {{Stellar feedback in galaxies and the origin of galaxy-scale winds}},
    year = {2012},
    journal = {Monthly Notices of the Royal Astronomical Society},
    author = {Hopkins, Philip F. and Quataert, Eliot and Murray, Norman},
    number = {4},
    month = {4},
    pages = {3522--3537},
    volume = {421},
    publisher = {Narnia},
    url = {https://academic.oup.com/mnras/article-lookup/doi/10.1111/j.1365-2966.2012.20593.x},
    doi = {10.1111/j.1365-2966.2012.20593.x},
    issn = {00358711}
}

@article{Stanway2016,
    title = {{Stellar population effects on the inferred photon density at reionization}},
    year = {2016},
    journal = {Monthly Notices of the Royal Astronomical Society},
    author = {Stanway, Elizabeth R. and Eldridge, J. J. and Becker, George D.},
    number = {1},
    month = {2},
    pages = {485--499},
    volume = {456},
    publisher = {Oxford University Press},
    url = {https://academic.oup.com/mnras/article/456/1/485/1068024},
    doi = {10.1093/mnras/stv2661},
    issn = {13652966},
    arxivId = {1511.03268}
}

@article{Martin-Alvarez2025,
    title = {{Stirring the cosmic pot: how black hole feedback shapes the matter power spectrum in the fable simulations}},
    year = {2025},
    journal = {Monthly Notices of the Royal Astronomical Society},
    author = {Martin-Alvarez, Sergio and Ir{\v{s}}i{\v{c}}, Vid and Koudmani, Sophie and Bourne, Martin A and Bigwood, Leah and Sijacki, Debora},
    number = {2},
    month = {4},
    pages = {1738--1755},
    volume = {539},
    publisher = {Oxford Academic},
    url = {https://dx.doi.org/10.1093/mnras/staf470},
    doi = {10.1093/MNRAS/STAF470},
    issn = {0035-8711}
}

@article{Iyer2024,
    title = {{Stochastic Modeling of Star Formation Histories. III. Constraints from Physically Motivated Gaussian Processes}},
    year = {2024},
    journal = {The Astrophysical Journal},
    author = {Iyer, Kartheik G and Speagle, Joshua S and Caplar, Neven and Forbes, John C and Gawiser, Eric and Leja, Joel and Tacchella, Sandro and Dunlap, David A},
    number = {1},
    month = {1},
    pages = {53},
    volume = {961},
    publisher = {IOP Publishing},
    url = {https://iopscience.iop.org/article/10.3847/1538-4357/acff64 https://iopscience.iop.org/article/10.3847/1538-4357/acff64/meta},
    doi = {10.3847/1538-4357/ACFF64},
    issn = {0004-637X},
    arxivId = {2208.05938}
}

@article{Carvajal-Bohorquez2025,
    title = {{Stochastic star formation activity of galaxies within the first billion years probed by JWST}},
    year = {2025},
    journal = {Astronomy {\&} Astrophysics},
    author = {Carvajal-Bohorquez, C. and Ciesla, L. and Laporte, N. and Boquien, M. and Buat, V. and Ilbert, O. and Aufort, G. and Shuntov, M. and Witten, C. and Oesch, P. A. and Covelo-Paz, A.},
    month = {12},
    pages = {A290},
    volume = {704},
    publisher = {EDP Sciences},
    url = {https://www.aanda.org/articles/aa/full_html/2025/12/aa56471-25/aa56471-25.html https://www.aanda.org/articles/aa/abs/2025/12/aa56471-25/aa56471-25.html},
    doi = {10.1051/0004-6361/202556471},
    issn = {14320746},
    arxivId = {2507.13160}
}

@article{Boylan-Kolchin2023,
    title = {{Stress testing {$\Lambda$}CDM with high-redshift galaxy candidates}},
    year = {2023},
    journal = {Nature Astronomy},
    author = {Boylan-Kolchin, Michael},
    number = {6},
    month = {4},
    pages = {731--735},
    volume = {7},
    publisher = {Nature Publishing Group},
    url = {https://www.nature.com/articles/s41550-023-01937-7},
    doi = {10.1038/s41550-023-01937-7},
    issn = {23973366},
    arxivId = {2208.01611}
}

@article{Morlino2012,
    title = {{Strong evidence for hadron acceleration in Tycho's supernova remnant}},
    year = {2012},
    journal = {Astronomy and Astrophysics},
    author = {Morlino, G. and Caprioli, D.},
    month = {2},
    pages = {A81},
    volume = {538},
    publisher = {EDP Sciences},
    url = {http://www.aanda.org/10.1051/0004-6361/201117855},
    doi = {10.1051/0004-6361/201117855},
    issn = {00046361}
}

@article{Efstathiou1992,
    title = {{Suppressing the formation of dwarf galaxies via photoionization}},
    year = {1992},
    journal = {Monthly Notices of the Royal Astronomical Society},
    author = {Efstathiou, G.},
    number = {1},
    month = {5},
    pages = {43P-47P},
    volume = {256},
    publisher = {Oxford Academic},
    url = {https://academic.oup.com/mnras/article/256/1/43P/1025191},
    doi = {10.1093/mnras/256.1.43P},
    issn = {13652966}
}

@article{Steinhardt2023,
    title = {{Templates for Fitting Photometry of Ultra-high-redshift Galaxies}},
    year = {2023},
    journal = {The Astrophysical Journal Letters},
    author = {Steinhardt, Charles L. and Kokorev, Vasily and Rusakov, Vadim and Garcia, Ethan and Sneppen, Albert},
    number = {2},
    month = {7},
    pages = {L40},
    volume = {951},
    publisher = {IOP Publishing},
    url = {https://iopscience.iop.org/article/10.3847/2041-8213/acdef6 https://iopscience.iop.org/article/10.3847/2041-8213/acdef6/meta},
    doi = {10.3847/2041-8213/ACDEF6},
    issn = {2041-8205},
    arxivId = {2208.07879}
}

@article{Mason2023,
    title = {{The brightest galaxies at cosmic dawn}},
    year = {2023},
    journal = {Monthly Notices of the Royal Astronomical Society},
    author = {Mason, Charlotte A. and Trenti, Michele and Treu, Tommaso},
    number = {1},
    month = {3},
    pages = {497--503},
    volume = {521},
    publisher = {Oxford Academic},
    url = {https://dx.doi.org/10.1093/mnras/stad035},
    doi = {10.1093/MNRAS/STAD035},
    issn = {0035-8711},
    arxivId = {2207.14808}
}

@article{Endsley2025,
    title = {{The Burstiness of Star Formation at z ∼ 6: A Huge Diversity in the Recent Star Formation Histories of Very UV-faint Galaxies}},
    year = {2025},
    journal = {The Astrophysical Journal},
    author = {Endsley, Ryan and Chisholm, John and Stark, Daniel P. and Topping, Michael W. and Whitler, Lily},
    number = {2},
    month = {7},
    pages = {189},
    volume = {987},
    publisher = {IOP Publishing},
    url = {https://iopscience.iop.org/article/10.3847/1538-4357/addc74 https://iopscience.iop.org/article/10.3847/1538-4357/addc74/meta},
    doi = {10.3847/1538-4357/addc74},
    issn = {15384357},
    arxivId = {2410.01905}
}

@article{Calzetti2007,
    title = {{The Calibration of Mid-Infrared Star Formation Rate Indicators*}},
    year = {2007},
    journal = {The Astrophysical Journal},
    author = {Calzetti, D. and Kennicutt, R. C. and Engelbracht, C. W. and Leitherer, C. and Draine, B. T. and Kewley, L. and Moustakas, J. and Sosey, M. and Dale, D. A. and Gordon, K. D. and Helou, G. X. and Hollenbach, D. J. and Armus, L. and Bendo, G. and Bot, C. and Buckalew, B. and Jarrett, T. and Li, A. and Meyer, M. and Murphy, E. J. and Prescott, M. and Regan, M. W. and Rieke, G. H. and Roussel, H. and Sheth, K. and Smith, J. D. T. and Thornley, M. D. and Walter, F.},
    number = {2},
    month = {9},
    pages = {870},
    volume = {666},
    publisher = {IOP Publishing},
    url = {https://iopscience.iop.org/article/10.1086/520082 https://iopscience.iop.org/article/10.1086/520082/meta},
    doi = {10.1086/520082},
    issn = {0004-637X},
    arxivId = {0705.3377}
}

@article{Curti2023,
    title = {{The chemical enrichment in the early Universe as probed by JWST via direct metallicity measurements at z ∼ 8}},
    year = {2022},
    journal = {Monthly Notices of the Royal Astronomical Society},
    author = {Curti, Mirko and D'Eugenio, Francesco and Carniani, Stefano and Maiolino, Roberto and Sandles, Lester and Witstok, Joris and Baker, William M. and Bennett, Jake S. and Piotrowska, Joanna M. and Tacchella, Sandro and Charlot, Stephane and Nakajima, Kimihiko and Maheson, Gabriel and Mannucci, Filippo and Amiri, Amirnezam and Arribas, Santiago and Belfiore, Francesco and Bonaventura, Nina R. and Bunker, Andrew J. and Chevallard, Jacopo and Cresci, Giovanni and Curtis-Lake, Emma and Hayden-Pawson, Connor and Jones, Gareth C. and Kumari, Nimisha and Laseter, Isaac and Looser, Tobias J. and Marconi, Alessandro and Maseda, Michael V. and Scholtz, Jan and Smit, Renske and {\"{U}}bler, Hannah and Wallace, Imaan E.B.},
    number = {1},
    month = {11},
    pages = {425--438},
    volume = {518},
    publisher = {Oxford Academic},
    url = {https://dx.doi.org/10.1093/mnras/stac2737},
    doi = {10.1093/MNRAS/STAC2737},
    issn = {0035-8711},
    arxivId = {2207.12375}
}

@article{Tumlinson2017,
    title = {{The Circumgalactic Medium}},
    year = {2017},
    journal = {Annual Review of Astronomy and Astrophysics},
    author = {Tumlinson, Jason and Peeples, Molly S. and Werk, Jessica K.},
    number = {Volume 55, 2017},
    month = {8},
    pages = {389--432},
    volume = {55},
    publisher = {Annual Reviews Inc.},
    url = {https://www.annualreviews.org/content/journals/10.1146/annurev-astro-091916-055240},
    doi = {10.1146/ANNUREV-ASTRO-091916-055240/1},
    issn = {00664146},
    arxivId = {1709.09180}
}

@article{Kennicutt1998,
    title = {{The Global Schmidt Law in Star‐forming Galaxies}},
    year = {1998},
    journal = {The Astrophysical Journal},
    author = {Kennicutt, Jr., Robert C.},
    number = {2},
    month = {5},
    pages = {541--552},
    volume = {498},
    publisher = {IOP Publishing},
    url = {http://stacks.iop.org/0004-637X/498/i=2/a=541},
    doi = {10.1086/305588},
    issn = {0004-637X}
}
\bibliographystyle{aasjournalv7}

\end{document}